\documentclass{nature_sub}

\usepackage{graphicx}
\usepackage{amsmath}
\usepackage{amssymb}
\usepackage{color}
\usepackage{hyperref}

\usepackage{multibib}

\newcites{main,meth}%
         {,%
          }
\bibliographystylemain{naturemag}
\bibliographystylemeth{naturemag}

\title{Collapsars as a major source of r-process elements}

\author{Daniel M.~Siegel$^{1,2,3,4,5}$, Jennifer Barnes$^{1,2,5}$ \& Brian D.~Metzger$^{1,2}$}

\begin{document}

\maketitle

\begin{affiliations}
 \item Department of Physics, Columbia University, New York, NY, USA
 \item Columbia Astrophysics Laboratory, Columbia University, New York, NY, USA
 \item Perimeter Institute for Theoretical Physics, Waterloo, Ontario, Canada
 \item Department of Physics, University of Guelph, Guelph, Ontario, Canada
 \item NASA Einstein Fellow
\end{affiliations}

\begin{abstract}
The production of elements by rapid neutron capture (r-process) in neutron-star mergers is expected theoretically and is supported by multimessenger observations\citemain{LIGO+17DISCOVERY,Coulter2017,Soares-Santos+17} of gravitational-wave event GW170817: this production route is in principle sufficient to account for most of the r-process elements in the Universe\citemain{Kasen+17}. Analysis of the kilonova that accompanied GW170817 identified\citemain{Cowperthwaite2017,Radice2018a} delayed outflows from a remnant accretion disk formed around the newly born black hole\citemain{Fernandez&Metzger13,Just+15,Siegel2017a,Fernandez+18} as the dominant source of heavy r-process material from that event\citemain{Siegel2017a,Siegel2018a}. Similar accretion disks are expected to form in collapsars (the supernova-triggering collapse of rapidly rotating massive stars), which have previously been speculated to produce r-process elements\citemain{MacFadyen&Woosley99,Kohri+05}. Recent observations of stars rich in such elements in the dwarf galaxy Reticulum II\citemain{Ji+16}, as well as the Galactic chemical enrichment of europium relative to iron over longer timescales\citemain{Cote2017b,Hotokezaka2018a}, are more consistent with rare supernovae acting at low stellar metallicities than with neutron-star mergers. Here we report simulations that show that collapsar accretion disks yield sufficient r-process elements to explain observed abundances in the Universe. Although these supernovae are rarer than neutron- star mergers, the larger amount of material ejected per event compensates for the lower rate of occurrence. We calculate that collapsars may supply more than 80 per cent of the r-process content of the Universe.
\end{abstract}

\begin{figure}
\centering
\includegraphics[width=0.85\linewidth]{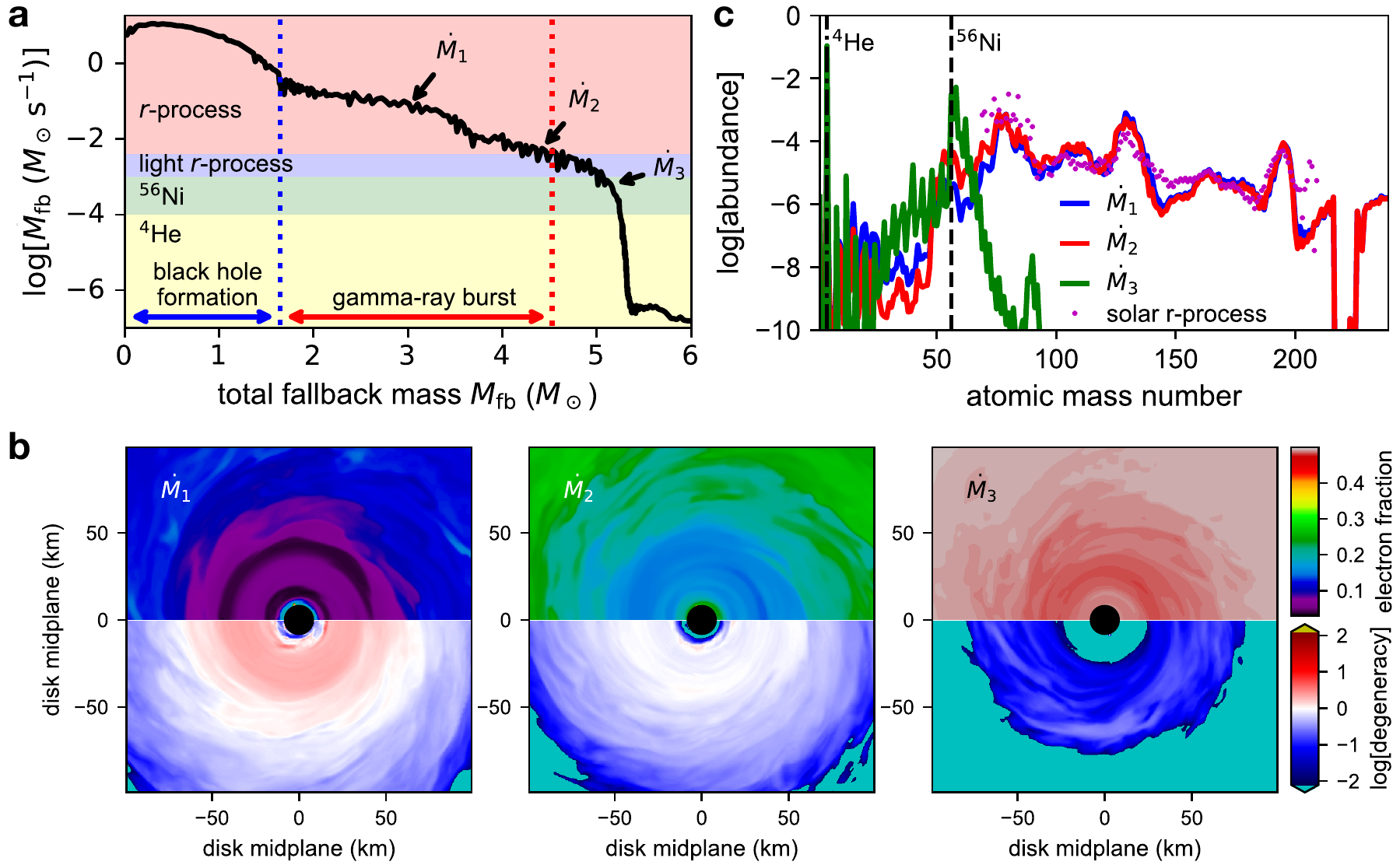}
\caption{{\bf Various stages of collapsar accretion and nucleosynthetic yields.} {\bf a}, Typical collapsar fallback accretion rate ($\dot{M}_{\rm fb}$, on a logarithmic scale) versus accreted mass, with arrows indicating the accretion stages $\dot{M}_1,\dot{M}_2,\dot{M}_3$ simulated here (see Methods). Vertical dotted lines indicate initial black hole formation and the part of the accretion process powering the gamma-ray burst. Horizontal bands indicate the different nucleosynthetic regimes of the disk outflows as identified from the simulations, and are labelled on the left. {\bf b}, Simulation snapshots of the disk's equatorial plane for the three different accretion stages, showing that above a critical threshold of $\dot{M}_{\rm ign}\approx 10^{-3}\,M_\odot {\rm s}^{-1}$ mild electron degeneracy $(\eta=\mu_e/kT\approx 1$; lower half of plots, colour scale on the right) is established, which drives the disk midplane neutron-rich (proton fraction $Y_e\ll 0.5$; upper half of plots, colour scale on the right). {\bf c}, Abundance distributions of nuclei synthesized in the disk outflows at the three different accretion stages (see key; dots represent the observed Solar System abundances). Above $\dot{M}_{\rm ign}$, a heavy r-process up to atomic mass numbers of around 195 is obtained ($\dot{M}_1,\dot{M_2}$), whereas below $\dot{M}_{\rm ign}$ a rapid transition to outflows rich in $^{56}$Ni and $^4$He is observed ($\dot{M}_3$).}
\label{fig:overview}
\end{figure}

\begin{figure}
\centering
\includegraphics[width=0.5\linewidth]{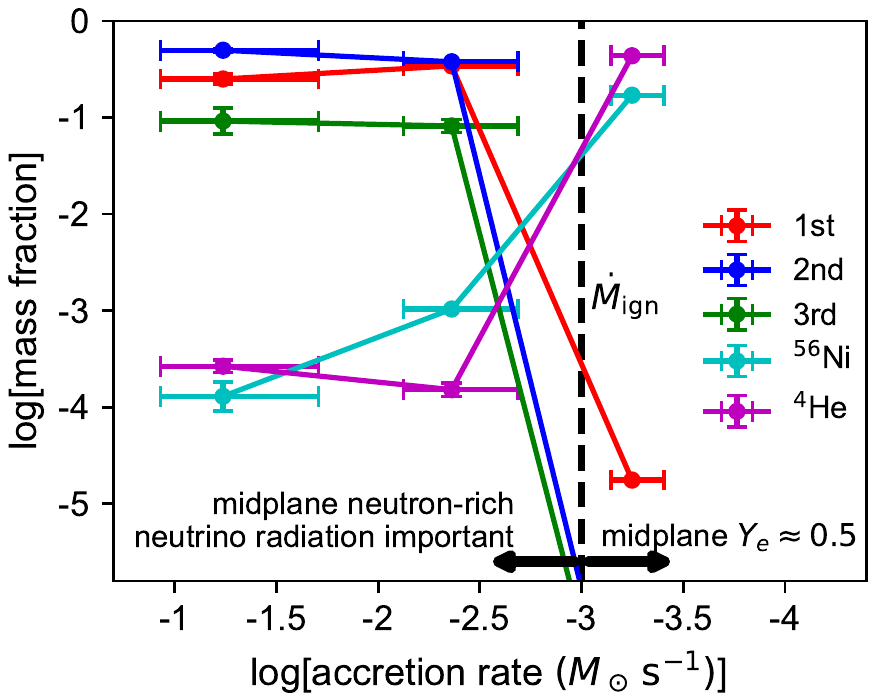}
\caption{{\bf Nucleosynthesis yields at the simulated collapsar accretion stages.} Mass fractions for first-peak (red; atomic mass number $A\approx 80$), second-peak (blue; $A\approx 130$), and third-peak (green; $A\approx 195$) r-process nuclei synthesized in the disk outflows are shown, as well as $^{56}$Ni (cyan) and helium ($^4$He; purple) mass fractions, for the three accretion regimes $\dot{M_1}$, $\dot{M}_2$, and $\dot{M}_3$ (from left to right). A sharp transition from a heavy r-process regime to $^{56}$Ni-rich outflows around a characteristic ignition threshold $\dot{M}_{\rm ign}\approx 10^{-3}\,M_\odot{\rm s}^{-1}$ (vertical dashed line) is apparent (see Methods). Uncertainties in the nucleosynthetic yields show the range of values obtained by using two different treatments of neutrino emission (see Methods). Uncertainties in the accretion rate are bracketed by the range of values attained during the simulation period used to monitor the accretion rate (see Methods; Extended Data Fig.~3).}
\label{fig:Xs}
\end{figure}

We have performed a suite of three-dimensional, general-relativistic magnetohydrodynamic (MHD) simulations of neutrino-cooled accretion disks with initial nuclear compositions characteristic of collapsars (Fig.~\ref{fig:overview}) in order to quantify the nucleosynthesis products of the unbound outflows from the disk (see Methods). We consider three different disk masses, in order to explore the outflow properties across the range of accretion rates experienced at different epochs in the collapsar disk evolution. We post-process the thermodynamic trajectories of tracer particles tracking the unbound matter using an r-process reaction network to determine their detailed nucleosynthetic yields, accounting for the absorption of electron neutrinos and antineutrinos from the disk on the proton fraction of the winds. For black hole accretion rates in the range $>\!0.003-0.1\,M_{\odot}{\rm s}^{-1}$ needed to explain the observed energetics and timescales of long gamma-ray bursts (GRBs; commonly invoked as resulting from relativistic jets driven by collapsar accretion\citemain{MacFadyen&Woosley99}), we find that the wind ejecta are neutron-rich and robustly synthesize both light and heavy r-process nuclei, extending up to the third abundance peak at atomic mass number $A \sim 195$ (Figs.~\ref{fig:overview}, \ref{fig:Xs}; see Methods). Previous studies of collapsar accretion disks did not find the synthesis of such nuclei in disk outflows either because simulations assumed equal number of protons and neutrons\citemain{MacFadyen&Woosley99} instead of self-consistently evolving the proton fraction under charged-current weak interactions, or parametrized models assumed the disk winds to be entirely neutrino-driven, in which case outflowing matter absorbed many electron neutrinos generated from the disk and experienced high rates of positron captures\citemain{Pruet+04,Surman+11} (see Methods).

Although the infalling progenitor star is comprised of approximately equal numbers of neutrons and protons, matter is driven once in the disk midplane to a neutron-rich state (proton fraction $\ll\!0.5$) by electron capture reactions on protons. (The proton fraction is defined as the ratio in number densities of protons to all baryons---protons and neutrons---and, for historical reasons, is referred to as the electron fraction, $Y_e$.) For sufficiently high accretion rates (see Methods), neutrino cooling regulates the electron chemical potential $\mu_e$ in the midplane to a mildly degenerate state ($\eta=\mu_e/k_{\rm B}T \approx 1$, where $k_{\rm B}$ is the Boltzmann constant and $T$ is the temperature; Fig.~\ref{fig:overview}); this electron degeneracy suppresses positron creation and thus reduces the opposing rate of positron captures on neutrons\citemain{Beloborodov03}. The large midplane neutron excess is preserved in the disk outflows, which expand to large radii sufficiently rapidly to avoid substantial neutron destruction by electron neutrinos. Neutrinos will have a much more pronounced effect during earlier phases following collapse, while the hot proto-neutron star is still present prior to black hole formation\citemain{Dessart+08,Moesta2018}; however, only a small fraction of the total accreted mass, and thus of the wind ejecta, occurs during this phase.  

Given their broadly similar physical conditions, it is reasonable to expect that the fraction of initially inflowing mass that becomes unbound from the inner region of collapsar disks is similar to that in neutron star mergers.  However, the total accreted mass in collapsars is typically $\sim\!30$ times larger than in mergers, as expected on theoretical grounds and supported empirically by the similarly larger observed jet energies of long GRBs compared to short GRBs\citemain{Ghirlanda+09} (the isotropic gamma-ray luminosities of the burst classes are comparable). The disk wind ejecta $\sim\!0.03-0.06M_{\odot}$ inferred from the GW170817 kilonova then translates into an average collapsar r-process yield of up to $\sim\!1M_{\odot}$ (see Methods).  We reach a similar conclusion by comparing a toy model for the mass accreted at different rates during the collapsar evolution to the range of accretion rates our disk simulations show give rise to r-process ejecta (see Methods). The collapsar and neutron star merger scenarios differ in that the collapsar disk outflows encounter ram pressure from infalling material of the stellar envelope; however, simple analytic arguments as well as previous simulations\citemain{MacFadyen&Woosley99} show that under typical conditions these outflows are powerful enough to escape (see Methods). The larger outflow yield per collapsar is more than sufficient to make up for their lower cosmically-averaged rate as compared to neutron star mergers, thus implicating collapsars as the dominant site of the second and third-peak r-process in our Galaxy and solar system (see Methods).

Observational signatures of the r-process in collapsars should be present in their accompanying supernovae. The luminosities of GRB supernovae are powered by the large quantity $\approx\!0.2-0.5\,M_{\odot}$ of radioactive nickel ($^{56}$Ni), synthesized mainly during the initial explosion itself. A moderate quantity $\sim\!10^{-3}-0.1M_{\odot}$ of $^{56}$Ni can be generated from the collapsar disk winds at late-times, but only once the accretion rate drops below the r-process threshold $<\!10^{-3}\,M_{\odot} {\rm s}^{-1}$ (see Methods). Near peak supernova light, the radioactive heating rate of $^{56}$Ni greatly exceeds that of r-process nuclei, making it possible to ``hide" the latter in GRB supernova light curves and spectra, depending on how efficiently the r-process products are mixed outwards into the high-velocity layers of the explosion (Fig.~3, see Methods). The presence of high-opacity lanthanide elements\citemain{Kasen2013} deeper within the ejecta could nevertheless be visible as excess near-infrared (NIR) emission at late times (Fig.~3). Although such an excess may, in principle, be degenerate with other effects, the presence of r-process elements may be revealed in combination with NIR spectra and future observations of neutron star mergers. Once the unique late NIR signatures of r-process elements are pinned down empirically by observations of future neutron star mergers (``pure" r-process sources), e.g., by the James Webb Space Telescope, similar (but narrower, due to the lower ejecta velocity) line features could be sought in late-time spectra of GRB supernovae. Indeed, the early-time `MHD supernova' phase of the explosion\citemain{Winteler+12} is probably already ruled out as a heavy r-process site; unlike in the delayed disk wind scenario described here, high-opacity r-process material generated during the explosion phase would necessarily be mixed to high velocities with the $^{56}$Ni in a way that would be incompatible with present observations of GRB supernovae (Fig.~3).  

Collapsars as the dominant sources of the Galactic r-process help alleviate several of the observational challenges for neutron star merger models. It remains a long-standing question whether the average delay required for binary neutron stars to coalesce after star formation is sufficiently short to explain the high r-process abundances in metal-poor halo stars, i.e. those polluted by just a few generations of stars. Though this tension may be relieved by more consistent Galactic chemical evolution histories which account for cosmic structure growth\citemain{Shen+15,vandeVoort+15}, or considering alternative formation channels at high redshift\citemain{RamirezRuiz+15}, the issue remains unsettled.  Collapsars, by contrast, occur primarily in low-metallicity environments\citemain{Stanek+06} and thus would be over-represented among the first generations of stars, providing a natural explanation for the observed carbon-enhanced metal-poor stars with high r-process enrichment\citemain{Sneden+03}. 

The most direct current evidence for a single r-process event comes from the ultra-faint dwarf galaxy Reticulum II, which was polluted early in its history by a rare, high-yield source\citemain{Ji+16}.  Though a neutron star merger indeed provides a high r-process yield, the supernova explosions giving birth to the two neutron stars would need to impart them with small natal kicks\citemain{Beniamini+16} in order to retain the binary in such a tiny host galaxy; the merger delay time would also need to be sufficiently short $<\!100\,{\rm Myr}$ compared to the brief timescale over which the stellar population was formed.  Collapsars, by contrast, originate from the core collapse of very massive stars, within a few Myr following star formation.  

Finally, the observed growth with metallicity of the abundance of the r-process element europium as compared to iron in Galactic stars appears in tension with the main r-process source originating from a source population with the $\propto t^{-1}$ delay time distribution expected for neutron star mergers\citemain{Cote2017b,Hotokezaka2018a}. However, we find that the observed europium versus iron evolution results naturally if the main r-process source directly tracks the star-formation rate (growth of $\alpha$-abundances), as would a rare subset of core collapse supernovae like collapsars (see Methods).


\begin{addendum}
 \item Resources supporting this work were provided by the NASA High-End Computing (HEC) Program through the NASA Advanced Supercomputing (NAS) Division at Ames Research Center. Support for this work was provided by the National Aeronautics and Space Administration through Einstein Postdoctoral Fellowships Award Numbers PF6-170159 and PF7-180162 issued by the Chandra X-ray Observatory Center, which is operated by the Smithsonian Astrophysical Observatory for and on behalf of the National Aeronautics Space Administration under contract NAS8-03060. D.M.S. acknowledges the support of the Natural Sciences and Engineering Research Council of Canada (NSERC). Research at Perimeter Institute is supported in part by the Government of Canada through the Department of Innovation, Science and Economic Development Canada and by the Province of Ontario through the Ministry of Colleges and Universities. B.D.M. acknowledges support from the National Aeronautics \& Space Administration, through the Astrophysics Theory Program (NNX16AB30G).
\end{addendum}

\bibliographymain{manuscript}

\newpage

\begin{figure}
\centering
\includegraphics[width=0.7\linewidth]{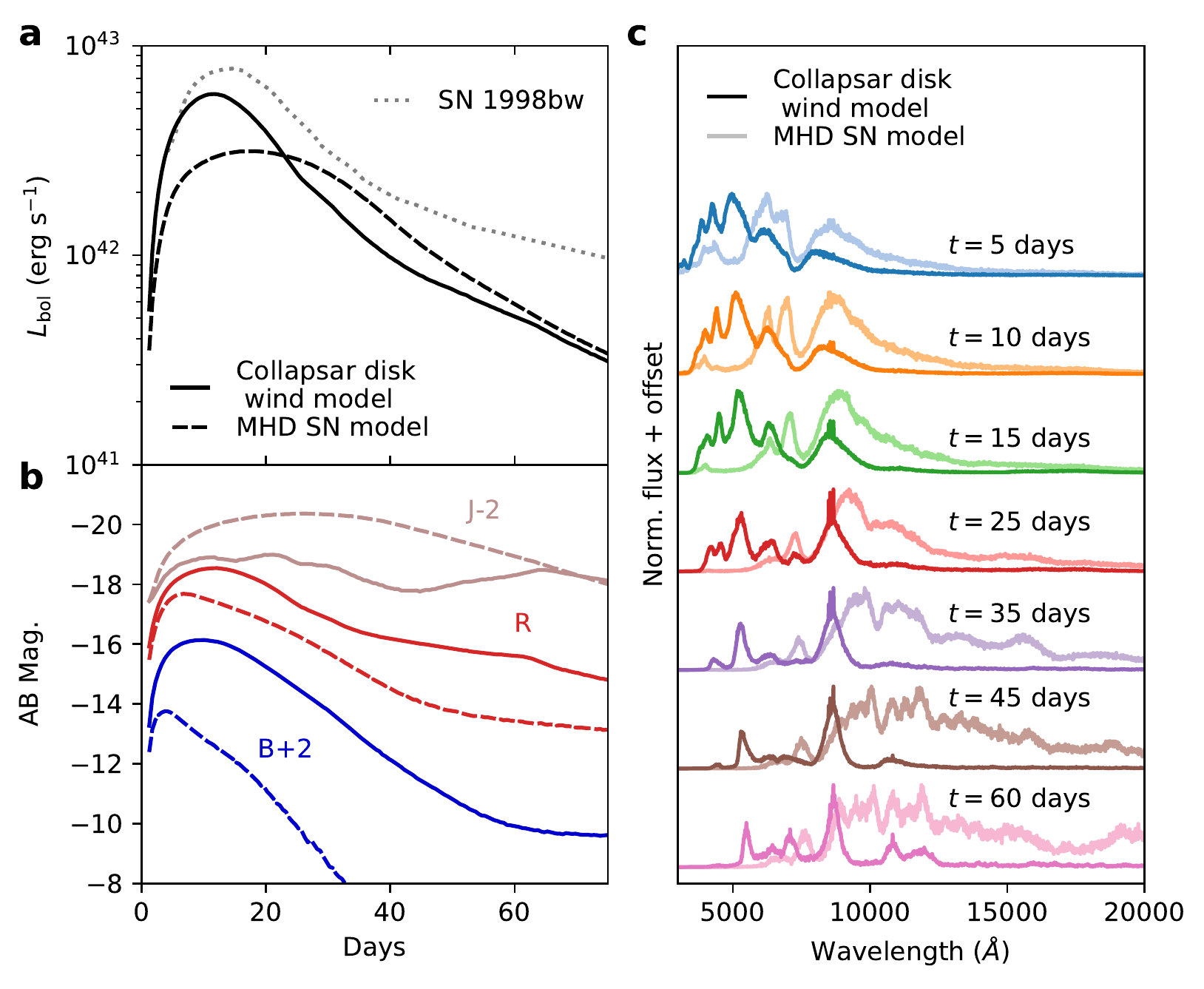}
\caption{{\bf The signatures of r-process nucleosynthesis in GRB supernovae.} Shown are synthetic photometry and spectra for models of r-process-enriched supernovae corresponding to an MHD supernova and a collapsar disk outflow scenario. Light curves and spectra depend sensitively on the outward mixing of r-process material with radioactive nickel $^{56}$Ni generated during the early explosion phase. In the MHD supernova case, the distribution of r-process material tracks the distribution of $^{56}$Ni, whereas in the collapsar wind model, the r-process material is embedded inside the $^{56}$Ni-rich ejecta. {\bf a}, Bolometric light curves for the two models, compared
to the bolometric light curve of SN 1998bw, a well sampled reference light curve for GRB supernovae. {\bf b}, Selected broadband light curves: J-band (offset by two magnitudes, `$J-2$'), R-band, B-band (offset by two magnitudes,
`B+2'). {\bf c}, Spectral evolution of both systems. At each time $t$, the curves coloured dark and light show the output of the collapsar disk wind model and the MHD supernova model, respectively. The spectra at 10 days correspond roughly to B-band maximum magnitude for the collapsar disk wind model.}
\label{fig:snRT}
\end{figure}

\clearpage

\begin{methods}

\subsection{Details of GRMHD disk simulations.} 

Simulations are performed with the general-relativistic magnetohydrodynamics (GRMHD) code described previously\citemain{Siegel2018a}$^,$\citemeth{Siegel2018b}, including weak interactions and approximate neutrino transport via a leakage scheme. It is based on the GRMHD code GRHydro\citemeth{Moesta2014a} and the Einstein Toolkit\citemeth{Loeffler2012} (http://einsteintoolkit.org). Neutrino emissivities are computed from charged-current $\beta$-processes, electron--positron pair annihilation, and plasmon decay. Neutrino opacities include absorption of electron and anti-electron neutrinos by nucleons, as well as coherent scattering on heavy nuclei and free nucleons. The microphysical description of matter is based on the Helmholtz equation of state\citemeth{Timmes1999,Timmes2000} (EOS), which includes nuclei as ideal gas with Coulomb corrections, electrons and positrons with an arbitrary degree of relativity and degeneracy, and photons in local thermodynamic equilibrium. The set of nuclei employed in our simulations comprises free neutrons and protons, and $\alpha$-particles, whose abundances at given rest-mass density $\rho$, temperature $T$, and electron fraction $Y_e$ are computed assuming nuclear statistical equilibrium. We include the nuclear binding energy release from $\alpha$-particle formation, which plays an important role in setting the asymptotic velocities of unbound outflows\citemain{Siegel2018a}. The choice of the Helmholtz EOS also minimizes thermodynamic inconsistencies between the simulations and the post-processing of tracer particles with the nuclear-reaction network (see below).

In order to investigate the accretion process of collapsar accretion disks, we perform three simulations---referred to as Run 1, 2, and 3---which capture the accretion process at different accretion rates, i.e., at different stages and times following core collapse. Performing a global 3D simulation of the entire accretion process over the relevant timescales including the fallback material from the stellar envelope are currently computationally prohibitive at this level of physical detail.

Initial data is obtained by relaxing axisymmetric equilibrium torus configurations around a rotating black hole, with constant specific angular momentum, a small constant specific entropy $s_0=8\,k_\mathrm{B}$ per baryon, and a constant electron fraction $Y_{\mathrm{e},0}=0.5$ corresponding to equal numbers of protons and neutrons for the nearly symmetric matter of the collapsing progenitor star.  Our results are not expected to be sensitive to the precise value of the initial entropy of the disk; its value is rapidly changed by turbulent heating and neutrino cooling, which come into balance for the highest accretion rates relevant to neutron-rich outflows\citemeth{DiMatteo+02}.  The black hole has a mass $M_\mathrm{BH}=3\,M_\odot$, similar to those expected in collapsars, and dimensionless spin $a=0.8$ as predicted by presupernova stellar models of collapsar progenitors (see model on mass fallback below). Runs 1, 2, and 3 correspond to initial torus masses of $M_{t,0}=(0.02,0.0016, 0.00016)\,M_\odot$, respectively. Weak magnetic seed fields are superimposed on the tori matter initially, defined by the vector potential $A^r=A^\theta = 0$, and $A^\phi =A_b\, \mathrm{max}\{p-p_\mathrm{cut},0\}$. Here, $p_\mathrm{cut}=1.3\times 10^{-2}p_\mathrm{max}$, where $p_\mathrm{max}$ is the pressure at maximum density in the respective torus. The parameter $A_b$ is adjusted such that the maximum magnetic-to-fluid pressure ratio in the tori is $<5\times 10^{-3}$. The tori are initially embedded in a uniform atmosphere with $T=10^6\,\mathrm{K}$, $Y_e=1$, and $\rho = (37, 3.7, 0.37)\,\mathrm{g}\,\mathrm{cm}^{-3}$ for Runs 1, 2, and 3, respectively, scaling the atmosphere density floor according to the decreasing maximum density of the tori. These atmosphere conditions are sufficiently cold and tenuous to impact neither the dynamics nor the composition of the disk outflows. We neglect any impact of the infalling star on the disk structure, motivated by the fact that, once the disk and outflows are established at the relatively late times of interest, the momentum flux of the disk outflows greatly exceed the ram pressure of the infalling stellar debris (see ram pressure of fallback material below). The relaxed, stationary state after 30\,ms of evolution effectively serves as initial data for the simulations; this initial period as well as all material ejected from the tori during the first 30\,ms is excluded from all further analysis.

We perform simulations in fixed spacetime and horizon-penetrating Kerr-Schild coordinates. The computational domain consists of a Cartesian grid hierarchy centered on the black hole with seven (Runs 2 \& 3) to eight (Run 1) refinement levels that extend out to 11400\,km and 15350\,km in all coordinate directions, respectively. The initial tori are entirely contained by the corresponding finest refinement level, which has a resolution of $\Delta_{x,y,z}=856\,\mathrm{m}$ and a diameter of 240\,km and 200\,km for Run 1 and Runs 2 \& 3, respectively. Simulations are performed in full 3D without imposing symmetries.

Detailed nucleosynthesis abundances of the outflows are obtained by conducting nuclear reaction network calculations on tracer particles with SkyNet\citemeth{Lippuner2017b}. The outflows are tracked by $10^4$ passive tracer particles of equal mass, which are advected with the fluid and placed within the initial torus with a probability proportional to the conserved rest-mass density. Nuclear reaction network calculations are restricted to unbound outflows (ejecta), that is, material with a non-vanishing escape velocity at infinity according to the Bernoulli criterion. Absorption of neutrinos is treated as described in previous work\citemain{Siegel2018a}. We estimate uncertainties in the nucleosynthetic yields (Fig.~\ref{fig:Xs}) as the difference resulting from using the two different emission geometries defined in previous work\citemain{Siegel2018a}.

\subsection{MHD turbulence, accretion Rates, effective viscosity.} 

Steady turbulent states of the disk are established in Runs 1, 2, and 3 after $t\approx30\,\mathrm{ms}$ (see Extended Data Fig.~\ref{fig:MHD_turbulence}). MHD turbulence is generated by the magnetorotational instability (MRI)\citemeth{Velikhov1959,Chandrasekhar1960,Balbus1991,Balbus1998,Balbus2003}, which is generally well resolved by at least 10 grid points per fastest-growing MRI wavelength (see top panel of Extended Data Fig.~\ref{fig:MHD_turbulence}); the latter can be estimated by\citemeth{Siegel2013,Kiuchi2015a,Kiuchi2018a}
\begin{equation}
  \lambda_\mathrm{MRI}\simeq\frac{2\pi}{\Omega} \frac{b}{\sqrt{4\pi\rho h
  + b^2}},
\end{equation}
where $\Omega=u^\phi/u^0$ is the angular frequency ($u^\mu$ being the four-velocity) and $b$ is the comoving magnetic field strength. The bottom panel of Extended Data Fig.~\ref{fig:MHD_turbulence} shows a `butterfly' diagram based on the radially averaged $y$-component of the magnetic field, indicating a fully operational dynamo and a steady turbulent state of the disk after $\approx20-30\,\mathrm{ms}$. At this point in time, the magnetic field has been self-consistently amplified to a saturated value by the MRI, independent of the initial magnetic field strength; Panel {\bf a} of Extended Data Fig.~\ref{fig:numerical_tests} indicates convergence of the maximum magnetic field strength at our fiducial resolution (referred to as $\Delta x$; see details of simulations above). Performing these simulations in full 3D without imposing symmetries is critical, as the anti-dynamo theorem in axisymmetry\citemeth{Cowling1933} does not allow for a steady turbulent state. Such a steady turbulent state, however, is important, as the amount of material unbound in outflows and its composition is critically affected by the precise balance between viscous heating from MHD turbulence and neutrino cooling. This prevented previous studies related to MHD collapsar jets\citemeth{Nakamura+15} from obtaining self-consistent initial data for r-process nucleosynthesis.

MHD turbulence mediates angular momentum transport, which, in turn, sets the accretion rate onto the black hole. Extended Data Fig.~\ref{fig:acc_rates} depicts the instantaneous accretion rates for Runs 1, 2, and 3. The overall accretion rates for these runs are roughly spaced by an order of magnitude each, $(5.8^{+6.2}_{-4.1}\times 10^{-2}, 4.3^{+3.3}_{-2.3}\times 10^{-3}, 5.6^{+2.1}_{-1.3}\times 10^{-4})\,M_\odot\,\mathrm{s}^{-1}$ for Runs 1, 2, and 3, respectively, where the nominal values correspond to a time average between $40-100\,\mathrm{ms}$, and where we bracket the uncertainties by the maximum and minimum accretion rate reached during that period. These accretion rates are indicated in Figs.~1 and 2 of the main text. The accretion rate and thus the release of gravitational binding energy is self-consistently set by MHD turbulence in our simulations. We have verified this by performing an additional simulation which is otherwise identical to Run 2, but starts with an initial magnetic field strength a few orders of magnitude smaller than the fiducial run, such that the subsequent amplification by the MRI cannot be captured (see Panel {\bf b} of Extended Data Fig.~\ref{fig:numerical_tests}). The accretion rate in this additional run is roughly an order of magnitude smaller, indicating that MHD turbulence indeed determines the accretion rate.

MHD turbulence effectively acts as a large-scale viscosity that can be parametrized by the Shakura-Sunyaev viscosity coefficient $\alpha$, which relates the accretion stress and the fluid pressure\citemeth{Shakura1973}. We estimate the radial profile of this parameter from our simulations according to
\begin{equation}
  \alpha_\mathrm{eff}(\varpi) \equiv \frac{\langle\langle\langle T^{r,\phi}\rangle_\phi\rangle_{\hat{D},z_H}\rangle_t}{\langle\langle\langle p\rangle_\phi\rangle_{\hat{D},z_H}\rangle_t},
\end{equation}
where $p$ denotes the fluid pressure and $T^{r,\phi}$ denotes the $r$-$\phi$ component of the stress-energy tensor in the frame comoving with the fluid. Here, $\langle\rangle_t$ refers to a time-average over a few neighboring simulation snapshots spanning a window of $\approx\!10\,\mathrm{ms}$. We have defined the azimuthal average of a quantity $\chi$ by
\begin{equation}
	\langle \chi \rangle_\phi \equiv \frac{\int_0^{2\pi} \chi \sqrt{\gamma_{\phi\phi}}\mathrm{d}\phi}{\int_0^{2\pi}\sqrt{\gamma_{\phi\phi}}\mathrm{d}\phi},
\end{equation}
where $\gamma_{\phi\phi}$ is the $\phi\phi$ component of the metric tensor.  Furthermore, we have introduced the rest-mass density average of a quantity $\chi$ over height $z$ as a function of the cylindrical coordinate radius $\varpi$, defined by
\begin{equation}
	\langle \chi \rangle_{\hat{D},z_H} \equiv \frac{\int_{-z_H}^{z_H} \chi\hat{D}\varpi\mathrm{d}z}{\int_{-z_H}^{z_H}\hat{D}\varpi\mathrm{d}z},
\end{equation}
where $\hat{D}=\sqrt{\gamma}\rho W$ is the conserved rest-mass density, with $\gamma$ the determinant of the spatial metric $\gamma_{ij}$ and $W$ the Lorentz factor. Here, we only integrate over the disk itself, defined by one scale height
\begin{equation}
  z_H(\varpi) \equiv \frac{\int |z| \langle \hat D\rangle_\phi \varpi\mathrm{d}z}{\int \langle \hat D\rangle_\phi \varpi \mathrm{d}z}.
\end{equation}
Extended Data Fig.~\ref{fig:alpha_visc} shows the radial profiles of $\alpha_\mathrm{eff}$ for Run 1 at different times during the evolution; the corresponding profiles for Runs 2 and 3 are very similar. For all runs we typically find $\alpha_\mathrm{eff}\approx 0.013$, averaged in radius over the inner disk $27\,\mathrm{km}<r<200\,\mathrm{km}$, excluding the innermost region with a rise of $\alpha_\mathrm{eff}$ around the innermost stable circular orbit toward the horizon due to increased mean magnetic field strengths\citemeth{Penna2013a}.

\subsection{Ignition accretion rate and transition to a thin disk.}

For high accretion rates, the midplane of the disk is sufficiently dense that the rate of neutrino cooling becomes fast relative to the radial advection of thermal energy\citemeth{Popham+99,Narayan+01,DiMatteo+02}.  This transition to an efficiently neutrino-cooled disk occurs at radii outside those of the innermost stable circular orbit for accretion rates above a critical ``ignition" rate.  The value of this critical accretion rate was previously estimated to be\citemeth{Chen&Beloborodov07,Metzger+08c} 
\begin{equation}
\dot{M}_{\rm ign} = K_{\rm ign}(a)\left(\frac{\alpha}{0.01}\right)^{5/3} \underset{a = 0.8}\approx 2\times 10^{-3}M_{\odot}{\rm s^{-1}}\left(\frac{\alpha}{0.02}\right)^{5/3},
\label{eq:Mdotign}
\end{equation}
where $\alpha$ is the dimensionless Shakura-Sunyaev viscosity coefficient.  Based on one-dimensional steady-state disk models, the value of the prefactor $K_{\rm ign}(a)$ was previously found to be $\approx\!1.5\times 10^{-3}M_{\odot}$ s$^{-1}$ for a non-spinning black hole ($a = 0$), decreasing to $\approx\!4.5\times 10^{-4} M_{\odot}$ s$^{-1}$ for $a = 0.95$ \citemeth{Chen&Beloborodov07}. In the final line of Eq.~\eqref{eq:Mdotign} we have adopted a value $K_{\rm ign} \approx 6\times 10^{-4} M_{\odot}$ s$^{-1}$ appropriate for $a \approx 0.8$.

The lower entropy and geometrically thinner structure of neutrino-cooled disks results in a higher electron degeneracy, which is self-sustained, and which suppresses the number of positrons relative to their number under non-degenerate conditions. This, in turn, favors electron capture reactions on protons over the opposing reaction (positron captures on neutrons), driving the midplane composition to be neutron-rich, with electron fraction\citemain{Beloborodov03}$^,$\citemain{Siegel2017a} $Y_e \sim 0.1$. As illustrated in Fig.~1, our two highest accretion rate simulations with $\dot{M} \approx 4\times 10^{-3}- 6\times 10^{-2}M_{\odot}$ s$^{-1}$ indeed reach states of low $Y_e\lesssim 0.2$, while our lowest $\dot{M} \lesssim 10^{-3}M_{\odot}$ s$^{-1}$ simulation instead maintains a high value of $Y_e$. This abrupt transition at $\dot{M} \sim \mathrm{few} \times 10^{-3}M_{\odot}$ s$^{-1}$ naturally corresponds to the ignition threshold effect described above (Eq.~\eqref{eq:Mdotign}), given the effective value of $\alpha_{\rm eff} \approx 0.013$ inferred from our MHD simulations (Extended Data Fig.~\ref{fig:alpha_visc}; see above). To our knowledge, the present study is the first to demonstrate the ignition threshold by means of self-consistent MHD simulations.

\subsection{Nickel disk winds at late times.}

At late stages in the collapsar evolution, the accretion rate decreases below a critical value $\dot{M}_{\rm ign} \approx 10^{-3}M_{\odot}$ s$^{-1}$ (Eq.~\eqref{eq:Mdotign}) for a neutrino-cooled, neutron-rich disk.  The winds at late times instead have a composition $Y_{e} \approx 0.5$ matching that of the progenitor star, making them a potential source of $^{56}$Ni \citemain{MacFadyen&Woosley99}$^,$\citemeth{MacFadyen+01}, in addition to that produced during a prompt supernova explosion phase. At least several tenths of a solar mass of $^{56}$Ni is needed to explain the luminosities of GRB supernovae\citemeth{Mazzali+06,Cano+16}. However, much of the outflowing material at these late times remains as $^{4}$He instead of forming heavy isotopes.  This is due to the relatively slow rate of the triple-$\alpha$ reaction bottleneck needed to create seed nuclei when $Y_{e} \approx 0.5$, as opposed to the faster neutron-aided reaction $^{4}$He($\alpha n,\gamma)^{9}$Be($\alpha,n)^{12}$C that operates and results in extremely efficient seed nucleus formation in the disk outflows when $Y_e \ll 0.5$. Here, we perform a simple estimate of how the $^{56}$Ni mass fraction in the disk wind varies with the accretion rate at late epochs, once $\dot{M} \ll \dot{M}_{\rm ign}$, which we then compare to our numerical results.

Working within the framework of an $\alpha$-disk model, we first estimate the mean temperature $T_{\rm m}$ and density $\rho_{\rm m}$ of the accretion flow. In steady-state, the surface density $\Sigma = 2H \rho_{\rm m}$ of a disk at radii well outside the innermost stable orbit is given by
\begin{equation}
\dot{M} \simeq 3\pi \nu \Sigma = 3\pi \alpha c_{\rm s}H \Sigma = 6\pi \alpha \rho_{\rm m} H^{3}\Omega ,
\end{equation}
where $\nu = \alpha c_{\rm s}H$ is the effective kinematic viscosity, $c_{\rm s} = H\Omega$ is the midplane sound speed, $\Omega = (GM_{\rm BH}/r^{3})^{1/2}$, and $H$ is the vertical scale height.  Scaling radii to $r_{\rm g} \equiv GM_{\rm BH}/c^{2}$, 
\begin{equation}
\rho_{\rm m} \approx 1.5\times 10^{8}{\rm g\,cm^{-3}}\left(\frac{r}{10 r_{\rm g}}\right)^{-3/2}\left(\frac{\dot{M}}{10^{-4}M_{\odot}\,s^{-1}}\right)\left(\frac{\alpha}{0.01}\right)^{-1}\left(\frac{M_{\rm BH}}{3M_{\odot}}\right)^{-1/2}\left(\frac{H}{r/3}\right)^{-3}.
\end{equation}
At low accretion rates of interest, the midplane pressure is dominated by radiation pressure, $\rho_{\rm m} c_{\rm s}^{2} = (11/12)a T_{\rm m}^{4}$, such that 
\begin{eqnarray}   T_{\rm m} &=&  \left(\frac{11}{72\pi}\frac{\dot{M}\Omega}{ \alpha a H }\right)^{1/4} \nonumber \\ &\approx& 1.84{\rm MeV}\, \left(\frac{r}{10 r_{\rm g}}\right)^{-5/8}\left(\frac{\dot{M}}{10^{-4}M_{\odot}\,s^{-1}}\right)^{1/4} \left(\frac{\alpha}{0.01}\right)^{-1/4}\left(\frac{M_{\rm BH}}{3M_{\odot}}\right)^{1/8}\left(\frac{H}{r/3}\right)^{-1/4}.
\end{eqnarray}
The entropy of radiation in the disk midplane (in units of $k_\text{B}$ per baryon) is then estimated to be
\begin{equation}
s_{\rm m,rad} = \frac{11\pi^{2}k^{3}m_\text{p}}{45 c^{3}\hbar^{3}}\frac{T_{\rm m}^{3}}{\rho_{\rm m}} \approx 21 \left(\frac{r}{10 r_{\rm g}}\right)^{-3/8}\left(\frac{\dot{M}}{10^{-4}M_{\odot}\,s^{-1}}\right)^{-1/4}\mskip-15mu\left(\frac{\alpha}{0.01}\right)^{1/4}\left(\frac{M_{\rm BH}}{3M_{\odot}}\right)^{7/8}\left(\frac{H}{r/3}\right)^{9/4},
\end{equation}
to which should be added the entropy in the non-relativistic nucleons,
\begin{equation}
s_{\rm m, N} \approx 13.3 + {\rm ln}\left(\frac{T_{\rm m, MeV}^{3/2}}{\rho_{\rm m,7}}\right).
\end{equation}
Here $T_{\rm m,MeV}$ indicates $T_{\rm m}$ in units of MeV, and $\rho_{\rm m,7}$ indicates $\rho_{\rm m}$ in units of $10^7$\,g\,cm$^{-3}$. For $Y_{e} \approx 0.5$ the dominant destruction process of $\alpha$-particles in the wind is\citemeth{Woosley1992a} $^{4}$He($\alpha,\gamma$)$^{12}$C, such that the $\alpha$-particle number fraction $Y_{\alpha}$ in the outflow evolves as\citemeth{Roberts2010}
\begin{equation}
\frac{dY_{\alpha}}{d\tau} \approx -14 \rho^{2}Y_{\alpha}^{3}\lambda_{3\alpha},
\end{equation}
where $\lambda_{3\alpha}(T)$ is the temperature-dependent triple-alpha rate coefficient and the factor of 14 comes from assuming $\alpha$ captures cease at $^{56}$Ni.  Defining a time variable $d \tau = -(\tau_{\rm d}/3T)dT$, where $\tau_{\rm d}$ is the expansion time of the wind around the point of $\alpha$-particle formation, and then integrating over the triple-alpha rate, the final abundance of seed particles is\citemeth{Roberts2010}
\begin{equation}
Y_{\rm seed} = \frac{1}{56}\left\{1 - [1 + 35(\tau_{\rm d}/{\rm ms})s_{\rm f}^{-2}]^{-1/2}\right\},
\label{eq:Yseed}
\end{equation}
where $s_{\rm f}$ is the final entropy in $k_{\rm B}$ per baryon. Assuming that the seed particles are mostly within the iron peak (charge numbers $24\le Z \le 28$; see also Fig.~1) with an abundance pattern roughly similar to Run 3, we can translate Eq.~\eqref{eq:Yseed} into a mass fraction $X_{^{56}\mathrm{Ni}}$ and $X_\mathrm{He}\approx 1 - X_{\rm seed}$. Extended Data Fig.~\ref{fig:XNi} shows these estimates for the $^{56}$Ni and $\mathrm{He}$ yield of collapsar disks as a function of $\dot{M} < \dot{M}_{\rm ign}$, taking $s_{\rm f} \approx s_{\rm m} = s_{\rm m,rad} + s_{\rm m,N}$, calculated for $\alpha = 0.013$ (see above; Extended Data Fig.~\ref{fig:alpha_visc}), a mean ejection radius of $20\,r_{\rm g}$ ($\approx 88\,\mathrm{km}$), and a mean scale-height of $H/r\approx 0.2$ at low $\dot{M}$.

The predicted value of $X_{\rm ^{56}Ni} \approx 0.1$ agrees well with that measured in the disk outflow for our lowest accretion rate simulation at $\dot{M} \lesssim 10^{-3}M_{\odot}$ s$^{-1}$. Extending the analytic scaling to all $\dot{M} < \dot{M}_{\rm ign}$, we estimate the $^{56}$Ni yield of collapsar disks in our toy fallback model (see below), as summarized in Supplementary Table 1. These typical Nickel masses $\lesssim 0.1M_{\odot}$ are generally less than those required to explain the luminosities of GRB supernovae\citemeth{Woosley&Bloom06}, suggesting that most of the $^{56}$Ni is instead produced by shock heating of the star during the early supernova explosion phase.  The early ejection of $^{56}$Ni, segregated from the late r-process production in disk winds, has important implications for the signatures of r-process elements in the supernova emission (see observational signatures below).

\subsection{Previous work on supernovae and collapsars as r-process sites.}  

Most previous work on the r-process in core collapse supernovae has focused on the neutrino-driven winds from the newly-formed proto-neutron star, which cools and deleptonizes in the seconds to minutes following a successful explosion\citemeth{Qian&Woosley96}). However, many works over the past twenty years have shown that such wind models encounter severe theoretical difficulties in achieving the requisite conditions of high entropy and low electron fraction needed for a successful second or third-peak r-process\citemeth{Qian&Woosley96,Thompson+01,Roberts+12,MartinezPinedo+12}. A higher entropy may be possible in some cases where the neutron star is highly magnetized\citemeth{Thompson03,Thompson&UdDoula18}.  However, even if a solution is found, the low predicted r-process yields of standard neutrino-wind models of $\sim 10^{-5}-10^{-4}M_{\odot}$ per event are in conflict with a growing number of astrophysical and terrestrial observations that instead favor high-yield rare events as the main Galactic r-process site\citemain{Ji+16}$^,$\citemeth{Wallner+15,Macias&RamirezRuiz18}.

Recent theoretical efforts have instead focused on the rare subset of core collapse events that give birth to rapidly-spinning and strongly-magnetized proto-neutron stars, so-called ``MHD supernovae"\citemain{Winteler+12}$^,$\citemeth{Thompson+04,Metzger+08}.  The rapid outwards acceleration in such winds by magneto-centrifugal effects can in principle result in the ejection of low-$Y_e$ material, resulting in an r-process even for moderate entropies. However, when consideration of the three-dimensional stability of the magnetized jets that give rise to the fast-expanding neutron-rich ejecta in these systems are taken into account\citemeth{Mosta+14}, such events are challenged to eject sufficiently large quantities of heavy r-process nuclei\citemain{Moesta2018}$^,$\citemeth{Halevi&Mosta18}. As in the winds from non-rotating proto-neutron stars, a main factor that prevents the escape of low-$Y_{e}$ material is irradiation of the outflowing matter by electron neutrinos from the proto-neutron star; the latter convert neutrons into protons via the reaction $\nu_e + n \rightarrow p + e^{-}$.  

Nevertheless, such a ``proto-magnetar" is a necessary prerequisite to the black hole accretion phase in collapsar models\citemain{Dessart+08}.  One cannot therefore exclude that these earlier phases of the explosion could give rise to a weak r-process, or even a moderate quantity $\lesssim 10^{-2}M_{\odot}$ of heavy r-process nuclei\citemain{Moesta2018}$^,$\citemeth{Halevi&Mosta18}).  However, as shown here, when fallback is sufficient to create a black hole surrounded by an accretion disk, then the early explosion/magnetar phase is likely to contribute only sub-dominantly relative to the {\it disk accretion/outflow} phase that follows over longer timescales of seconds to minutes and can eject up to $\sim\!1\,M_{\odot}$ in heavy nucleosynthesis products.  

Some works\citemeth{Fujimoto+07,Ono+12, Nakamura+15,Hayakawa&Maeda18,Soker2017} have proposed that low-$Y_{e}$ conditions are achieved in the relativistic jets from collapsars, giving rise to a moderate quantity of r-process ejecta ($\lesssim 0.01-0.1M_{\odot}$).  However, the simulations employed by these works assume the presence of large-scale ordered magnetic fields prior to and thus immediately following the initial collapse, i.e. they are put in by hand as an additional assumption.  They also assume axisymmetry and therefore do not properly capture the destabilizing effects of the non-axisymmetric kink instability\citemeth{Mosta+14}. Additionally, as a result of the anti-dynamo theorem\citemeth{Cowling1933} two-dimensional simulations also cannot self-consistently follow the growth of the magnetic field due to the MRI and thus obtain initial compositions and outflow masses for r-process nucleosynthesis that are not self-consistent. By contrast, the three-dimensional GRMHD simulations presented here demonstrate that neutron-rich disk winds arise naturally from MRI-driven disks, as would be present even if the initial magnetic field of the progenitor star was (arbitrarily) weak. The disk winds we consider, which are driven largely by thermal pressure and alpha-particle formation, are therefore not strongly dependent of the presence of a large-scale ordered magnetic field.  An ordered field, by providing stronger magneto-centrifugal acceleration, would only act to increase the quantity of low-$Y_{e}$ ejecta\citemain{Fernandez+18}.

In summary, although MHD supernovae have been widely discussed previously as r-process sources, there are important differences from the scenario of late-time accretion disk outflows proposed here, which arguably make this mechanism more robust and important.  First, the disk wind mechanism acts independently of two major uncertainties: (1) the initial strength and topology of the magnetic field in the stellar core; (2) the effect of neutrino absorption reaction on the electron fraction of the outgoing wind, which is much less pronounced after black hole formation than at early times when the proto-neutron star is still present and its neutrino luminosity is greatest.  Second, the disk wind scenario, calibrated directly to observations of GW170817 (see rates of r-process production below), predicts a larger quantity of ejecta per event than even the most optimistic previous findings from the early supernova explosion phase.  Finally, as we show below (see observational signatures), the inevitable mixing of r-process material during the MHD supernova phase with the $^{56}$Ni synthesized by the same explosion (see below), leads to predictions for the properties of GRB supernovae--the clearest examples of "MHD supernovae" in nature--which are inconsistent with current observations.

A wide range of observational evidence connects long duration GRBs with the hyper-energetic explosions of massive stars stripped of their outer hydrogen envelopes\citemeth{Woosley&Bloom06,Hjorth&Bloom12a}. The dynamics of the collapse to form a torus surrounded by an accretion shock, and its connection to the production of a successful relativistic GRB jet and supernova explosion, have been the focus of numerous analytical and numerical studies\citemeth{MacFadyen+01,Fujimoto+06,Uzdensky&MacFadyen07,Morsony+07,Bucciantini+08,Lazzati+08,Kumar2008b,Nagakura+11,Lindner+12, Lopez-Camara+13,Batta&Lee14}. The high optical luminosities of GRB supernovae are widely attributed to the radioactive decay of large quantities $\approx 0.2-0.6M_{\odot}$ of $^{56}$Ni synthesized in the explosion\citemeth{Mazzali+14,Cano+16}. These large $^{56}$Ni masses could originate from the shock-heating of the ejecta by the hyper-energetic supernova explosion or gamma-ray burst jet\citemeth{Maeda&Nomoto03,Fryer+06,Maeda&Tominaga09,Barnes+18}. Alternatively, it has been proposed that hot outflows from the black hole accretion disk, such as those studied here, are also a potential source of $^{56}$Ni \citemain{MacFadyen&Woosley99}$^,$\citemeth{MacFadyen+01}. However, the precise isotopes which are synthesized in the disk outflows depend sensitively on the electron fraction; in particular, $^{56}$Ni is only synthesized in relatively proton-rich ejecta with $Y_{e} > 0.495$.  The large $^{56}$Ni yields previously reported\citemain{MacFadyen&Woosley99} are likely a product of artificially fixing $Y_{e} = 0.5$ in the simulations, rather than following the evolution of $Y_e$ under the influence of charged-current weak interactions in the electron-degenerate disk.  Our simulations demonstrate that, when the accretion rate is sufficiently high $>10^{-3}M_{\odot}$ s$^{-1}$ (Eq.~\eqref{eq:Mdotign}), the midplane composition is driven to be neutron rich, $Y_{e} \ll 0.5$ within tens of gravitational radii from the black hole (Fig.~1, main text).  

For lower accretion rates $< 10^{-3}M_{\odot}$ s$^{-1}$, we find that some $^{56}$Ni is generated in the disk winds.  This is because weak interactions are slower due to the lower disk temperature, such that the electron fraction is not appreciably changed from the value $Y_{e} \approx 0.5$ inherited from the infalling star.  As such low infall rates are only achieved at late times after collapse (well after the prompt GRB phase), by combining a range of models for collapsar accretion (see model on mass fallback below) with our estimate of the $^{56}$Ni yield (see model for nickel winds at late times above), we find that at most a total $\sim\!10^{-3}-10^{-1}\,M_{\odot}$ of $^{56}$Ni ejecta could originate from late-time disk winds (Supplementary Table 1).  This indicates that additional sources of $^{56}$Ni, such as shock heating of the infalling progenitor core during the earliest stages in the explosion or GRB jet\citemeth{Barnes+18}, are required to produce most of the inferred $^{56}$Ni in GRB supernovae.

Our finding that a large fraction of the disk ejecta generates the r-process is consistent with previous numerical simulation work that focused on the context of binary neutron star mergers\citemain{Fernandez&Metzger13,Just+15,Siegel2017a}, which indeed was one motivation for this study. Some previous works, using steady-state wind models\citemain{Pruet+04,Surman+11}, found that collapsar disks mainly produce $^{56}$Ni, due to the outflowing matter experiencing high rate of positron captures\citemain{Pruet+04} or absorbing electron neutrinos generated from the disk\citemain{Surman+11}, while one-dimensional disk models also led to the speculation about r-process production\citemain{Kohri+05}. Other works finding the production of r-process material\citemeth{Caballero2012} had to assume extremely low initial electron fractions $Y_{e}\ll 0.1$. While we find that neutrinos play some role in raising $Y_e$, the outflows of our fully time-dependent MHD simulations still have sufficiently low $Y_{e} < 0.5$ for an r-process. A key difference between accretion disk outflows and purely neutrino-driven winds (e.g.,~as those from spherical proto-neutron stars) is that the latter start at effectively zero velocity from the base of the wind and are heated and unbound from the gravitational potential well exclusively by neutrinos themselves. By contrast, outflows from accretion disks are driven mainly by turbulent (or ``viscous") heating in the upper disk atmosphere\citemain{Fernandez&Metzger13,Siegel2017a,Siegel2018a}. Steady-state wind calculations which assume the matter accelerates smoothly starting from a disk midplane assumed to be hydrostatic at depth\citemain{Pruet+04}$^,$\citemeth{Metzger+08c} can therefore greatly overestimate the time matter spends in the non-degenerate regions and thus overestimate the impact of positron captures or neutrinos in raising the final electron fraction.  True MRI-active accretion disks are nowhere hydrostatic, even in the midplane, and the unbound matter spends comparatively little time in non-degenerate regions of the disk before escaping to infinity.

Finally, we note that it has been alternatively postulated that long GRBs are powered by the electromagnetic spin-down of millisecond magnetars, rather than accreting black holes\citemeth{Thompson+04}. Although our observationally-calibrated disk wind model is obviously predicated on the assumption that long GRBs are accretion-powered, our conclusions may also apply if the central compact object remains a neutron star. While at early times after the supernova, the high neutrino luminosity of the proto-neutron star can strongly raise the wind electron fraction and thus reduce its r-process yield\citemain{Moesta2018}$^,$\citemeth{Halevi&Mosta18}, as the neutron star cools on a timescale of seconds and its neutrino luminosity drops, its effect on the wind properties as compared to the black hole case will diminish. On the other hand, if long GRBs are powered entirely by magnetar spin-down, with no accretion disk present, then the total r-process yields of long GRBs come entirely from the magnetar wind\citemeth{Vlasov+17} and is likely to be substantially less than we predicted here. In this way, future constraints on r-process production in collapsars could also be used to test magnetar versus accretion-powered models for long GRBs.

\subsection{Model for the evolution of mass fall-back in collapsars.}

In order to translate the different ranges of accretion rates explored by our suite of numerical simulations into the physical conditions expected in collapsars, we model the fallback history of stellar material onto the accretion disk and the black hole similarly to previous work\citemeth{Kumar2008b}. Using their Eqs.~(1)--(7) we obtain the mass fallback rates by numerically solving these equations for presupernova stellar models\citemeth{Heger2000a}; one example is shown in Fig.~1, Panel  a. We employ the radial angular velocity profiles provided in these models and assume rigid rotation on spherical shells, $\Omega(r,\theta)=\Omega(r)$, i.e.,
\begin{equation}
  j(r,\theta) = j(r) \sin(\theta),
\end{equation}
where $j$ denotes specific angular momentum. A fraction of the stellar core may not have sufficient angular momentum to form an accretion disk and, instead, directly fall into the black hole. In particular, the initial black hole is formed by the innermost part of the pre-collapse star that circularizes inside the innermost stable circular orbit (ISCO) of a black hole corresponding to this amount of mass. The radius of this innermost part of the star is determined by the condition
\begin{equation}
 r_\mathrm{ISCO}(M_r,a_r) \Omega^2_k(r) = r \Omega^2(r), \label{eq:RISCO_r}
\end{equation}
where $M_r$ is the enclosed mass inside of radius $r$, and
\begin{equation}
  a_r = \frac{c J_r}{GM_r^2}
\end{equation}
is the black hole spin parameter. Here $J_r$ denotes the total enclosed angular momentum, $G$ is the gravitational constant, and $c$ is the speed of light. Furthermore,
\begin{equation}
  \Omega_k(r) = \left(\frac{GM_r}{r^3}\right)^{1/2}
\end{equation}
is the angular velocity of a Keplerian orbit around an object of mass $M_r$. The ISCO is given by\citemeth{Bardeen1972}
\begin{equation}
  R_\mathrm{ISCO}(M,a) = \frac{GM}{c^2}\left\{ 3 + z_2 - \left[ (3-z_1)(3+z_1+2z_2)\right]^{1/2}\right\},
\end{equation}
where
\begin{eqnarray}
  z_1 &=& 1 + (1-a^2)^{1/3}\left[ (1+a)^{1/3} + (1-a)^{1/3}\right], \\
  z_2 &=& (3a^2 + z_1^2)^{1/2}. \label{eq:z2}
\end{eqnarray}
We solve Eqns.~\eqref{eq:RISCO_r}--\eqref{eq:z2} numerically for the presupernova models to find the radius $r_\mathrm{ISCO}$ of the presupernova core that forms the initial black hole, its mass $M_\mathrm{BH,in}$, and the associated time since collapse $t_\mathrm{BH}$, and the critical mass fallback rate $\dot{M}_\mathrm{BH}$ after/below which an accretion disk forms. For model E15, this regime is indicated in Fig.~1 (Panel a, blue dotted line).

Only a limited time window exists within which the accretion flow can form a jet capable of generating a GRB similar to those observed (more specifically, GRBs used to derive our r-process production rate estimates, see below). The timescale $\tau_{\dot{M}_\mathrm{fb}}$ over which the fallback rate $\dot{M}_\mathrm{fb}$ changes increases with time, typically as a power-law $\tau_{\dot{M}_\mathrm{fb}} \propto t^\alpha$, where $\alpha\simeq 1$. More rigorously, we define
\begin{equation}
  \tau_{\dot{M}_{\mathrm{fb}}} \equiv \left(\frac{\mathrm{d}\ln \dot{M}_\mathrm{fb}}{\mathrm{d}t}\right)^{-1}. \label{eq:tau_Mdot}
\end{equation}
Fall-back accretion can only explain those GRBs observed as long as the evolution time of the fallback rate is smaller or equal the typical time required to generate a GRB, i.e., $\tau_{\dot{M}_\mathrm{fb}}\le \tau_\mathrm{GRB}$. We denote the time relative to collapse at which this equality $\tau_{\dot{M}_\mathrm{fb}} = \tau_\mathrm{GRB}$ is reached by $t_\mathrm{GRB}$.  In order to generate a GRB at all, the stellar model must also have sufficient angular momentum to satisfy the condition
\begin{equation}
t_\mathrm{GRB} > t_\mathrm{BH} . \label{eq:GRB_criterium}
\end{equation}
The observed duration of a GRB in the rest frame of the star, $\tau_\gamma$, is given by the difference of the engine time, $\tau_\mathrm{GRB}$, and the time $\tau_\mathrm{b}$ required for the jet to drill through the stellar envelope\citemeth{Bromberg2012a,Sobacchi2017}:
\begin{equation}
	\tau_\gamma = \tau_\mathrm{GRB} - \tau_\mathrm{b}.
\end{equation}
Assuming the typical breakout time of $\tau_\mathrm{b} = 57^{+13}_{-10}\,\mathrm{s}$ \citemeth{Sobacchi2017}, and a typical observed GRB duration of $\tau_\gamma = T_{90}/(1+z)=9\,\mathrm{s}$, where we used a characteristic $T_{90} \simeq 27\,\mathrm{s}$ and redshift $z\simeq 2$ \citemeth{Bhat2016}, we deduce
\begin{equation}
  \tau_\mathrm{GRB} \approx 66\,\mathrm{s}.
\end{equation}
The time range up to $t_\mathrm{GRB}$ for model E15 is indicated in Panel a of Fig.~1 (red dotted line).

All models examined here are capable of powering a GRB according to the criterion Eq.~\eqref{eq:GRB_criterium}. The fallback rate typically remains above the critical value $\dot{M}_\mathrm{fb} \gtrsim 10^{-3}M_{\odot}$ s$^{-1} \sim \dot{M}_{\rm ign}$ (Fig.~1; Eq.~\eqref{eq:Mdotign}) for the production of r-process elements throughout the GRB phase. We conclude that collapsar progenitors resulting in accretion flow properties which are consistent with those required to produce observed long GRBs also produce significant quantities of r-process wind ejecta.

Supplementary Table 1 shows the total amount of fallback material that is fed into unbound disk outflows during different epochs in accretion rate, which we have translated into different nucleosynthesis quantities based on the results of our numerical simulations (Figs.~1 and 2). Under the assumption that $\sim 30\%$ of the accreted mass is unbound in outflows (consistent with numerical simulations\citemain{Siegel2017a,Siegel2018a,Fernandez+18}) we find that our models produce $\sim\!0.02 - 2\,M_{\odot}$ in heavy r-process material, additional $\sim\!0.03-0.4\,M_{\odot}$ in light r-process material, and $0.002-0.17\,M_{\odot}$ in radioactive $^{56}$Ni. The latter may contribute in part to the supernova light curve, while the signatures of the r-process nuclei could largely be hidden until late times (see observational signatures below).

\subsection{Ram pressure of fallback material.}

In order to escape the disk outflows must overcome the ram pressure of the infalling material from the stellar mantle, $p_\mathrm{fb}\simeq \rho_\mathrm{fb}v_\mathrm{ff}^2$, where $v_\mathrm{ff}=\sqrt{GM_\mathrm{BH}/r}$ is the free-fall velocity and $\rho_\mathrm{fb}$ is determined by the mass falling onto the outflows per solid angle and unit time, $\mathrm{d}M_\mathrm{fb,w}/(\mathrm{d}\Omega\mathrm{d}t)=\rho_\mathrm{fb}r^2 v_\mathrm{ff}$. The outflows produce a ram pressure $p_\mathrm{w}\simeq \rho_\mathrm{w} v_\mathrm{w}^2\simeq f_\mathrm{w} v_\mathrm{w}\mathrm{d}M_\mathrm{fb}/(\mathrm{d}\Omega\mathrm{d}t)$, where $v_w\approx 0.1c$ is the wind speed and $f_w\approx 0.3-0.4$ is the fraction of matter inflowing through the inner few tens of gravitational radii of the black hole becoming unbound in winds\citemain{Siegel2017a,Siegel2018a,Fernandez+18} (see also model for mass fallback above). We assume these winds are channeled into an outflow at intermediate polar angles $\sim 30^\circ-45^\circ$, outside of which inflows onto the black hole and the disk may still persist on timescales of interest\citemain{MacFadyen&Woosley99}. Using our model for mass fallback (see below) we find that for typical presupernova stellar models the fraction of total fallback material $M_\mathrm{fb}$ into this solid angle and into the innermost $r\le 500\,\mathrm{km}$ of the accretion disk where most of the neutron-rich outflows are generated is typically $f_\mathrm{fb}<0.1$. The criterion for successful outflows $p_\mathrm{w}/p_\mathrm{fb}\ge 1$ then translates into
\begin{equation}
	r \gtrsim 30\,\mathrm{km} \left(\frac{M_\mathrm{BH}}{3M_\odot}\right) \left(\frac{0.1}{v_\mathrm{w}}\right)^2 \left(\frac{0.4}{f_\mathrm{w}}\right)^2 \left(\frac{f_\mathrm{fb}}{0.1}\right)^2.
\end{equation}
This analytic estimate shows that once launched the outflows are indeed powerful enough to overcome the ram pressure of infalling material. This is further confirmed by previous global simulations, which find that under typical conditions such outflows can indeed successfully escape the infalling material\citemain{MacFadyen&Woosley99}.

\subsection{Rates of Galactic r-process production in collapsars versus mergers.}

If one multiplies the r-process yield inferred from the kilonova associated with GW170817 of $\approx\!0.03-0.06M_{\odot}$ \citemain{Kasen+17,Cowperthwaite2017}$^,$\citemeth{Drout+17,Tanvir+17} with the binary neutron star merger rate of $R_{\rm BNS} = 1540^{+3200}_{-1220}$ Gpc$^{-3}$ yr$^{-1}$ inferred from the LIGO/Virgo discovery\citemain{LIGO+17DISCOVERY}, and converting the latter to a per-galaxy rate, one obtains an r-process production rate in the Milky Way that is broadly consistent with that required to explain their abundances in our Galaxy and solar system\citemain{Kasen+17,Cowperthwaite2017,Hotokezaka2018a}$^,$\citemeth{Cote2018a}. However, being based on just a single event, the statistical errors on this estimate alone exceed an order of magnitude.  Furthermore, several additional systematic uncertainties enter this estimate, such as the detailed star-formation history of the Milky Way\citemain{Hotokezaka2018a}, as well as the exact quantity of ejecta in GW170817 (due to uncertainties in the nuclear heating rate\citemeth{Barnes+16}) and details of the composition (e.g. the relative abundance of light versus heavy r-process nuclei).  Thus, although GW170817 resoundingly confirmed that neutron star mergers are an important r-process site\citemain{Kasen+17}, it does not prove they are the dominant one.

Given these still large uncertainties in {\it absolute} rate estimates, an arguably better approach is to compare the {\it relative} r-process contribution from collapsars to that of neutron star mergers (though we also provide an absolute rate estimate for collapsars below).  We assume that in both mergers and collapsars, a fixed fraction $f_{\rm w} \lesssim 1$ of the mass which is flowing through the inner several gravitational radii of the accretion disk is unbound in winds as r-process nuclei, with the remaining fraction $1-f_{\rm w}$ being accreted by the black hole and potentially used to power the GRB jet.  The relative total contribution of collapsars versus mergers to the solar system or present-day Galactic r-process can be roughly estimated from the ratio of the product of their volumetric rates ($\int R(z) dz$, integrated over cosmic time/redshift) and the average amount of mass accreted in each event, respectively.  

We make the further assumption that the accreted mass (and thus wind ejecta mass) is directly proportional to the radiated gamma-ray energy, i.e. that the energy radiated by the GRB jet is proportional to the total accreted mass, $m_{\rm acc}$.  Although the physics that gives rise to the collimated relativistic outflows and gamma-ray radiation is complex, we are only making the more limited assumption that the {\it same physical processes} operate in both types of bursts. This is motivated by the strikingly similar temporal and spectral properties of the prompt gamma-ray emission in short and long GRBs\citemain{Ghirlanda+09}$^,$\citemeth{Li+16}. Although short GRBs show similar isotropic gamma-ray luminosities to long GRBs, thus indicating similar black hole accretion rates (there is, however, an alternative proposition\citemeth{Tchekhovskoy2015} ), their total isotropic gamma-ray energies, $E_{\rm iso}$, are on average $\approx 10^{1.6} \approx 40$ times smaller than those of long GRBs\citemain{Ghirlanda+09}$^,$\citemeth{Li+16}, mainly due to their shorter average $T_{90}$ durations ($\approx\!0.2\,{\rm s}$ versus $\approx\!10\,{\rm s}$, respectively).

Following the above method, the relative contributions of collapsars versus mergers to the r-process mass in the Galaxy, $m_{\rm r}$, may be crudely estimated by the ratio
\begin{equation}
\frac{m_{\rm r, coll}}{m_{\rm r,merger}} \sim \frac{m_{\rm acc}^{\rm LGRB}\int R_{\rm coll}(z)dz}{m_{\rm acc}^{\rm SGRB}\int R_{\rm merger}(z)dz} > \frac{E^{\rm LGRB}_{\rm iso}R_{\rm LGRB}(z = 0)}{E^{\rm SGRB}_{\rm iso}R_{\rm SGRB}(z = 0)} \approx 4-30 \label{eq:Macc},
\end{equation}
where we have used the local $z\approx 0$ rates of short GRBs of\citemeth{Wanderman2015} $R_{\rm SGRB}(z = 0) \approx 4.1_{-1.9}^{+2.3}$ Gpc$^{-3}$ yr$^{-1}$ and long GRBs of\citemeth{Wanderman&Piran10} $R_{\rm LGRB}(z = 0) \approx 1.3_{-0.7}^{0.6}$ Gpc$^{-3}$ yr$^{-1}$. This approximation gives a conservative lower limit on the ratio because the ratio of long to short GRBs increases with redshift; long GRBs approximately track star formation, which peaks at $z \approx 2-3$, while short GRBs are consistent with a sizable delay time\citemeth{Berger14,Wanderman2015}. This estimate suggests that collapsars could well contribute more total r-process production in the Galaxy than neutron star mergers (i.e., at least 80\%; see Extended Data Fig.~\ref{fig:cartoon} for a schematic summary), which is consistent with requirements to explain the late-time Galactic enrichment of europium versus iron (see below).

In addition to normal (`classical') long GRBs, there exists a class of `low-luminosity' GRBs\citemeth{Liang+07} (LLGRB; luminosities $L<10^{50}\,\mathrm{erg}\,\mathrm{s}^{-1}$), which have much lower radiated gamma-ray energies, produced by a less energetic central engine and/or a GRB jet which fails to successfully break out of the progenitor star. We note that the above rate estimate for GRBs does not include this class\citemeth{Wanderman&Piran10}. In fact, most GRB supernovae discovered in the local universe are associated with LLGRB\citemeth{Melandri2014}. Although the observed rate of LLGRBs is substantially higher than classical GRBs, their gamma-ray beaming fraction may also be larger, such that their volumetric rate is at most only $\sim\!10-30$ times larger. If LLGRBs are responsible for r-process production, then their required per-event r-process yield would thus be a factor $\sim\!10$ times less than those of the most energetic long GRBs.

We also perform a rough {\it absolute} estimate of the r-process ejecta mass needed per collapsar in order to explain their solar system abundances.  Depending on whether one is considering abundances which extend in atomic mass number down to the 1st or 2nd r-process peak, the Solar mass fraction of r-process nuclei is $X_{\rm r} = 4\times 10^{-7}$ or $6\times 10^{-8}$, respectively\citemeth{Arnould2007} (see Galactic chemical evolution below). The r-process mass per burst needed to explain the solar system abundances is given by
\begin{equation}
  m_{\rm r, coll} \sim \frac{X_{\rm r} \int^{t_Z}\dot{\Psi}_{\rm SF}\,{\rm d}t}{V_{\rm MW} \int^{t(z=0)}\dot{N}_{\rm coll} \,{\rm d}t},
\end{equation}
where $\dot{\Psi}_{\rm SF}$ is the Galactic star-formation rate in mass per unit time (see Galactic chemical evolution below), $\dot{N}_{\rm coll}=R_{\rm LGRB}/f_{\rm b}$ is the volumetric rate of collapsar events, with $f_{\rm b}$ being the long GRB beaming fraction, and $V_{\rm MW}$ is the volume of Milky-Way equivalent galaxies (see Galactic chemical evolution below). Furthermore, $t_Z$ denotes the characteristic time after which long GRBs no longer occur in the Milky Way due to their suppression above a metallicity threshold (see below). If the rate of long GRBs tracks the star-formation rate, then the r-process mass per burst needed to explain the solar system abundances may be very roughly approximated as
\begin{equation}
m_{\rm r,coll} \sim X_{\rm r}f_{Z}^{-1}\frac{\dot{\rho}_{\rm SF}(z = 0)f_{\rm b}}{R_{\rm LGRB}(z = 0)} \approx 0.08-0.3 M_{\odot}\left(\frac{f_{Z}}{0.25}\right)^{-1}\left(\frac{X_{\rm r}}{4\times 10^{-7}}\right)\left(\frac{f_{b}}{5\times 10^{-3}}\right), \label{eq:Mr_min}
\end{equation}
where $\dot{\rho}_{\rm SF}(z = 0) \approx 2\times 10^{7}M_{\odot}$ yr$^{-1}$ Gpc$^{-3}$ is the local star-formation rate\citemeth{Kistler+08}, $f_{\rm b} \approx 5\times 10^{-3}$ is a recent estimate of the long GRB beaming fraction\citemeth{Goldstein+16}, and the prefactor $f_{\rm Z}=\int^{t_Z}\dot{\Psi}_{\rm SF}\,{\rm d}t/ \int^{t(z=0)}\dot{\Psi}_{\rm SF}\,{\rm d}t$ is a conservative limit on the fraction of star formation in the Milky Way that occurred below the critical metallicity threshold required for collapsars (see below).  As previous GRMHD simulations show that a fraction $f_{\rm w} \approx 0.3-0.4$ of the matter inflowing through the inner few tens of gravitational radii of the black hole is unbound in winds\citemain{Siegel2017a,Siegel2018a,Fernandez+18}, we conclude that a total mass $m_{\rm r, acc} = m_{\rm r, coll}/f_{\rm w} \lesssim 0.2-1 M_{\odot}$ must be accreted per collapsar to explain their solar system abundances.  This is well within the range predicted by theoretical models\citemain{MacFadyen&Woosley99} (see model for mass fallback above).

The prefactor $f_{\rm Z} < 1$ in Eq.~(\ref{eq:Mr_min}) accounts for the fact that host galaxy studies show that long GRBs may occur preferentially below a certain stellar metallicity\citemain{Stanek+06} and thus may shut-off in recent Galactic history.  It has been argued\citemeth{Perley+16} that this ``metallicity threshold" is between 12 + log$_{10}$(O/H) $<$ 8.64 and 12 + log$_{10}$(O/H) $<$ 8.94, independent of redshift.  While only a small fraction of the star formation occurs at such low metallicity in the local Universe, the fraction was higher at larger redshift. We can therefore obtain a lower bound on the value of $f_Z$ by looking at stellar metallicities at $z=0$. Assuming that the local population of stars is representative of the entire population of stars in the Milky Way, we can employ survey data from the Apache Point Observatory Galactic Evolution Experiment (APOGEE) to roughly estimate the fraction of stars below the aforementioned oxygen thresholds. Extended Data Fig.~\ref{fig:Apogee} shows $[\mathrm{O/H}]$ versus $[\mathrm{Fe/H}]$ (that is, metallicity) for all high signal-to-noise ($>\!200$) stars with oxygen abundance measurements of the $>2.7\times 10^{5}$ stars of the APOGEE DR14 data release\citemeth{Abolfathi+18}. Here, $[X/Y] = \log(N_X/N_Y) - \log(N_X/N_Y)_\odot$, where $N_X$ and $N_Y$ denote the number densities of elements $X$ and $Y$, respectively, and where $\odot$ refers to the solar value. From this sample, we estimate that $f_Z$ is of order unity, $f_Z \approx 0.23$ and $f_Z \approx 0.74$ for the above thresholds, respectively, motivating a fiducial value $f_{Z} = 0.25$ used in Eq.~(\ref{eq:Mr_min}).

\subsection{Observational signatures of r-process ejecta in GRB supernovae.}

The broad-lined type Ic (Ic-BL) supernovae observed in conjunction with long GRBs\citemeth{Woosley&Bloom06} rise to their peak luminosities on timescales of a few weeks, roughly corresponding to the time required for radiation to diffuse out of the interior of the ejecta. The evolution of the supernova light curve depends at all epochs on the energy injected into the system by radioactivity, which sets the overall luminosity budget for the supernova, and on the opacity, which controls the diffusion of thermal radiation through the ejecta. 

The introduction of r-process material in the ejecta has little impact on the radioactive energy supplied to the supernova. The luminosities of supernovae Ic-BL imply the synthesis of between 0.2 and 0.5\,$M_\odot$ of $^{56}$Ni \citemeth{Cano+16}),  comparable to the mass of r-process material our simulations suggest may be liberated from the collapsar accretion disk. However, as shown in Extended Data Fig.~\ref{fig:heating}, the specific radioactive heating due to the decay chain $^{56}$Ni $\rightarrow ^{56}$Co $\rightarrow ^{56}$Fe dominates that from the decay of freshly-synthesized r-process nuclei from around 1 day to several hundred days\citemeth{Metzger+10}, spanning the entire lifetime of the supernova's photospheric phase. Heating from r-process decay is therefore a minor perturbation to the heating due to the $^{56}$Ni decay chain, even if the mass of the r-processed outflow exceeds the mass of $^{56}$Ni by a factor of several.

The high opacity of r-process material compared to elements found in a typical supernova composition has a much greater potential to impact the supernova light curve and spectra. While the optical properties of the lighter r-process elements are similar to those of the iron-group elements\citemain{Kasen+17}$^,$\citemeth{Tanaka+18} that supply the bulk of the supernova opacity in traditional supernova models, a subset of the heavier r-process elements---in particular lanthanides and actinides---have a much higher opacity at wavelengths extending from the UV and optical out into the near infrared (NIR)\citemain{Kasen2013}. The increased opacity inhibits the diffusion of radiation through the ejecta, producing dimmer bolometric light curves\citemeth{Barnes&Kasen13,Tanaka&Hotokezaka13,Wollaeger+18}. The enhanced opacity at optical wavelengths also pushes the emission of lanthanide- and actinide-rich compositions out to redder regions of the electromagnetic spectrum. Even a small amount of lanthanide or actinide contamination can have a substantial effect on the overall opacity of the material, and strongly impact the resulting transient emission. How effectively r-process material can be ``hidden'' inside supernova ejecta is therefore sensitive to the mass of r-process material relative to the total ejecta mass, $M_\mathrm{ej}$, as well as its distribution within the ejecta cloud.

While detailed hydrodynamic studies are needed to determine the precise distribution of r-process material and $^{56}$Ni in the supernova ejecta, the two proposed scenarios by which GRB-supernovae may synthesize r-process material serve as instructive limiting cases. In MHD supernovae models\citemeth{Halevi&Mosta18} neutron-rich matter is ejected at high velocities during the supernova explosion phase. The mixing of this material with the $^{56}$Ni synthesized as early shocks heat and explode the progenitor star is a seemingly inevitable feature of this model. In contrast, in our model, outflows from the accretion disk are delayed relative to the explosion, and naturally concentrate r-processed material at low velocities, interior to the location of the bulk of the $^{56}$Ni, limiting the ability of the high-opacity r-process material to affect the supernova light curves and spectra.

To understand the effect of these differences on supernova observables, we carried out radiation transport simulations for models of supernovae Ic-BL with different patterns of r-process enrichment. In the model corresponding to the MHD supernova scenario, 0.025\,$M_\odot$ of r-process material is mixed uniformly through the ejecta out to a velocity of $0.15c$, then tapers smoothly to zero over $0.1c < v < 0.18c$. The distribution of $^{56}$Ni tracks the r-process matter, but is scaled up by a factor of ten so that $M_{56} = 0.25\,M_\odot$. The collapsar disk wind model contains a much higher mass of r-process material ($0.25\,M_\odot$) but concentrates it in the center of the ejecta, at $v <0.015c$. Exterior to the embedded r-process core, $0.25\,M_\odot$ of $^{56}$Ni is distributed uniformly out to $v = 0.15c$.

Radiation transport calculations were carried out using the time-dependent Monte Carlo LTE radiation transport code SEDONA\citemeth{Kasen+06} on a one-dimensional grid that extends to a maximum velocity $v=0.2c$.  Both models had a total ejecta mass $M_\mathrm{ej} = 2.5\,M_\odot$, a kinetic energy of $6.7\times 10^{51}$ erg, and $M_{56} = 0.25 \,M_\odot$ of $^{56}$Ni. We adopted as our density profile the polar angle-averaged density profile of a two-dimensional supernova Ic-BL model\citemeth{Barnes+18}, with the same composition of the stellar progenitor. We assume that the supernova explosion mixes r-process matter and/or $^{56}$Ni into the ejecta, but does not change the relative abundances of the non-radioactive species present in the progenitor.

The atomic data for calculating the bound-bound opacity of non-lanthanide species is taken from previous work\citemeth{Kurucz&Bell95}. Synthetic data for the ions of the lanthanide neodymium (Nd) was determined using atomic structure calculations, as previously described\citemain{Kasen2013}\citemeth{Barnes&Kasen13}. We use a simplified composition to stand in for the full range of r-process elements and boost the lanthanide opacity accordingly, following the procedure previously outlined\citemeth{Barnes&Kasen13}. The disk winds are assumed to contain 3\% lanthanides and actinides and 29\% light r-process (\emph{d}-block) elements by mass, with the remainder composed of a low-opacity filler. Radioactivity from the $^{56}$Ni decay chain takes the form of $\gamma$-rays, which are explicitly propagated through the ejecta, and positrons, which are assumed to thermalize instantaneously. Since even energetic supernovae have ejecta that are dense compared to a typical kilonova, we assume that the radioactive energy from r-process decay has a thermalization efficiency comparable to that of the denser (more massive/slower) kilonova models previously studied\citemeth{Barnes+16}.

The synthetic light curves and spectral snapshots for the two models are presented in Fig.~3. As expected, the effects of r-process material on the supernova observables is strongly dependent on the degree of mixing. Even though the collapsar wind model contains ten times more r-process matter by mass than the MHD supernova model, the impact of the r-process is much more dramatic in the latter.

While the bolometric light curve for the collapsar wind model generally conforms to those of observed supernovae Ic-BL, the light curve of the MHD supernova model is dimmer, and exhibits an extended plateau rather than a clearly-defined peak. Differences are also apparent in the spectral energy distribution (SED) of the two models. Relative to the collapsar wind model, the emission of the MHD supernova model is shifted away from the blue and optical bands, and into the near infrared. This is due to the high opacity of the lanthanide-polluted ejecta, which inhibits the diffusion of optical photons and redistributes radiation to redder wavelengths.

In the collapsar wind model, the high-opacity material is interior to the $^{56}$Ni, and only the thermal energy associated with r-process decay is reprocessed as described above. Since r-process radioactivity represents just a small fraction of the total radioactivity at times from $t\sim 1$ day to $t\sim 1$ year for models with $M_{\rm rp} \approx M_{56}$, the effects on the light curve and spectra in this case are minor. The spectra of collapsar wind model looks fairly typical of supernovae Ic-BL out to $t > 1$ month. Even beyond this point, the energy emitted with $\lambda > 10,000$ \AA\ is minimal. In contrast, in the MHD supernova model, the thermal energy produced by $^{56}$Ni- and $^{56}$Co-decay is also processed through high-opacity ejecta and reddened. As a result, the effects of high-opacity r-process material manifest even before the light curve peaks; by $t = 45$ days, the spectrum has lost any identifiable features of a supernova.

The emission of any supernova that generates comparable amounts of r-process material and $^{56}$Ni will eventually be dominated by its r-process component, as the energy generated from r-process decay dominates that from the decay of $^{56}$Ni/$^{56}$Co on very long timescales (more than a few hundred days). Systems that undergo a minimal amount of mixing, such that some of the r-process material is ``backlit'' by the decay of $^{56}$Ni and $^{56}$Co may exhibit r-process features sooner. Regardless of the exact degree of mixing, signs of the r-process will eventually appear, either in the late photospheric or the nebular phase. Late-time observations can therefore probe the production of r-process elements in GRB-supernovae.

Relatively few NIR spectra are available for GRB supernovae, in part because emission from the non-thermal GRB afterglow dominates the NIR bands. This is particularly true of the most powerful GRB jets (classical/cosmological GRBs), which in our disk wind r-process scenario might be associated with the greatest yield of r-process elements and thus the most conspicuous spectral features. For instance, NIR spectra of the supernova that accompanied GRB 100316D were taken\citemeth{Bufano+12} out to 30 days, and showed no clear NIR excess indicative of r-process material. In addition, our centrally-concentrated r-process spectral models in Fig.~3 show that such features can be hidden to at least $t \gtrsim 65$ days. However, GRB 100316D belongs to the class of low-luminosity GRBs (LLGRBs; see above discussion under rates of Galactic r-process production) and the radiated $\gamma$-ray energy was much lower than those of cosmological long GRBs, which are the focus of this paper. Since the per-event rate of r-process material in LLGRBs would be a factor $\sim\!10$ times less than those of cosmological GRBs (see above), constraints from lightcurves and spectra of LLGRBs on our model will be less stringent, and late-time data on cosmological GRBs will be needed.

Although other effects may also cause late-time NIR excess, the presence of r-process material may be inferred with higher certainty in combination with spectra. Future late-time optical/NIR observations of neutron star mergers (a ``pure'' r-process source), will reveal the dominant r-process emission lines and spectral features of r-process materials. These can then be sought in late-time or nebular spectra of particularly nearby GRB-supernovae. Future instruments with greater sensitivity, such as the James Webb Space Telescope\citemeth{Villar+18,Wu+18}, will allow a more comprehensive survey of these events' late-time NIR emission.

\subsection{Galactic chemical evolution.}

In order to explore the implications of r-process enrichment due to collapsars for Galactic chemical evolution, we have developed a simple one-zone Galactic model, similar to previous models\citemain{Hotokezaka2018a}$^,$\citemeth{Cote2017a}. Although a one-zone model is inadequate for describing early phases in the Galactic chemical evolution, when hierarchical structure growth and incomplete mixing greatly complicate the picture\citemain{Shen+15,vandeVoort+15}$^,$\citemeth{Komiya2016,Cescutti2015,Wehmeyer2015,Hirai2015,Ishimaru2015}, here we instead focus on the late-time chemical evolution, when the homogeneous approximation is well-motivated. We assume that the abundances of $\alpha$-elements (such as magnesium) and europium trace the histories of core-collapse supernovae and r-process production\citemeth{Burris2000,Battistini2016}, respectively. We assume that the [Eu/Fe] abundances are the result of the combined enrichment processes by neutron-star mergers and core-collapse supernovae. Both core-collapse supernovae and type Ia supernovae contribute to the iron abundance distribution. As usual, we adopt the definition $[X/Y] = \log(N_X/N_Y) - \log(N_X/N_Y)_\odot$, where $N_X$ and $N_Y$ denote the number densities of elements $X$ and $Y$, respectively, and where $\odot$ refers to the proto-solar value, i.e., to the solar value at the time of formation of the Sun 4.568\,Gyr ago.

We assume that the Galactic star-formation history follows the cosmic star-formation history\citemeth{Madau2017}
\begin{equation}
  \dot{\rho}_\mathrm{SF}(z) = 0.01 \frac{(1 + z)^{2.6}}{1+[(1+z)/3.2]^{6.2}} M_\odot \mathrm{Mpc}^{-3} \mathrm{yr}^{-1}.
\end{equation}
Our conclusions are largely independent of the assumed star-formation history; we reach similar conclusions for drastically different star-formation histories, e.g., a constant star-formation history, $\dot{\rho}_\mathrm{SF}(z) = \text{const}$. The rates of core-collapse supernovae and collapsar events are assumed to follow the star-formation history with negligible time delay, while type Ia supernovae and neutron-star mergers are assumed to follow with significant time delay:
\begin{eqnarray}
\mskip-25mu R_\mathrm{CCSN}(t) \mskip-10mu &=& \mskip-10mu c_\mathrm{CCSN} \Psi_\mathrm{SF}(t), \label{eq:R_CCSNe}\\
R_\mathrm{coll}(t) \mskip-10mu &=& \mskip-10mu c_\mathrm{coll} \Psi_\mathrm{SF}(t), \label{eq:R_coll}\\
  R_\mathrm{SNIa}(t) \mskip-10mu &=& \mskip-10mu c_\mathrm{SNIa} \int^t_0 D(t-t';t_\mathrm{min,SN Ia},b_\mathrm{SNIa})\Psi_\mathrm{SF}(t')\mathrm{d}t', \label{eq:R_SNIa}\\
  R_\mathrm{NS}(t) \mskip-10mu &=& \mskip-10mu c_\mathrm{NS} \int^t_0 D(t-t';t_\mathrm{min,NS},b_\mathrm{NS})\Psi_\mathrm{SF}(t')\mathrm{d}t'. \label{eq:R_NS}
\end{eqnarray}
Here, $\Psi_\mathrm{SF}$ denotes the Galactic star-formation rate in mass per unit time. In converting volumetric into Galactic rates, we assume $0.01$ Milky-Way equivalent galaxies per $\mathrm{Mpc}^{-3}$ \citemeth{Kopparapu2008}. While the results for chemical evolution are somewhat dependent on this normalization, further motivation arises from the fact that the model discussed here must reproduce solar metallicity 4.6\,Gyr ago. In order to relate cosmic time $t$ to redshift, we assume a cosmic concordance cosmology with parameters ($\Omega_M$, $\Omega_\Lambda$, $\Omega_b$, $h$) = (0.3, 0.7, 0.046, 0.7). Furthermore,
\begin{equation}
  D(t;t_\mathrm{min},b) = \frac{\Theta(t-t_\mathrm{min})}{t^b}
\end{equation}
defines a delay time distribution, with $\Theta$ the Heaviside function and $t_\mathrm{min}$ the minimum delay time. We adjust $c_\mathrm{CCSN}$ such that $R_\mathrm{CCSN}$ at $z=0$ corresponds to the local observed rate\citemeth{Li2011} of $7.05^{+1.43}_{-1.25}\times 10^{-5}\,\mathrm{Mpc}^{-3}\mathrm{yr}^{-1}$. For collapsar events, we adjust $c_\mathrm{coll}$ such that $R_\mathrm{coll}(z=0)$ corresponds to $R_\mathrm{LGRB}(z=0)/f_b$, where $R_{\rm LGRB}(z = 0) \approx 1.3_{-0.7}^{0.6}$ Gpc$^{-3}$ yr$^{-1}$ is the rate of local long GRBs\citemeth{Wanderman&Piran10} and $f_{b} \approx 5\times 10^{-3}$ is the beaming fraction\citemeth{Goldstein+16}; this does not include the separate class of low-luminosity GRBs, which are not the focus of our argument here (see also above discussion under rates of r-process production). We explore different $t_\mathrm{min,SNIa}$ between $40\,\mathrm{Myr}$ and $1\,\mathrm{Gyr}$ and fix $b_\mathrm{SNIa}=1.0$ to the observationally inferred value\citemeth{Maoz2017}. For each choice of the minimum delay time, we calibrate $c_\mathrm{SNIa}$ to the Hubble-time integrated type Ia supernova production efficiency of $(1.3\pm 0.1)\times 10^{-3}$ per $M_\odot$ of stellar mass formed\citemeth{Maoz2017}. Furthermore, we fix $t_\mathrm{min,NS} = 20\,\mathrm{Myr}$ and $b_\mathrm{NS}=1.0$ as inferred from observations of the redshift distribution and peak flux of short GRBs\citemeth{Wanderman2015} as well as from population synthesis modeling\citemeth{Dominik2012,Chruslinska2018}. We tune $c_\mathrm{NS}$ such that $R_\mathrm{NS}(z=0)$ corresponds to the local rate of binary neutron-star mergers inferred from LIGO/Virgo\citemain{LIGO+17DISCOVERY}, $R_{\rm NSNS} = 1540^{+3200}_{-1220}\, \mathrm{Gpc}^{-3} \mathrm{yr}^{-1}$.

Using the aforementioned assumptions, the evolution of certain chemical elements in the interstellar medium (ISM) can be written as
\begin{eqnarray}
\frac{\mathrm{d}M_\mathrm{Mg}}{\mathrm{d}t} = m_\mathrm{Mg} R_\mathrm{CCSN}(t) - M_\mathrm{Mg} f(t), \label{eq:M_Mg}\\
\frac{\mathrm{d}M_\mathrm{Fe}}{\mathrm{d}t} = m_\mathrm{Fe,CCSN} R_\mathrm{CCSN}(t) + m_\mathrm{Fe,SNIa} R_\mathrm{SNIa}(t)- M_\mathrm{Fe} f(t), \label{eq:M_Fe}\\
\frac{\mathrm{d}M_\mathrm{Eu}}{\mathrm{d}t} = m_\mathrm{Eu,coll} f_{Z,{\rm cut}}(t)R_\mathrm{coll}(t) + m_\mathrm{Eu,NS} f_{\mathrm{NS}}R_\mathrm{NS}(t)- M_\mathrm{Eu} f(t). \label{eq:M_Eu}
\end{eqnarray}
Here, $M_X$ and $m_X$ denote the total mass of element $X$ in the ISM and the mass of element $X$ produced in each event type, respectively.  We employ the observationally inferred values $m_\mathrm{Mg}=0.12\,M_\odot/\mathrm{event}$, $m_\mathrm{Fe,CCSN}=0.074\,M_\odot/\mathrm{event}$, and $m_\mathrm{Fe,SNIa}=0.7\,M_\odot/\mathrm{event}$\citemeth{Maoz2017}, which are consistent with the choice of delay time distributions and cosmic star-formation history used here. Regarding the production of europium in neutron-star mergers, we employ the observational value of $m_{\rm r,NS}\approx 0.05\,M_\odot$ based on the total r-process production inferred from the GW170817 kilonova\citemain{Cowperthwaite2017}$^,$\citemeth{Villar2017} and assume that this is a representative amount for all neutron-star mergers. This value is consistent with outflows from the remnant accretion disk\citemain{Siegel2017a,Siegel2018a,Fernandez+18}. According to our estimate in Eq.~\eqref{eq:Macc}, the amount of r-process material ejected by collapsars is higher by a factor of $\sim\!10$, and thus we set $m_\mathrm{\rm r,coll}=0.5\,M_\odot$. 

In translating the total r-process mass $m_r$ into the europium mass per event, $m_{\rm Eu}$, we assume the Solar abundance distribution of r-process elements\citemeth{Arnould2007} starting at mass number $A=69$. Note that for the solar r-process abundance pattern and this choice of minimum atomic mass number, the lanthanide mass fraction of the ejecta is consistent with that inferred for the red kilonova of GW170817\citemain{Kasen+17}. 

The function $f(t)$ in Eqs.~\eqref{eq:M_Mg}--\eqref{eq:M_Eu} represents the mass-loss rate due to star formation and Galactic outflow, which we specify as in previous work\citemain{Hotokezaka2018a}. We introduce an additional factor $0<f_\mathrm{NS}<1$ in Eq.~\eqref{eq:M_Eu} which is degenerate with the neutron-star merger rate, in order to explore smaller merger rates within the large observational error bars and to account for the fact that not all neutron-star mergers contribute equally to r-process enrichment of the galaxy (e.g., due to their natal kicks); we employ a fiducial value of $f_\mathrm{NS}\approx 0.5$. Finally, we introduce a collapsar cutoff factor $f_{Z,{\rm cut}}(t)$ that accounts for the fact that collapsar events only occurred below a certain metallicity threshold in our galaxy (see rates of $r$-process production above; Extended Data Fig.~\ref{fig:Apogee}). It smoothly interpolates between unity and zero in-between $12 + \log({\rm O/H})=8.64$ and $12 + \log({\rm O/H})=8.94$ (see rates of $r$-process production above).

Extended Data Fig.~\ref{fig:chemical_evolution} compares the model predictions with observational data of stars from the Stellar Abundances for Galactic Archeology (SAGA) database\citemeth{Suda2008} (http://sagadatabase.jp). Although we show evolution across the entire metallicity range, we focus on the late-time evolution $[\text{Fe/H}] > -1$, as noted above. For this purpose, we specifically highlight observational Eu data obtained for stars in the Galactic disk\citemeth{Battistini2016}.

The mean [Mg/Fe] abundance is observed to be roughly constant with increasing [Fe/H] up to a `knee' at $[\text{Fe/H}]\approx -1$ (corresponding to $z\approx 2$; see Extended Data Fig.~\ref{fig:chemical_evolution}a). Such a knee is also obtained with our model calculations, favoring larger values of the minimum delay time for type Ia supernovae. Within our model this behavior is the result of a sharp increase in the relative type Ia supernova contribution to the iron production (thus decreasing [Mg/Fe]) due to the power-law tail of the type Ia supernova delay time distribution\citemeth{Maoz2017}. 

Similarly, the mean value of [Eu/Fe] is observed to decrease for $[\text{Fe/H}] > -1$. As shown in Extended Data Fig.~\ref{fig:chemical_evolution}c, such a decline is difficult to obtain if only neutron-star mergers contribute to r-process nucleosynthesis, independent of the assumed star-formation history\citemain{Cote2017b,Hotokezaka2018a}$^,$\citemeth{Cote2018b}. The fact that the evolution of [Eu/Fe] is much shallower than [Mg/Fe] or even increasing for comparable minimum delay times of type Ia supernovae and mergers is a result of mergers being delayed relative to star formation by a delay time distribution $\propto t^{-1}$. Such a flat or even increasing [Eu/Fe] evolution for [Fe/H] $> -1$ is thus a generic feature for merger-only r-process scenarios. While much steeper slopes of the neutron-star merger delay time distribution $\propto t^{-2}$ would alleviate this problem\citemain{Cote2017b,Hotokezaka2018a}$^,$\citemeth{Cote2018b}, they would be inconsistent with values derived from observations of the redshift distribution of short GRBs\citemeth{Fong+13} and population synthesis models\citemeth{Wanderman2015,Dominik2012}.

The observed mean [Eu/Fe] evolution at [Fe/H] $> -1$ can be obtained more naturally if collapsars contribute to r-process element production (cf.~Extended Data Fig.~\ref{fig:chemical_evolution}). The fact that a population of r-process enrichment events follows the star-formation history more closely results in a knee in the [Eu/Fe] evolution similar to that of $\alpha$ elements (if the overall r-process contribution is dominant), which necessarily leads to a decline of [Eu/Fe] at late times. This point is illustrated in more detail in Extended Data Fig.~\ref{fig:chemical_evolution_eta_merger}, which shows scenarios for varying relative contributions of neutron-star mergers and collapsars to the total r-process element production in the galaxy. The neutron-star merger contribution is altered by renormalizing the neutron-star merger rates, tuning $f_\mathrm{NS} R_\mathrm{NS}(z=0)$ (cf.~Eq.~\eqref{eq:M_Eu}) by a factor between 0.3 and 100. All scenarios produce sufficient r-process material to explain the total amount of r-process material in the galaxy, which can be roughly estimated to $\sim\!2.6\times 10^{4}\,M_\odot$, assuming a total stellar mass of the Milky Way of $\approx\! 6.4\times 10^{10}\,M_\odot$\citemeth{McMillan2011} and a mean r-process abundance of stars in the Galactic disk (for $A\ge 69$) similar to the solar value\citemeth{Battistini2016}. A best fit to the mean abundances is obtained if collapsars contribute well over half of the total Galactic r-process material (red and purple lines).

\end{methods}

\bibliographymeth{manuscript}

\renewcommand{\figurename}{Extended Data Figure}
\setcounter{figure}{0}

\begin{figure}
\centering
\includegraphics[width=0.7\linewidth]{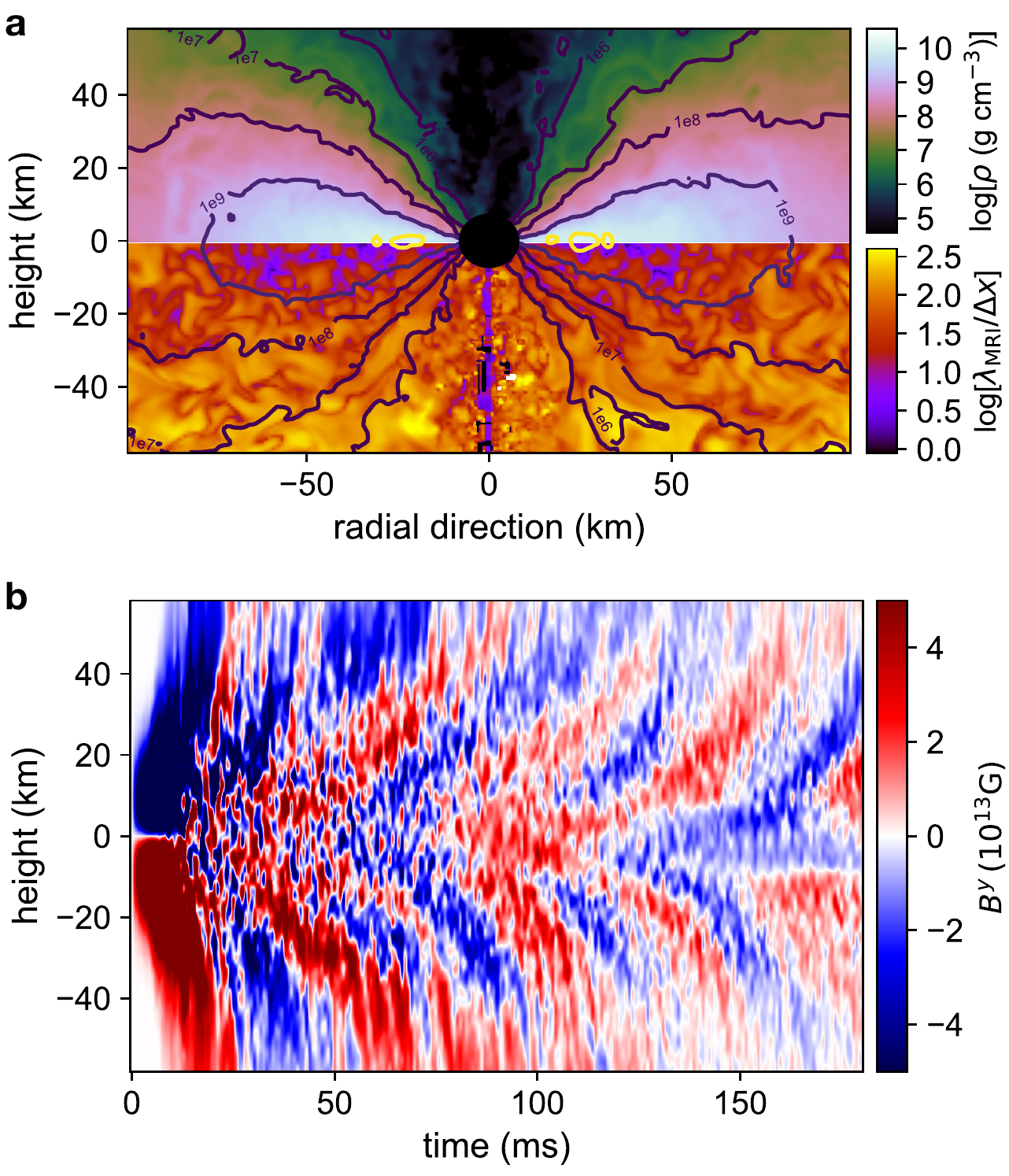}
\caption{{\bf MHD characteristics of simulations.} {\bf a}, Simulation snapshot of the meridional plane for Run 2, showing the rest-mass density ($\rho$; upper half; contours at $10^6$, $10^7$, $10^8$, $10^9$, $10^{10}$\,g\,cm$^{-3}$) and number of grid points per wavelength of the fastest-growing MRI mode ($\lambda_{\rm MRI}/\Delta x$; lower half) once the stationary state has been reached after 30\,ms. Note that the MRI is well resolved. {\bf b}, Spacetime diagram of the $y$ component of the magnetic field for Run 2, radially averaged between 45 and 70\,km from the rotation axis in the $x$-$z$ (meridional) plane, as a function of height $z$ relative to the equatorial plane, and time, indicating a fully operational dynamo and a steady turbulent state of the disk after about $20-30$\,ms.}
\label{fig:MHD_turbulence}
\end{figure}

\begin{figure}
\centering
\includegraphics[width=0.75\linewidth]{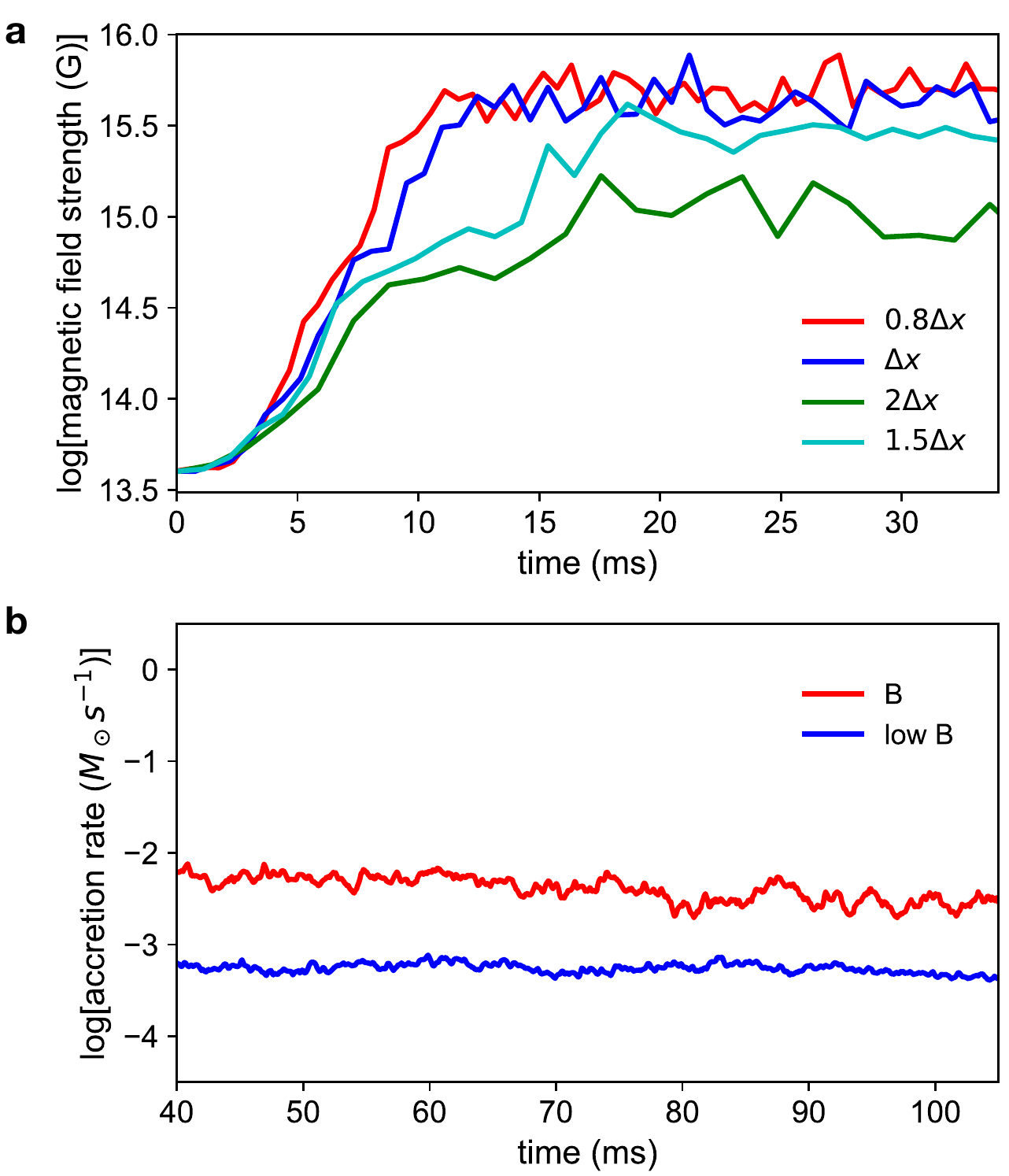}
\caption{{\bf Numerical characteristics of simulations.} {\bf a}, Resolution study showing the maximum magnetic field strength in the disk midplane for Run 2 and additional runs with varying resolution (but otherwise identical), indicating that magnetic field amplification has converged for the fiducial run with finest grid spacing, $\Delta x$ (see key). {\bf b}, Comparison of the accretion rate of Run 2 (`$B$' in key) to a run with much lower initial magnetic field (but otherwise identical; `low $B$' in key), showing that angular momentum transport and viscous heating are set by MHD turbulence.}
\label{fig:numerical_tests}
\end{figure}

\begin{figure}
\centering
\includegraphics[width=0.75\linewidth]{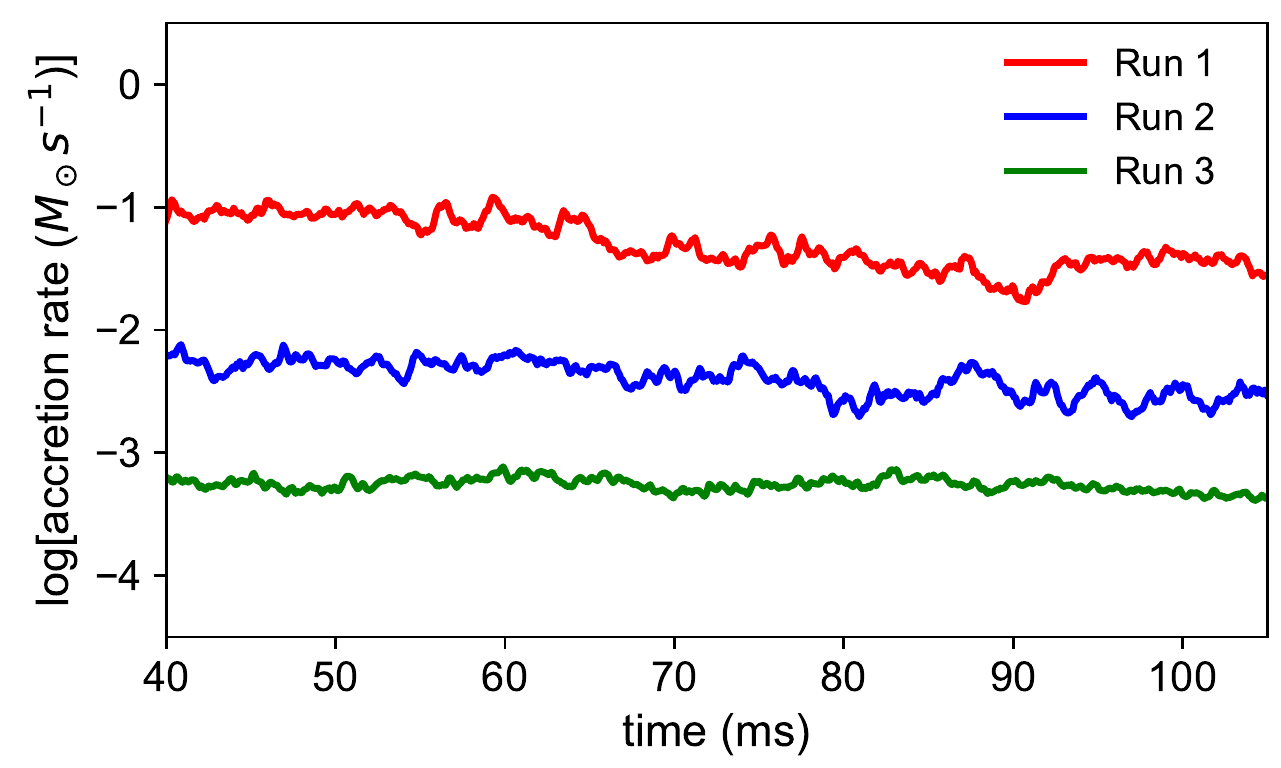}
\caption{{\bf Black hole accretion rate.} Shown are the black hole accretion rates as a function of time for the three main runs (1, 2, and 3), which represent the state of a collapsar accretion flow at consecutively later times following the core collapse of the star (see Fig.~1).}
\label{fig:acc_rates}
\end{figure}

\begin{figure}
\centering
\includegraphics[width=0.75\linewidth]{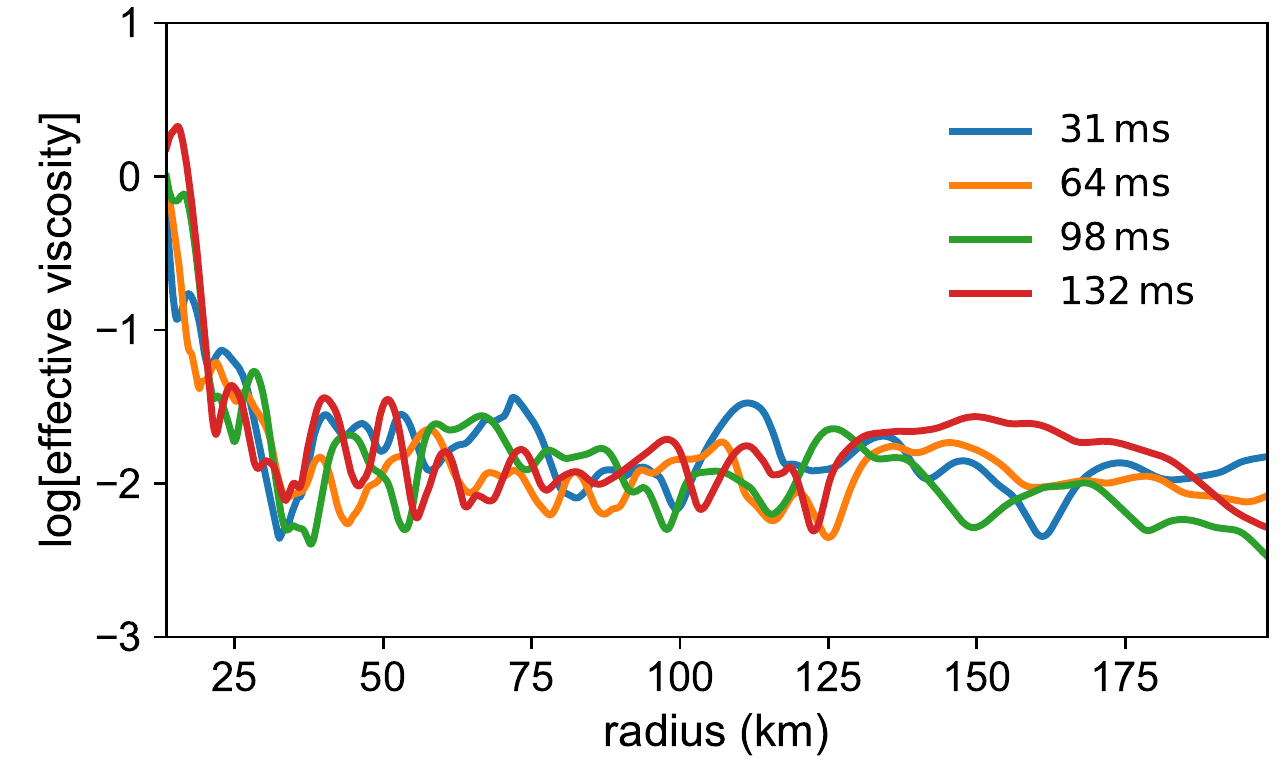}
\caption{{\bf Effective MHD viscosity of the collapsar accretion disk.} Shown are radial profiles of the effective $\alpha$-viscosity parameter for Run 1 at different times (see key) spanning 100\,ms of evolution.}
\label{fig:alpha_visc}
\end{figure}

\begin{figure}
\centering
\includegraphics[width=0.75\linewidth]{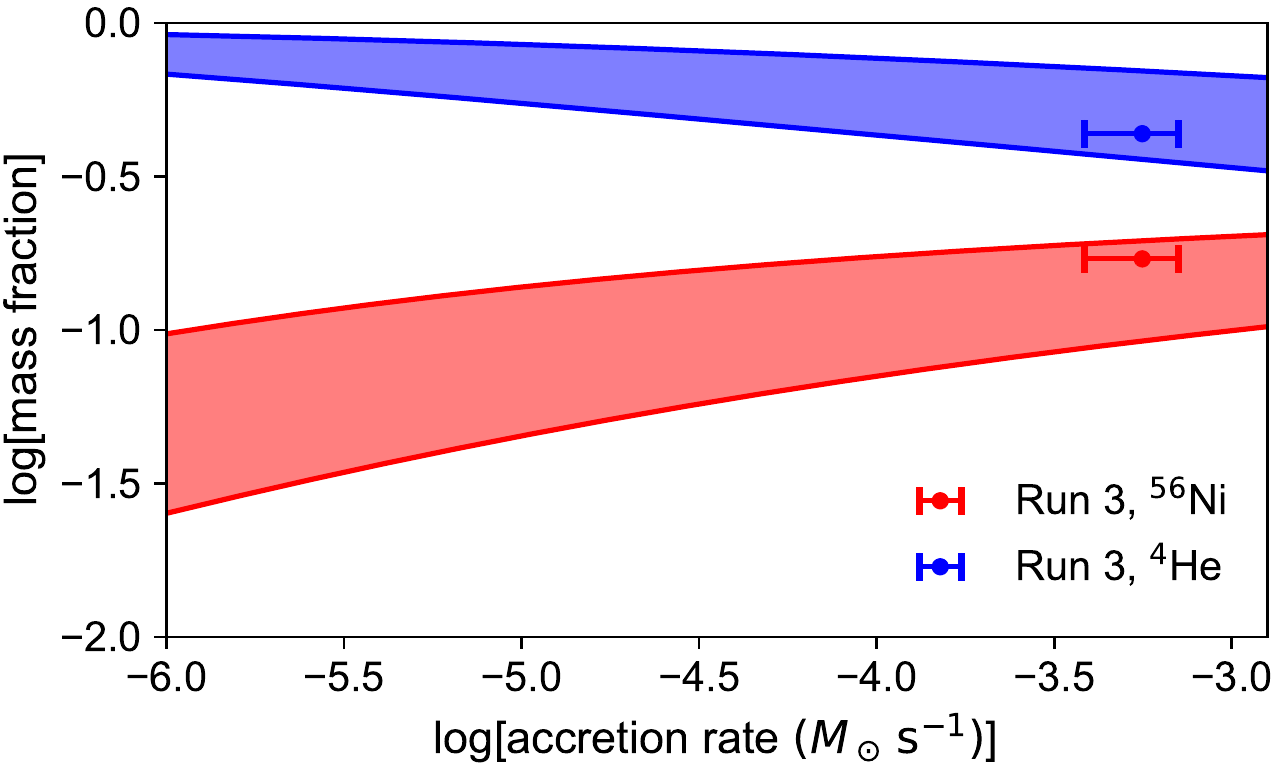}
\caption{{\bf Nickel and Helium production in the collapsar disk outflows.} Shown are estimated $^{56}$Ni (red) and $^4$He (blue) mass fractions based on Eq.~\eqref{eq:Yseed} along with extracted mass fractions from Run 3 (uncertainties in the accretion rate defined as in Fig.~1), at accretion rates below the ignition threshold (Eq.~\eqref{eq:Mdotign}). The colored bands correspond to estimates bracketing the distribution of expansion timescales between 5 and 30\,ms.}
\label{fig:XNi}
\end{figure}

\begin{figure}
\centering
\includegraphics[width=0.75\linewidth]{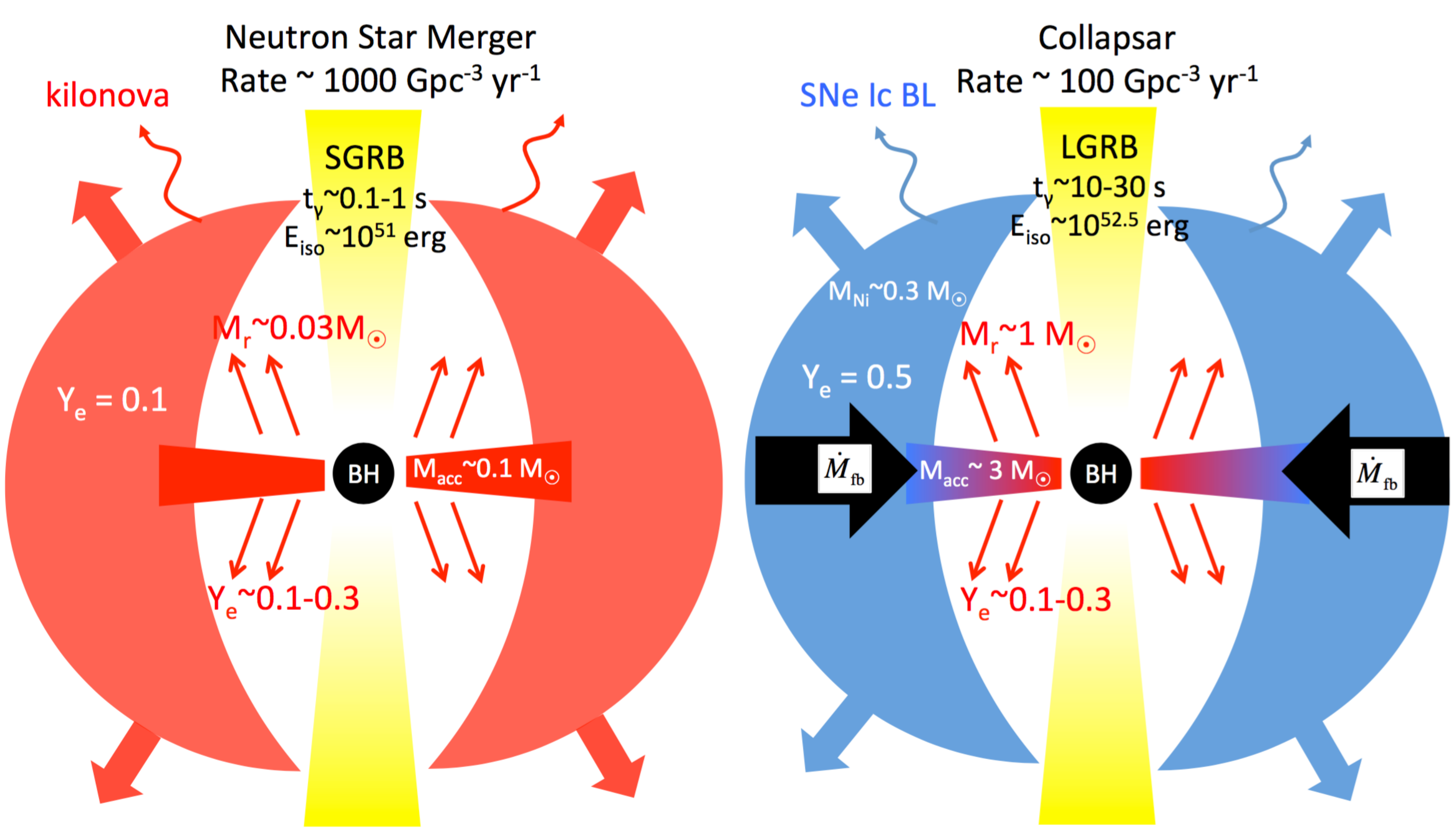}
\caption{{\bf r-process enrichment through neutron-star mergers (left) and collapsars (right).} Although collapsars are somewhat less frequent than mergers over cosmic time, their higher r-process yields (by a factor of about 30, if calibrated using the energetics of long versus short GRB jets) make them an important and likely dominant r-process site (Eq.~\eqref{eq:Macc}). See Methods for nomenclature.}
\label{fig:cartoon}
\end{figure}

\begin{figure}
\centering
\includegraphics[width=0.75\linewidth]{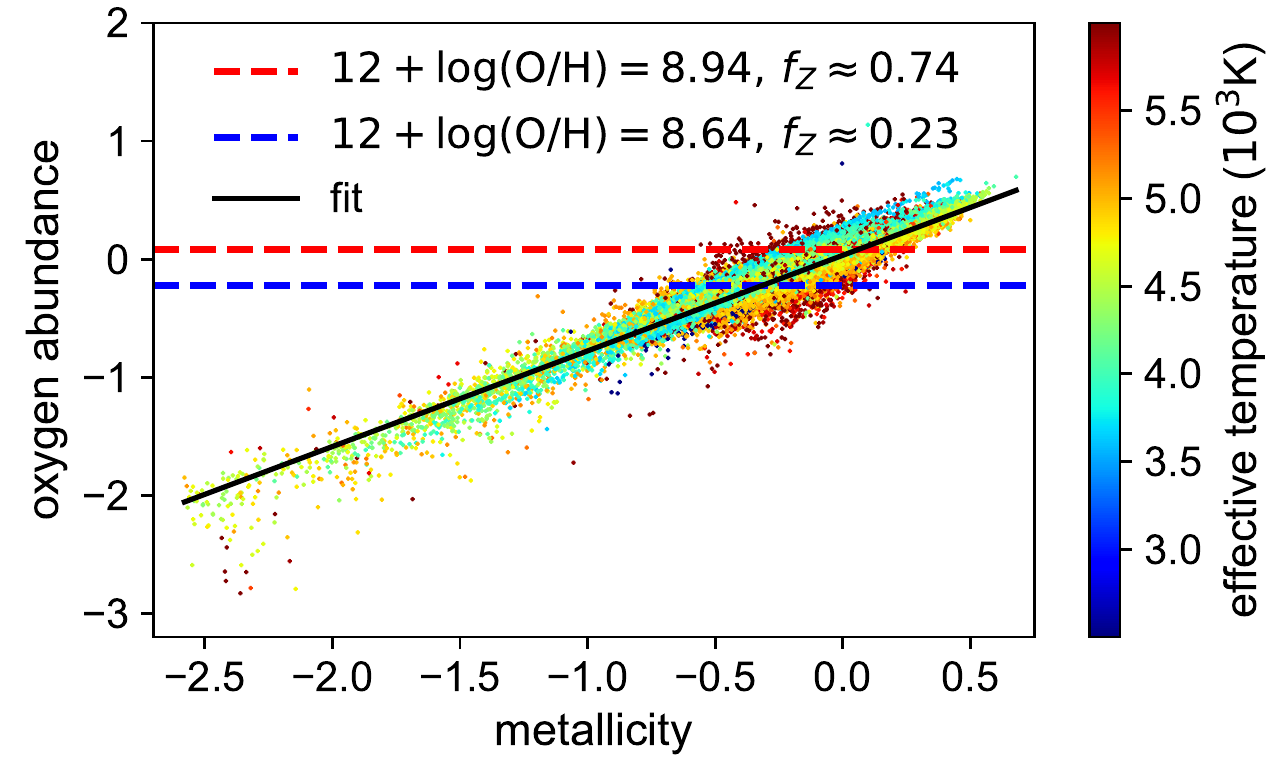}
\caption{{\bf Oxygen abundance and metallicity of Milky Way stars.} Oxygen abundances for high-signal-to-noise ($>\!200$) stars from the full APOGEE DR14 sample\protect\citemeth{Abolfathi+18} are plotted versus metallicity, with individual stars being colour-coded by their effective temperature (colour scale at right). Shown for comparison with dashed lines are the range of oxygen thresholds for GRB generation\protect\citemeth{Perley+16}, with the fraction $f_Z$ of stars below the threshold indicated. An order-unity fraction of stars in the Milky Way were formed at metallicities below the threshold needed for collapsar production.}
\label{fig:Apogee}
\end{figure}

\begin{figure}
\centering
\includegraphics[width=.75\linewidth]{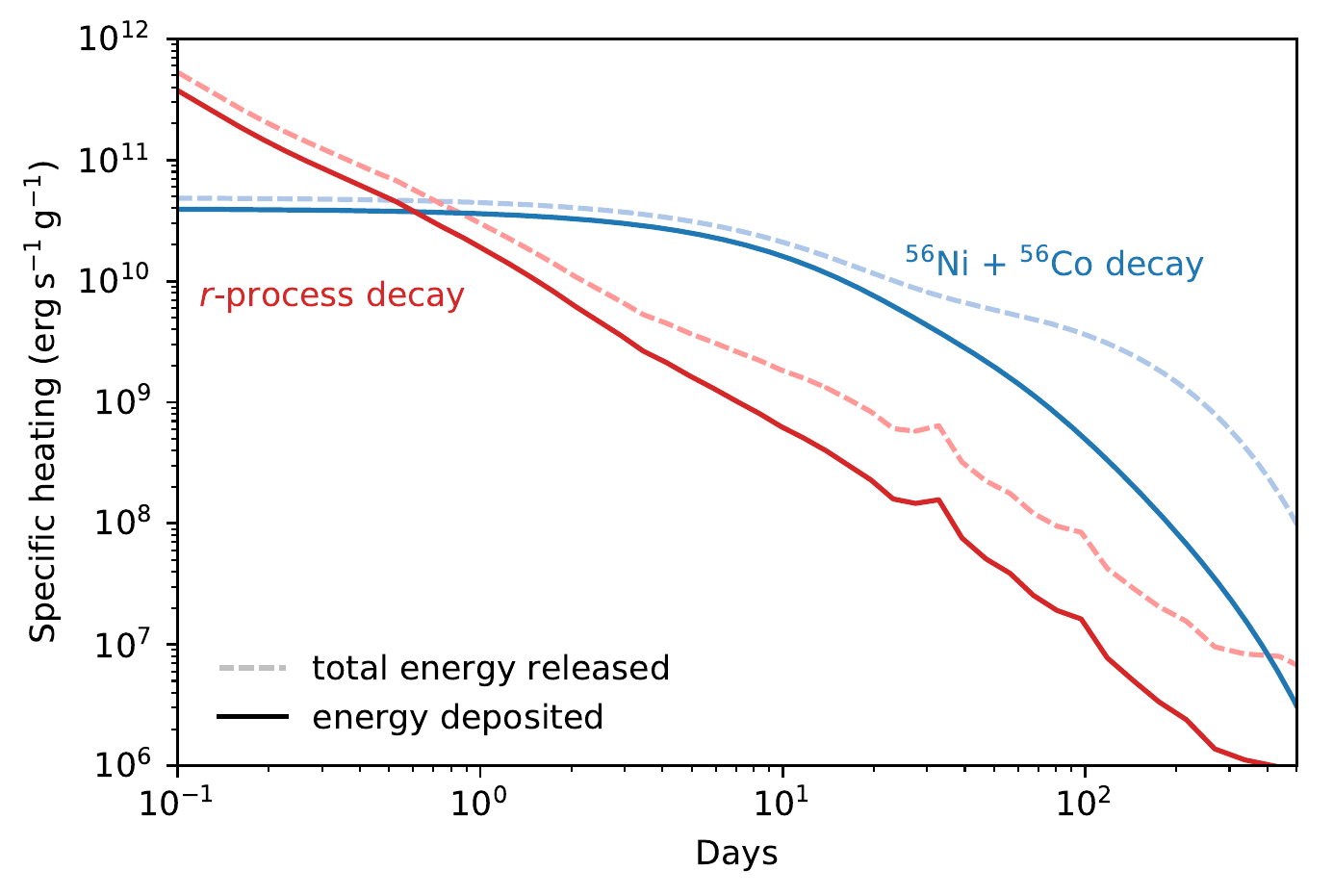}
\caption{{\bf Radioactive heating rate from the r-process and the nickel burned in the GRB supernova.} The specific radioactive heating rate of the $^{56}$Ni/$^{56}$Co chain from the associated supernova (blue) exceeds that of r-process nuclei (red) for 1 day $\lesssim t \lesssim$ 600 days. This makes it possible to conceal large quantities of r-process material from collapsars in the centre of long GRB supernovae until very late times $t \gtrsim 100$\,days. The difference between the released (dashed lines) and deposited (solid lines) energy reflects energy lost to neutrinos, to incomplete deposition of $\gamma$-ray energy from $^{56}$Ni/$^{56}$Co decay\protect\citemeth{Colgate.ea.1980_SN.lum.g_ray.dep}, and to inefficient thermalization of r-process decay products (as previously calculated\protect\citemeth{Barnes+16}). Figure adapted from previous work\protect\citemeth{Metzger+10}.}
\label{fig:heating}
\end{figure}

\begin{figure}
\centering
\includegraphics[width=\linewidth]{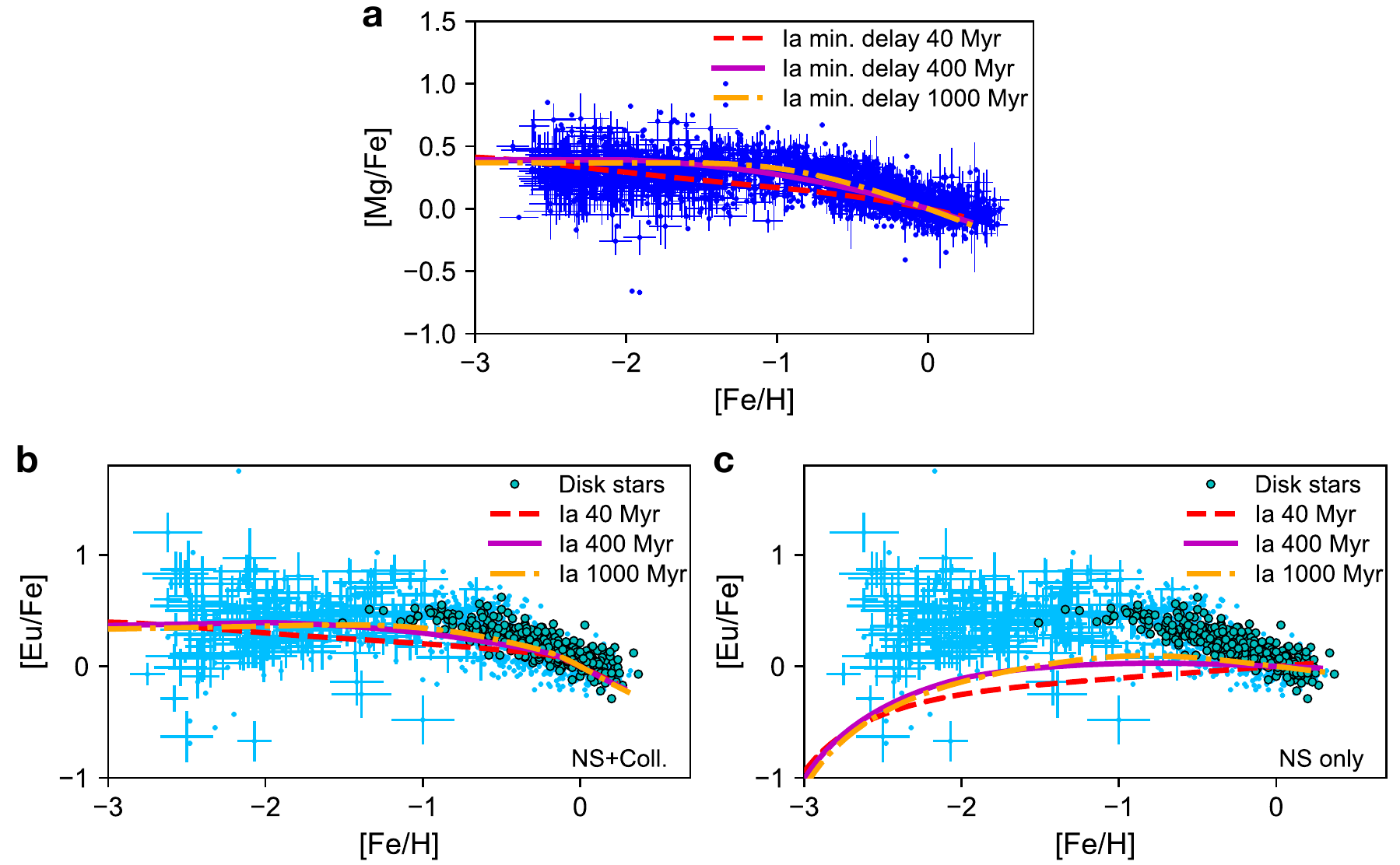}
\caption{{\bf Galactic chemical evolution.} Shown are comparisons of model predictions with observational data for magnesium and europium abundances from the SAGA database\protect\citemeth{Suda2008} (http://sagadatabase.jp) and europium abundances of Galactic disk stars\protect\citemeth{Battistini2016}. For definitions of error bars we refer to the SAGA database\protect\citemeth{Suda2008}. Model predictions are shown for different minimum delay times (see keys) of type Ia supernovae with respect to star formation.
{\bf a}, Comparison for magnesium as a representative $\alpha$-element. {\bf b}, Comparison for europium as an r-process tracer, assuming both neutron-star mergers and collapsars (NS+Coll.) contribute to Galactic r-process nucleosynthesis. Note that the decreasing trend of [Eu/Fe] at high metallicity can be obtained. {\bf c}, As {\bf b} but assuming that only neutron-star mergers (NS only) contribute to Galactic r-process nucleosynthesis, showing that merger-only models cannot explain the [Eu/Fe] trend of stars in the Galactic disk.}
\label{fig:chemical_evolution}
\end{figure}

\begin{figure}
\centering
\includegraphics[width=0.75\linewidth]{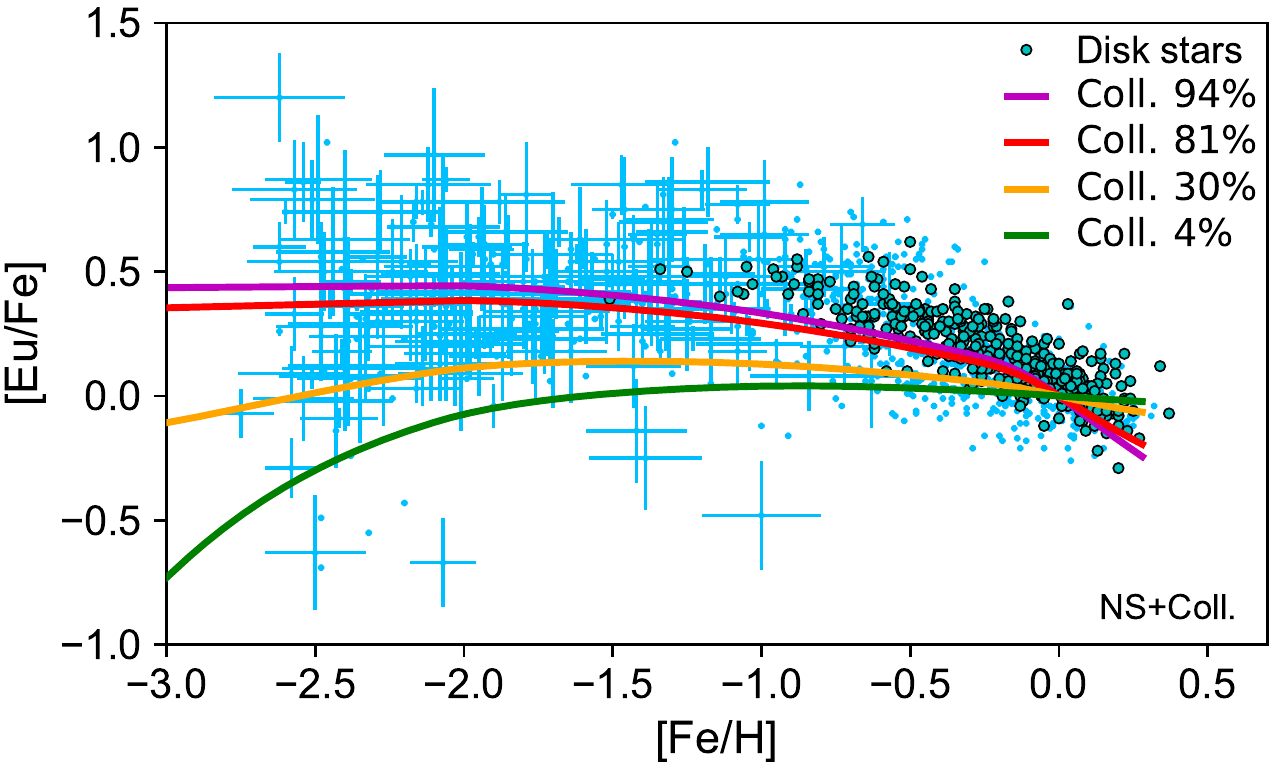}
\caption{{\bf Galactic chemical evolution with varying collapsar contribution to the r-process.} The figure shows a comparison of model predictions with observational data for europium abundances as in Extended Data Fig.~\ref{fig:chemical_evolution} (light blue points with error bars refer to the SAGA dataset\protect\citemeth{Suda2008} (http://sagadatabase.jp), that is, stars in the Milky Way disk plus halo, while cyan points refer to disk stars only\protect\citemeth{Battistini2016}), assuming a minimum type Ia supernova delay time of 400\,Myr and varying the contribution of neutron-star mergers. For definitions of error bars we refer to the SAGA database\protect\citemeth{Suda2008}. The curves are labeled (see key) by the fraction of overall r-process material contributed to the galaxy by collapsars at the time of formation of the Solar System. The neutron-star merger contribution is altered by renormalizing the neutron-star merger rates, tuning $f_\mathrm{NS} R_\mathrm{NS}(z=0)$ (compare Eq.~\ref{eq:M_Eu}) by a factor between 0.3 and 100. The fiducial model in Extended Data Fig.~\ref{fig:chemical_evolution} corresponds to the red curve. A dominant contribution to the total Galactic r-process from collapsars improves the evolution of r-process enrichment at high metallicity relative to merger-only models.}
\label{fig:chemical_evolution_eta_merger}
\end{figure}

\clearpage

\thispagestyle{empty}

\renewcommand{\tablename}{Supplementary Table}

\begin{table}
\caption{{\bf Collapsar accretion disk ejecta.} Listed are the amount of collapsar disk wind ejecta (in units of $M_\odot$) from circularization of fallback material during the different accretion stages (see Fig. 1; Methods) for various presupernova models\protect\citemeth{Heger2000a}, with names and masses listed in the two leftmost columns. Masses are calculated assuming a fraction $f_{w} = 0.3$ of inflowing mass is ejected in disk winds. The $^{56}$Ni masses are calculated using the model for nickel disk winds at late times (see Methods).}
\label{tab:fb_masses}
\centering
\begin{tabular}{ccccccc}
\hline\hline 
model & $M_\mathrm{in}$ & GRB & $r$-process & light $r$-process & $^{56}$Ni \\
\hline
E15 & 10.86 & 0.87 & 0.90 & 0.12 & 0.059 \\
E20 & 11.01 & 1.77 & 1.93 & 0.10 & 0.024 \\
E25 & 5.45 & 1.61 & 1.56 & 0.03 & 0.002 \\
E15B & 13.26 & 0.10 & 0.02 & 0.31 & 0.17 \\
E20B & 15.19 & 0.70 & 0.84 & 0.40 & 0.12 \\
F15B & 12.89 & 0.30 & 0.17 & 0.38 & 0.15 \\
F20B & 14.76 & 0.69 & 0.99 & 0.37 & 0.11 \\
\hline 
\end{tabular}
\end{table}


\begin{thebibliography}{10}
\expandafter\ifx\csname url\endcsname\relax
  \def\url#1{\texttt{#1}}\fi
\expandafter\ifx\csname urlprefix\endcsname\relax\def\urlprefix{URL }\fi
\providecommand{\bibinfo}[2]{#2}
\providecommand{\eprint}[2][]{\url{#2}}

\bibitem{LIGO+17DISCOVERY}
\bibinfo{author}{{Abbott}, B.~P.} \emph{et~al.}
\newblock \bibinfo{title}{{GW170817: Observation of Gravitational Waves from a
  Binary Neutron Star Inspiral}}.
\newblock \emph{\bibinfo{journal}{Physical Review Letters}}
  \textbf{\bibinfo{volume}{119}}, \bibinfo{pages}{161101}
  (\bibinfo{year}{2017}).

\bibitem{Coulter2017}
\bibinfo{author}{{Coulter}, D.~A.} \emph{et~al.}
\newblock \bibinfo{title}{{Swope Supernova Survey 2017a (SSS17a), the optical
  counterpart to a gravitational wave source}}.
\newblock \emph{\bibinfo{journal}{Science}} \textbf{\bibinfo{volume}{358}},
  \bibinfo{pages}{1556--1558} (\bibinfo{year}{2017}).

\bibitem{Soares-Santos+17}
\bibinfo{author}{{Soares-Santos}, M.} \emph{et~al.}
\newblock \bibinfo{title}{{The Electromagnetic Counterpart of the Binary
  Neutron Star Merger LIGO/Virgo GW170817. I. Discovery of the Optical
  Counterpart Using the Dark Energy Camera}}.
\newblock \emph{\bibinfo{journal}{\apjl}} \textbf{\bibinfo{volume}{848}},
  \bibinfo{pages}{L16} (\bibinfo{year}{2017}).

\bibitem{Kasen+17}
\bibinfo{author}{{Kasen}, D.}, \bibinfo{author}{{Metzger}, B.},
  \bibinfo{author}{{Barnes}, J.}, \bibinfo{author}{{Quataert}, E.} \&
  \bibinfo{author}{{Ramirez-Ruiz}, E.}
\newblock \bibinfo{title}{{Origin of the heavy elements in binary neutron-star
  mergers from a gravitational-wave event}}.
\newblock \emph{\bibinfo{journal}{\nat}} \textbf{\bibinfo{volume}{551}},
  \bibinfo{pages}{80--84} (\bibinfo{year}{2017}).

\bibitem{Cowperthwaite2017}
\bibinfo{author}{{Cowperthwaite}, P.~S.} \emph{et~al.}
\newblock \bibinfo{title}{{The Electromagnetic Counterpart of the Binary
  Neutron Star Merger LIGO/Virgo GW170817. II. UV, Optical, and Near-infrared
  Light Curves and Comparison to Kilonova Models}}.
\newblock \emph{\bibinfo{journal}{\apjl}} \textbf{\bibinfo{volume}{848}},
  \bibinfo{pages}{L17} (\bibinfo{year}{2017}).

\bibitem{Radice2018a}
\bibinfo{author}{{Radice}, D.}, \bibinfo{author}{{Perego}, A.},
  \bibinfo{author}{{Zappa}, F.} \& \bibinfo{author}{{Bernuzzi}, S.}
\newblock \bibinfo{title}{{GW170817: Joint Constraint on the Neutron Star
  Equation of State from Multimessenger Observations}}.
\newblock \emph{\bibinfo{journal}{\apjl}} \textbf{\bibinfo{volume}{852}},
  \bibinfo{pages}{L29} (\bibinfo{year}{2018}).

\bibitem{Fernandez&Metzger13}
\bibinfo{author}{{Fern{\'a}ndez}, R.} \& \bibinfo{author}{{Metzger}, B.~D.}
\newblock \bibinfo{title}{{Delayed outflows from black hole accretion tori
  following neutron star binary coalescence}}.
\newblock \emph{\bibinfo{journal}{Mon. Not. R. Astron. Soc.}}
  \textbf{\bibinfo{volume}{435}}, \bibinfo{pages}{502--517}
  (\bibinfo{year}{2013}).

\bibitem{Just+15}
\bibinfo{author}{{Just}, O.}, \bibinfo{author}{{Bauswein}, A.},
  \bibinfo{author}{{Pulpillo}, R.~A.}, \bibinfo{author}{{Goriely}, S.} \&
  \bibinfo{author}{{Janka}, H.-T.}
\newblock \bibinfo{title}{{Comprehensive nucleosynthesis analysis for ejecta of
  compact binary mergers}}.
\newblock \emph{\bibinfo{journal}{Mon. Not. R. Astron. Soc.}}
  \textbf{\bibinfo{volume}{448}}, \bibinfo{pages}{541--567}
  (\bibinfo{year}{2015}).

\bibitem{Siegel2017a}
\bibinfo{author}{{Siegel}, D.~M.} \& \bibinfo{author}{{Metzger}, B.~D.}
\newblock \bibinfo{title}{{Three-Dimensional General-Relativistic
  Magnetohydrodynamic Simulations of Remnant Accretion Disks from Neutron Star
  Mergers: Outflows and r -Process Nucleosynthesis}}.
\newblock \emph{\bibinfo{journal}{\prl}} \textbf{\bibinfo{volume}{119}},
  \bibinfo{pages}{231102} (\bibinfo{year}{2017}).

\bibitem{Fernandez+18}
\bibinfo{author}{{Fern{\'a}ndez}, R.}, \bibinfo{author}{{Tchekhovskoy}, A.},
  \bibinfo{author}{{Quataert}, E.}, \bibinfo{author}{{Foucart}, F.} \&
  \bibinfo{author}{{Kasen}, D.}
\newblock \bibinfo{title}{{Long-term GRMHD Simulations of Neutron Star Merger
  Accretion Disks: Implications for Electromagnetic Counterparts}}.
\newblock \emph{\bibinfo{journal}{\mnras}}  (\bibinfo{year}{2018}).

\bibitem{Siegel2018a}
\bibinfo{author}{{Siegel}, D.~M.} \& \bibinfo{author}{{Metzger}, B.~D.}
\newblock \bibinfo{title}{{Three-dimensional GRMHD simulations of
  neutrino-cooled accretion disks from neutron star mergers}}.
\newblock \emph{\bibinfo{journal}{\apj}} \textbf{\bibinfo{volume}{858}},
  \bibinfo{pages}{52} (\bibinfo{year}{2018}).

\bibitem{MacFadyen&Woosley99}
\bibinfo{author}{{MacFadyen}, A.~I.} \& \bibinfo{author}{{Woosley}, S.~E.}
\newblock \bibinfo{title}{{Collapsars: Gamma-Ray Bursts and Explosions in
  ``Failed Supernovae''}}.
\newblock \emph{\bibinfo{journal}{\apj}} \textbf{\bibinfo{volume}{524}},
  \bibinfo{pages}{262--289} (\bibinfo{year}{1999}).

\bibitem{Kohri+05}
\bibinfo{author}{{Kohri}, K.}, \bibinfo{author}{{Narayan}, R.} \&
  \bibinfo{author}{{Piran}, T.}
\newblock \bibinfo{title}{{Neutrino-dominated Accretion and Supernovae}}.
\newblock \emph{\bibinfo{journal}{\apj}} \textbf{\bibinfo{volume}{629}},
  \bibinfo{pages}{341--361} (\bibinfo{year}{2005}).

\bibitem{Ji+16}
\bibinfo{author}{{Ji}, A.~P.}, \bibinfo{author}{{Frebel}, A.},
  \bibinfo{author}{{Chiti}, A.} \& \bibinfo{author}{{Simon}, J.~D.}
\newblock \bibinfo{title}{{R-process enrichment from a single event in an
  ancient dwarf galaxy}}.
\newblock \emph{\bibinfo{journal}{Nature}} \textbf{\bibinfo{volume}{531}},
  \bibinfo{pages}{610--613} (\bibinfo{year}{2016}).

\bibitem{Cote2017b}
\bibinfo{author}{{C{\^o}t{\'e}}, B.} \emph{et~al.}
\newblock \bibinfo{title}{{Advanced LIGO Constraints on Neutron Star Mergers
  and r-process Sites}}.
\newblock \emph{\bibinfo{journal}{\apj}} \textbf{\bibinfo{volume}{836}},
  \bibinfo{pages}{230} (\bibinfo{year}{2017}).

\bibitem{Hotokezaka2018a}
\bibinfo{author}{{Hotokezaka}, K.}, \bibinfo{author}{{Beniamini}, P.} \&
  \bibinfo{author}{{Piran}, T.}
\newblock \bibinfo{title}{{Neutron star mergers as sites of r-process
  nucleosynthesis and short gamma-ray bursts}}.
\newblock \emph{\bibinfo{journal}{Intern. J. Mod. Phys. D}}
  \textbf{\bibinfo{volume}{27}}, \bibinfo{pages}{1842005}
  (\bibinfo{year}{2018}).

\bibitem{Pruet+04}
\bibinfo{author}{{Pruet}, J.}, \bibinfo{author}{{Thompson}, T.~A.} \&
  \bibinfo{author}{{Hoffman}, R.~D.}
\newblock \bibinfo{title}{{Nucleosynthesis in Outflows from the Inner Regions
  of Collapsars}}.
\newblock \emph{\bibinfo{journal}{\apj}} \textbf{\bibinfo{volume}{606}},
  \bibinfo{pages}{1006--1018} (\bibinfo{year}{2004}).

\bibitem{Surman+11}
\bibinfo{author}{{Surman}, R.}, \bibinfo{author}{{McLaughlin}, G.~C.} \&
  \bibinfo{author}{{Sabbatino}, N.}
\newblock \bibinfo{title}{{Nucleosynthesis of Nickel-56 from Gamma-Ray Burst
  Accretion Disks}}.
\newblock \emph{\bibinfo{journal}{\apj}} \textbf{\bibinfo{volume}{743}},
  \bibinfo{pages}{155} (\bibinfo{year}{2011}).

\bibitem{Beloborodov03}
\bibinfo{author}{{Beloborodov}, A.~M.}
\newblock \bibinfo{title}{{Nuclear Composition of Gamma-Ray Burst Fireballs}}.
\newblock \emph{\bibinfo{journal}{\apj}} \textbf{\bibinfo{volume}{588}},
  \bibinfo{pages}{931--944} (\bibinfo{year}{2003}).

\bibitem{Dessart+08}
\bibinfo{author}{{Dessart}, L.}, \bibinfo{author}{{Burrows}, A.},
  \bibinfo{author}{{Livne}, E.} \& \bibinfo{author}{{Ott}, C.~D.}
\newblock \bibinfo{title}{{The Proto-Neutron Star Phase of the Collapsar Model
  and the Route to Long-Soft Gamma-Ray Bursts and Hypernovae}}.
\newblock \emph{\bibinfo{journal}{\apjl}} \textbf{\bibinfo{volume}{673}},
  \bibinfo{pages}{L43--L46} (\bibinfo{year}{2008}).

\bibitem{Moesta2018}
\bibinfo{author}{{M{\"o}sta}, P.} \emph{et~al.}
\newblock \bibinfo{title}{{r-process Nucleosynthesis from Three-dimensional
  Magnetorotational Core-collapse Supernovae}}.
\newblock \emph{\bibinfo{journal}{\apj}} \textbf{\bibinfo{volume}{864}},
  \bibinfo{pages}{171} (\bibinfo{year}{2018}).

\bibitem{Ghirlanda+09}
\bibinfo{author}{{Ghirlanda}, G.}, \bibinfo{author}{{Nava}, L.},
  \bibinfo{author}{{Ghisellini}, G.}, \bibinfo{author}{{Celotti}, A.} \&
  \bibinfo{author}{{Firmani}, C.}
\newblock \bibinfo{title}{{Short versus long gamma-ray bursts: spectra,
  energetics, and luminosities}}.
\newblock \emph{\bibinfo{journal}{\aap}} \textbf{\bibinfo{volume}{496}},
  \bibinfo{pages}{585--595} (\bibinfo{year}{2009}).

\bibitem{Kasen2013}
\bibinfo{author}{{Kasen}, D.}, \bibinfo{author}{{Badnell}, N.~R.} \&
  \bibinfo{author}{{Barnes}, J.}
\newblock \bibinfo{title}{{Opacities and Spectra of the r-process Ejecta from
  Neutron Star Mergers}}.
\newblock \emph{\bibinfo{journal}{\apj}} \textbf{\bibinfo{volume}{774}},
  \bibinfo{pages}{25} (\bibinfo{year}{2013}).

\bibitem{Winteler+12}
\bibinfo{author}{{Winteler}, C.} \emph{et~al.}
\newblock \bibinfo{title}{{Magnetorotationally Driven Supernovae as the Origin
  of Early Galaxy $r$-process Elements?}}
\newblock \emph{\bibinfo{journal}{Astrophys. J. Lett.}}
  \textbf{\bibinfo{volume}{750}}, \bibinfo{pages}{L22} (\bibinfo{year}{2012}).

\bibitem{Shen+15}
\bibinfo{author}{{Shen}, S.} \emph{et~al.}
\newblock \bibinfo{title}{{The History of $r$-Process Enrichment in the Milky
  Way}}.
\newblock \emph{\bibinfo{journal}{Astrophys. J.}}
  \textbf{\bibinfo{volume}{807}}, \bibinfo{pages}{115} (\bibinfo{year}{2015}).

\bibitem{vandeVoort+15}
\bibinfo{author}{{van de Voort}, F.}, \bibinfo{author}{{Quataert}, E.},
  \bibinfo{author}{{Hopkins}, P.~F.}, \bibinfo{author}{{Kere{\v s}}, D.} \&
  \bibinfo{author}{{Faucher-Gigu{\`e}re}, C.-A.}
\newblock \bibinfo{title}{{Galactic $r$-process enrichment by neutron star
  mergers in cosmological simulations of a {M}ilky {W}ay-mass galaxy}}.
\newblock \emph{\bibinfo{journal}{Mon. Not. R. Astron. Soc.}}
  \textbf{\bibinfo{volume}{447}}, \bibinfo{pages}{140--148}
  (\bibinfo{year}{2015}).

\bibitem{RamirezRuiz+15}
\bibinfo{author}{{Ramirez-Ruiz}, E.} \emph{et~al.}
\newblock \bibinfo{title}{{Compact Stellar Binary Assembly in the First Nuclear
  Star Clusters and r-process Synthesis in the Early Universe}}.
\newblock \emph{\bibinfo{journal}{\apjl}} \textbf{\bibinfo{volume}{802}},
  \bibinfo{pages}{L22} (\bibinfo{year}{2015}).

\bibitem{Stanek+06}
\bibinfo{author}{{Stanek}, K.~Z.} \emph{et~al.}
\newblock \bibinfo{title}{{Protecting Life in the Milky Way: Metals Keep the
  GRBs Away}}.
\newblock \emph{\bibinfo{journal}{\actaa}} \textbf{\bibinfo{volume}{56}},
  \bibinfo{pages}{333--345} (\bibinfo{year}{2006}).

\bibitem{Sneden+03}
\bibinfo{author}{{Sneden}, C.} \emph{et~al.}
\newblock \bibinfo{title}{{The Extremely Metal-poor, Neutron Capture-rich Star
  CS 22892-052: A Comprehensive Abundance Analysis}}.
\newblock \emph{\bibinfo{journal}{\apj}} \textbf{\bibinfo{volume}{591}},
  \bibinfo{pages}{936--953} (\bibinfo{year}{2003}).

\bibitem{Beniamini+16}
\bibinfo{author}{{Beniamini}, P.}, \bibinfo{author}{{Hotokezaka}, K.} \&
  \bibinfo{author}{{Piran}, T.}
\newblock \bibinfo{title}{{Natal Kicks and Time Delays in Merging Neutron Star
  Binaries: {I}mplications for $r$-process Nucleosynthesis in Ultra-faint
  Dwarfs and in the Milky Way}}.
\newblock \emph{\bibinfo{journal}{Astrophys. J. Lett.}}
  \textbf{\bibinfo{volume}{829}}, \bibinfo{pages}{L13} (\bibinfo{year}{2016}).

\end{thebibliography}


\begin{thebibliography}{100}
\expandafter\ifx\csname url\endcsname\relax
  \def\url#1{\texttt{#1}}\fi
\expandafter\ifx\csname urlprefix\endcsname\relax\def\urlprefix{URL }\fi
\providecommand{\bibinfo}[2]{#2}
\providecommand{\eprint}[2][]{\url{#2}}

\bibitem{Siegel2018b}
\bibinfo{author}{{Siegel}, D.~M.}, \bibinfo{author}{{M{\"o}sta}, P.},
  \bibinfo{author}{{Desai}, D.} \& \bibinfo{author}{{Wu}, S.}
\newblock \bibinfo{title}{{Recovery Schemes for Primitive Variables in
  General-relativistic Magnetohydrodynamics}}.
\newblock \emph{\bibinfo{journal}{\apj}} \textbf{\bibinfo{volume}{859}},
  \bibinfo{pages}{71} (\bibinfo{year}{2018}).

\bibitem{Moesta2014a}
\bibinfo{author}{{M{\"o}sta}, P.} \emph{et~al.}
\newblock \bibinfo{title}{{GRHydro: a new open-source general-relativistic
  magnetohydrodynamics code for the Einstein toolkit}}.
\newblock \emph{\bibinfo{journal}{\cqg}} \textbf{\bibinfo{volume}{31}},
  \bibinfo{pages}{015005} (\bibinfo{year}{2014}).

\bibitem{Loeffler2012}
\bibinfo{author}{{L{\"o}ffler}, F.} \emph{et~al.}
\newblock \bibinfo{title}{{The Einstein Toolkit: a community computational
  infrastructure for relativistic astrophysics}}.
\newblock \emph{\bibinfo{journal}{\cqg}} \textbf{\bibinfo{volume}{29}},
  \bibinfo{pages}{115001} (\bibinfo{year}{2012}).

\bibitem{Timmes1999}
\bibinfo{author}{{Timmes}, F.~X.} \& \bibinfo{author}{{Arnett}, D.}
\newblock \bibinfo{title}{{The Accuracy, Consistency, and Speed of Five
  Equations of State for Stellar Hydrodynamics}}.
\newblock \emph{\bibinfo{journal}{\apjs}} \textbf{\bibinfo{volume}{125}},
  \bibinfo{pages}{277--294} (\bibinfo{year}{1999}).

\bibitem{Timmes2000}
\bibinfo{author}{{Timmes}, F.~X.} \& \bibinfo{author}{{Swesty}, F.~D.}
\newblock \bibinfo{title}{{The Accuracy, Consistency, and Speed of an
  Electron-Positron Equation of State Based on Table Interpolation of the
  Helmholtz Free Energy}}.
\newblock \emph{\bibinfo{journal}{\apjs}} \textbf{\bibinfo{volume}{126}},
  \bibinfo{pages}{501--516} (\bibinfo{year}{2000}).

\bibitem{DiMatteo+02}
\bibinfo{author}{{Di Matteo}, T.}, \bibinfo{author}{{Perna}, R.} \&
  \bibinfo{author}{{Narayan}, R.}
\newblock \bibinfo{title}{{Neutrino Trapping and Accretion Models for Gamma-Ray
  Bursts}}.
\newblock \emph{\bibinfo{journal}{\apj}} \textbf{\bibinfo{volume}{579}},
  \bibinfo{pages}{706--715} (\bibinfo{year}{2002}).

\bibitem{Lippuner2017b}
\bibinfo{author}{{Lippuner}, J.} \& \bibinfo{author}{{Roberts}, L.~F.}
\newblock \bibinfo{title}{{SkyNet: A Modular Nuclear Reaction Network
  Library}}.
\newblock \emph{\bibinfo{journal}{\apjs}} \textbf{\bibinfo{volume}{233}},
  \bibinfo{pages}{18} (\bibinfo{year}{2017}).

\bibitem{Velikhov1959}
\bibinfo{author}{{Velikhov}, E.~P.}
\newblock \bibinfo{title}{Stability of an ideally conducting liquid flowing
  between cylinders rotating in a magnetic field}.
\newblock \emph{\bibinfo{journal}{Sov. Phys. JETP}}
  \textbf{\bibinfo{volume}{36}}, \bibinfo{pages}{995--998}
  (\bibinfo{year}{1959}).

\bibitem{Chandrasekhar1960}
\bibinfo{author}{{Chandrasekhar}, S.}
\newblock \bibinfo{title}{{The Stability of Non-Dissipative Couette Flow in
  Hydromagnetics}}.
\newblock \emph{\bibinfo{journal}{Proc. Natl. Acad. Sci.}}
  \textbf{\bibinfo{volume}{46}}, \bibinfo{pages}{253--257}
  (\bibinfo{year}{1960}).

\bibitem{Balbus1991}
\bibinfo{author}{{Balbus}, S.~A.} \& \bibinfo{author}{{Hawley}, J.~F.}
\newblock \bibinfo{title}{{A powerful local shear instability in weakly
  magnetized disks. I - Linear analysis. II - Nonlinear evolution}}.
\newblock \emph{\bibinfo{journal}{\apj}} \textbf{\bibinfo{volume}{376}},
  \bibinfo{pages}{214--233} (\bibinfo{year}{1991}).

\bibitem{Balbus1998}
\bibinfo{author}{{Balbus}, S.~A.} \& \bibinfo{author}{{Hawley}, J.~F.}
\newblock \bibinfo{title}{{Instability, turbulence, and enhanced transport in
  accretion disks}}.
\newblock \emph{\bibinfo{journal}{\rmph}} \textbf{\bibinfo{volume}{70}},
  \bibinfo{pages}{1--53} (\bibinfo{year}{1998}).

\bibitem{Balbus2003}
\bibinfo{author}{{Balbus}, S.~A.}
\newblock \bibinfo{title}{{Enhanced Angular Momentum Transport in Accretion
  Disks}}.
\newblock \emph{\bibinfo{journal}{\araa}} \textbf{\bibinfo{volume}{41}},
  \bibinfo{pages}{555--597} (\bibinfo{year}{2003}).

\bibitem{Siegel2013}
\bibinfo{author}{{Siegel}, D.~M.}, \bibinfo{author}{{Ciolfi}, R.},
  \bibinfo{author}{{Harte}, A.~I.} \& \bibinfo{author}{{Rezzolla}, L.}
\newblock \bibinfo{title}{{Magnetorotational instability in relativistic
  hypermassive neutron stars}}.
\newblock \emph{\bibinfo{journal}{\prd}} \textbf{\bibinfo{volume}{87}},
  \bibinfo{pages}{121302(R)} (\bibinfo{year}{2013}).

\bibitem{Kiuchi2015a}
\bibinfo{author}{{Kiuchi}, K.} \emph{et~al.}
\newblock \bibinfo{title}{{High resolution magnetohydrodynamic simulation of
  black hole-neutron star merger: Mass ejection and short gamma ray bursts}}.
\newblock \emph{\bibinfo{journal}{\prd}} \textbf{\bibinfo{volume}{92}},
  \bibinfo{pages}{064034} (\bibinfo{year}{2015}).

\bibitem{Kiuchi2018a}
\bibinfo{author}{{Kiuchi}, K.}, \bibinfo{author}{{Kyutoku}, K.},
  \bibinfo{author}{{Sekiguchi}, Y.} \& \bibinfo{author}{{Shibata}, M.}
\newblock \bibinfo{title}{{Global simulations of strongly magnetized remnant
  massive neutron stars formed in binary neutron star mergers}}.
\newblock \emph{\bibinfo{journal}{\prd}} \textbf{\bibinfo{volume}{97}},
  \bibinfo{pages}{124039} (\bibinfo{year}{2018}).

\bibitem{Cowling1933}
\bibinfo{author}{Cowling, T.~G.}
\newblock \bibinfo{title}{The magnetic field of sunspots}.
\newblock \emph{\bibinfo{journal}{\mnras}} \textbf{\bibinfo{volume}{94}},
  \bibinfo{pages}{39--48} (\bibinfo{year}{1933}).

\bibitem{Nakamura+15}
\bibinfo{author}{{Nakamura}, K.}, \bibinfo{author}{{Kajino}, T.},
  \bibinfo{author}{{Mathews}, G.~J.}, \bibinfo{author}{{Sato}, S.} \&
  \bibinfo{author}{{Harikae}, S.}
\newblock \bibinfo{title}{{r-process nucleosynthesis in the MHD+neutrino-heated
  collapsar jet}}.
\newblock \emph{\bibinfo{journal}{\aap}} \textbf{\bibinfo{volume}{582}},
  \bibinfo{pages}{A34} (\bibinfo{year}{2015}).

\bibitem{Shakura1973}
\bibinfo{author}{{Shakura}, N.~I.} \& \bibinfo{author}{{Sunyaev}, R.~A.}
\newblock \bibinfo{title}{{Black holes in binary systems. Observational
  appearance.}}
\newblock \emph{\bibinfo{journal}{Astron. Astrophys.}}
  \textbf{\bibinfo{volume}{24}}, \bibinfo{pages}{337--355}
  (\bibinfo{year}{1973}).

\bibitem{Penna2013a}
\bibinfo{author}{{Penna}, R.~F.}, \bibinfo{author}{{S{\c a}dowski}, A.},
  \bibinfo{author}{{Kulkarni}, A.~K.} \& \bibinfo{author}{{Narayan}, R.}
\newblock \bibinfo{title}{{The Shakura-Sunyaev viscosity prescription with
  variable {$\alpha$} (r)}}.
\newblock \emph{\bibinfo{journal}{\mnras}} \textbf{\bibinfo{volume}{428}},
  \bibinfo{pages}{2255--2274} (\bibinfo{year}{2013}).

\bibitem{Popham+99}
\bibinfo{author}{{Popham}, R.}, \bibinfo{author}{{Woosley}, S.~E.} \&
  \bibinfo{author}{{Fryer}, C.}
\newblock \bibinfo{title}{{Hyperaccreting Black Holes and Gamma-Ray Bursts}}.
\newblock \emph{\bibinfo{journal}{Astrophys. J.}}
  \textbf{\bibinfo{volume}{518}}, \bibinfo{pages}{356--374}
  (\bibinfo{year}{1999}).

\bibitem{Narayan+01}
\bibinfo{author}{{Narayan}, R.}, \bibinfo{author}{{Piran}, T.} \&
  \bibinfo{author}{{Kumar}, P.}
\newblock \bibinfo{title}{{Accretion Models of Gamma-Ray Bursts}}.
\newblock \emph{\bibinfo{journal}{\apj}} \textbf{\bibinfo{volume}{557}},
  \bibinfo{pages}{949--957} (\bibinfo{year}{2001}).

\bibitem{Chen&Beloborodov07}
\bibinfo{author}{{Chen}, W.-X.} \& \bibinfo{author}{{Beloborodov}, A.~M.}
\newblock \bibinfo{title}{{Neutrino-cooled Accretion Disks around Spinning
  Black Holes}}.
\newblock \emph{\bibinfo{journal}{\apj}} \textbf{\bibinfo{volume}{657}},
  \bibinfo{pages}{383--399} (\bibinfo{year}{2007}).

\bibitem{Metzger+08c}
\bibinfo{author}{{Metzger}, B.~D.}, \bibinfo{author}{{Piro}, A.~L.} \&
  \bibinfo{author}{{Quataert}, E.}
\newblock \bibinfo{title}{{Time-dependent models of accretion discs formed from
  compact object mergers}}.
\newblock \emph{\bibinfo{journal}{\mnras}} \textbf{\bibinfo{volume}{390}},
  \bibinfo{pages}{781--797} (\bibinfo{year}{2008}).

\bibitem{MacFadyen+01}
\bibinfo{author}{{MacFadyen}, A.~I.}, \bibinfo{author}{{Woosley}, S.~E.} \&
  \bibinfo{author}{{Heger}, A.}
\newblock \bibinfo{title}{{Supernovae, Jets, and Collapsars}}.
\newblock \emph{\bibinfo{journal}{\apj}} \textbf{\bibinfo{volume}{550}},
  \bibinfo{pages}{410--425} (\bibinfo{year}{2001}).

\bibitem{Mazzali+06}
\bibinfo{author}{{Mazzali}, P.~A.} \emph{et~al.}
\newblock \bibinfo{title}{{Models for the Type Ic Hypernova SN 2003lw
  associated with GRB 031203}}.
\newblock \emph{\bibinfo{journal}{\apj}} \textbf{\bibinfo{volume}{645}},
  \bibinfo{pages}{1323--1330} (\bibinfo{year}{2006}).

\bibitem{Cano+16}
\bibinfo{author}{{Cano}, Z.}, \bibinfo{author}{{Johansson Andreas}, K.~G.} \&
  \bibinfo{author}{{Maeda}, K.}
\newblock \bibinfo{title}{{A self-consistent analytical magnetar model: the
  luminosity of {$\gamma$}-ray burst supernovae is powered by radioactivity}}.
\newblock \emph{\bibinfo{journal}{\mnras}} \textbf{\bibinfo{volume}{457}},
  \bibinfo{pages}{2761--2772} (\bibinfo{year}{2016}).

\bibitem{Woosley1992a}
\bibinfo{author}{{Woosley}, S.~E.} \& \bibinfo{author}{{Hoffman}, R.~D.}
\newblock \bibinfo{title}{{The alpha-process and the r-process}}.
\newblock \emph{\bibinfo{journal}{\apj}} \textbf{\bibinfo{volume}{395}},
  \bibinfo{pages}{202--239} (\bibinfo{year}{1992}).

\bibitem{Roberts2010}
\bibinfo{author}{{Roberts}, L.~F.}, \bibinfo{author}{{Woosley}, S.~E.} \&
  \bibinfo{author}{{Hoffman}, R.~D.}
\newblock \bibinfo{title}{{Integrated Nucleosynthesis in Neutrino-driven
  Winds}}.
\newblock \emph{\bibinfo{journal}{\apj}} \textbf{\bibinfo{volume}{722}},
  \bibinfo{pages}{954--967} (\bibinfo{year}{2010}).

\bibitem{Woosley&Bloom06}
\bibinfo{author}{{Woosley}, S.~E.} \& \bibinfo{author}{{Bloom}, J.~S.}
\newblock \bibinfo{title}{{The Supernova Gamma-Ray Burst Connection}}.
\newblock \emph{\bibinfo{journal}{\araa}} \textbf{\bibinfo{volume}{44}},
  \bibinfo{pages}{507--556} (\bibinfo{year}{2006}).

\bibitem{Qian&Woosley96}
\bibinfo{author}{{Qian}, Y.} \& \bibinfo{author}{{Woosley}, S.~E.}
\newblock \bibinfo{title}{{Nucleosynthesis in Neutrino-driven Winds. {I}. The
  Physical Conditions}}.
\newblock \emph{\bibinfo{journal}{Astrophys. J.}}
  \textbf{\bibinfo{volume}{471}}, \bibinfo{pages}{331} (\bibinfo{year}{1996}).

\bibitem{Thompson+01}
\bibinfo{author}{{Thompson}, T.~A.}, \bibinfo{author}{{Burrows}, A.} \&
  \bibinfo{author}{{Meyer}, B.~S.}
\newblock \bibinfo{title}{{The Physics of Proto-Neutron Star Winds:
  Implications for $r$-Process Nucleosynthesis}}.
\newblock \emph{\bibinfo{journal}{Astrophys. J.}}
  \textbf{\bibinfo{volume}{562}}, \bibinfo{pages}{887--908}
  (\bibinfo{year}{2001}).

\bibitem{Roberts+12}
\bibinfo{author}{{Roberts}, L.~F.}, \bibinfo{author}{{Reddy}, S.} \&
  \bibinfo{author}{{Shen}, G.}
\newblock \bibinfo{title}{{Medium modification of the charged-current neutrino
  opacity and its implications}}.
\newblock \emph{\bibinfo{journal}{Phys. Rev. C}} \textbf{\bibinfo{volume}{86}},
  \bibinfo{pages}{065803} (\bibinfo{year}{2012}).

\bibitem{MartinezPinedo+12}
\bibinfo{author}{{Mart{\'{\i}}nez-Pinedo}, G.}, \bibinfo{author}{{Fischer},
  T.}, \bibinfo{author}{{Lohs}, A.} \& \bibinfo{author}{{Huther}, L.}
\newblock \bibinfo{title}{{Charged-Current Weak Interaction Processes in Hot
  and Dense Matter and its Impact on the Spectra of Neutrinos Emitted from
  Protoneutron Star Cooling}}.
\newblock \emph{\bibinfo{journal}{Phys. Rev. Lett.}}
  \textbf{\bibinfo{volume}{109}}, \bibinfo{pages}{251104}
  (\bibinfo{year}{2012}).

\bibitem{Thompson03}
\bibinfo{author}{{Thompson}, T.~A.}
\newblock \bibinfo{title}{{Magnetic Protoneutron Star Winds and $r$-Process
  Nucleosynthesis}}.
\newblock \emph{\bibinfo{journal}{Astrophys. J. Lett.}}
  \textbf{\bibinfo{volume}{585}}, \bibinfo{pages}{L33--L36}
  (\bibinfo{year}{2003}).

\bibitem{Thompson&UdDoula18}
\bibinfo{author}{{Thompson}, T.~A.} \& \bibinfo{author}{{ud-Doula}, A.}
\newblock \bibinfo{title}{{High-entropy ejections from magnetized proto-neutron
  star winds: implications for heavy element nucleosynthesis}}.
\newblock \emph{\bibinfo{journal}{\mnras}} \textbf{\bibinfo{volume}{476}},
  \bibinfo{pages}{5502--5515} (\bibinfo{year}{2018}).

\bibitem{Wallner+15}
\bibinfo{author}{{Wallner}, A.} \emph{et~al.}
\newblock \bibinfo{title}{{Abundance of live $^{244}$Pu in deep-sea reservoirs
  on Earth points to rarity of actinide nucleosynthesis}}.
\newblock \emph{\bibinfo{journal}{Nature Commun.}}
  \textbf{\bibinfo{volume}{6}}, \bibinfo{pages}{5956} (\bibinfo{year}{2015}).

\bibitem{Macias&RamirezRuiz18}
\bibinfo{author}{{Macias}, P.} \& \bibinfo{author}{{Ramirez-Ruiz}, E.}
\newblock \bibinfo{title}{{A Stringent Limit on the Mass Production Rate of
  r-process Elements in the Milky Way}}.
\newblock \emph{\bibinfo{journal}{\apj}} \textbf{\bibinfo{volume}{860}},
  \bibinfo{pages}{89} (\bibinfo{year}{2018}).

\bibitem{Thompson+04}
\bibinfo{author}{{Thompson}, T.~A.}, \bibinfo{author}{{Chang}, P.} \&
  \bibinfo{author}{{Quataert}, E.}
\newblock \bibinfo{title}{{Magnetar Spin-Down, Hyperenergetic Supernovae, and
  Gamma-Ray Bursts}}.
\newblock \emph{\bibinfo{journal}{Astrophys. J.}}
  \textbf{\bibinfo{volume}{611}}, \bibinfo{pages}{380--393}
  (\bibinfo{year}{2004}).

\bibitem{Metzger+08}
\bibinfo{author}{{Metzger}, B.~D.}, \bibinfo{author}{{Thompson}, T.~A.} \&
  \bibinfo{author}{{Quataert}, E.}
\newblock \bibinfo{title}{{On the Conditions for Neutron-rich Gamma-Ray Burst
  Outflows}}.
\newblock \emph{\bibinfo{journal}{\apj}} \textbf{\bibinfo{volume}{676}},
  \bibinfo{pages}{1130--1150} (\bibinfo{year}{2008}).

\bibitem{Mosta+14}
\bibinfo{author}{{M{\"o}sta}, P.} \emph{et~al.}
\newblock \bibinfo{title}{{Magnetorotational Core-collapse Supernovae in Three
  Dimensions}}.
\newblock \emph{\bibinfo{journal}{Astrophys. J. Lett.}}
  \textbf{\bibinfo{volume}{785}}, \bibinfo{pages}{L29} (\bibinfo{year}{2014}).

\bibitem{Halevi&Mosta18}
\bibinfo{author}{{Halevi}, G.} \& \bibinfo{author}{{M{\"o}sta}, P.}
\newblock \bibinfo{title}{{r-Process nucleosynthesis from three-dimensional
  jet-driven core-collapse supernovae with magnetic misalignments}}.
\newblock \emph{\bibinfo{journal}{\mnras}} \textbf{\bibinfo{volume}{477}},
  \bibinfo{pages}{2366--2375} (\bibinfo{year}{2018}).

\bibitem{Fujimoto+07}
\bibinfo{author}{{Fujimoto}, S.-i.}, \bibinfo{author}{{Hashimoto}, M.-a.},
  \bibinfo{author}{{Kotake}, K.} \& \bibinfo{author}{{Yamada}, S.}
\newblock \bibinfo{title}{{Heavy-Element Nucleosynthesis in a Collapsar}}.
\newblock \emph{\bibinfo{journal}{\apj}} \textbf{\bibinfo{volume}{656}},
  \bibinfo{pages}{382--392} (\bibinfo{year}{2007}).

\bibitem{Ono+12}
\bibinfo{author}{{Ono}, M.}, \bibinfo{author}{{Hashimoto}, M.},
  \bibinfo{author}{{Fujimoto}, S.}, \bibinfo{author}{{Kotake}, K.} \&
  \bibinfo{author}{{Yamada}, S.}
\newblock \bibinfo{title}{{Explosive Nucleosynthesis in Magnetohydrodynamical
  Jets from Collapsars. II --- Heavy-Element Nucleosynthesis of s, p,
  r-Processes}}.
\newblock \emph{\bibinfo{journal}{Prog. Theor. Phys.}}
  \textbf{\bibinfo{volume}{128}}, \bibinfo{pages}{741--765}
  (\bibinfo{year}{2012}).

\bibitem{Hayakawa&Maeda18}
\bibinfo{author}{{Hayakawa}, T.} \& \bibinfo{author}{{Maeda}, K.}
\newblock \bibinfo{title}{{A Collapsar Model with Disk Wind: Implications for
  Supernovae Associated with Gamma-Ray Bursts}}.
\newblock \emph{\bibinfo{journal}{\apj}} \textbf{\bibinfo{volume}{854}},
  \bibinfo{pages}{43} (\bibinfo{year}{2018}).

\bibitem{Soker2017}
\bibinfo{author}{{Soker}, N.} \& \bibinfo{author}{{Gilkis}, A.}
\newblock \bibinfo{title}{{Magnetar-powered Superluminous Supernovae Must First
  Be Exploded by Jets}}.
\newblock \emph{\bibinfo{journal}{\apj}} \textbf{\bibinfo{volume}{851}},
  \bibinfo{pages}{95} (\bibinfo{year}{2017}).

\bibitem{Hjorth&Bloom12a}
\bibinfo{author}{{Hjorth}, J.} \& \bibinfo{author}{{Bloom}, J.~S.}
\newblock \bibinfo{title}{{The Gamma-Ray Burst - Supernova Connection}}.
\newblock In \bibinfo{editor}{{Kouveliotou}, C.}, \bibinfo{editor}{{Wijers},
  R.~A.~M.~J.} \& \bibinfo{editor}{{Woosley}, S.} (eds.)
  \emph{\bibinfo{booktitle}{Gamma-Ray Bursts}}, Cambridge Astrophysics Series
  51, \bibinfo{pages}{169--190} (\bibinfo{publisher}{Cambridge University
  Press}, \bibinfo{address}{Cambridge, UK}, \bibinfo{year}{2012}).

\bibitem{Fujimoto+06}
\bibinfo{author}{{Fujimoto}, S.-i.}, \bibinfo{author}{{Kotake}, K.},
  \bibinfo{author}{{Yamada}, S.}, \bibinfo{author}{{Hashimoto}, M.-a.} \&
  \bibinfo{author}{{Sato}, K.}
\newblock \bibinfo{title}{{Magnetohydrodynamic Simulations of a Rotating
  Massive Star Collapsing to a Black Hole}}.
\newblock \emph{\bibinfo{journal}{\apj}} \textbf{\bibinfo{volume}{644}},
  \bibinfo{pages}{1040--1055} (\bibinfo{year}{2006}).

\bibitem{Uzdensky&MacFadyen07}
\bibinfo{author}{{Uzdensky}, D.~A.} \& \bibinfo{author}{{MacFadyen}, A.~I.}
\newblock \bibinfo{title}{{Magnetar-Driven Magnetic Tower as a Model for
  Gamma-Ray Bursts and Asymmetric Supernovae}}.
\newblock \emph{\bibinfo{journal}{\apj}} \textbf{\bibinfo{volume}{669}},
  \bibinfo{pages}{546--560} (\bibinfo{year}{2007}).

\bibitem{Morsony+07}
\bibinfo{author}{{Morsony}, B.~J.}, \bibinfo{author}{{Lazzati}, D.} \&
  \bibinfo{author}{{Begelman}, M.~C.}
\newblock \bibinfo{title}{{Temporal and Angular Properties of Gamma-Ray Burst
  Jets Emerging from Massive Stars}}.
\newblock \emph{\bibinfo{journal}{\apj}} \textbf{\bibinfo{volume}{665}},
  \bibinfo{pages}{569--598} (\bibinfo{year}{2007}).

\bibitem{Bucciantini+08}
\bibinfo{author}{{Bucciantini}, N.}, \bibinfo{author}{{Quataert}, E.},
  \bibinfo{author}{{Arons}, J.}, \bibinfo{author}{{Metzger}, B.~D.} \&
  \bibinfo{author}{{Thompson}, T.~A.}
\newblock \bibinfo{title}{{Relativistic jets and long-duration gamma-ray bursts
  from the birth of magnetars}}.
\newblock \emph{\bibinfo{journal}{\mnras}} \textbf{\bibinfo{volume}{383}},
  \bibinfo{pages}{L25--L29} (\bibinfo{year}{2008}).

\bibitem{Lazzati+08}
\bibinfo{author}{{Lazzati}, D.}, \bibinfo{author}{{Perna}, R.} \&
  \bibinfo{author}{{Begelman}, M.~C.}
\newblock \bibinfo{title}{{X-ray flares, neutrino-cooled discs and the dynamics
  of late accretion in gamma-ray burst engines}}.
\newblock \emph{\bibinfo{journal}{\mnras}} \textbf{\bibinfo{volume}{388}},
  \bibinfo{pages}{L15--L19} (\bibinfo{year}{2008}).

\bibitem{Kumar2008b}
\bibinfo{author}{{Kumar}, P.}, \bibinfo{author}{{Narayan}, R.} \&
  \bibinfo{author}{{Johnson}, J.~L.}
\newblock \bibinfo{title}{{Mass fall-back and accretion in the central engine
  of gamma-ray bursts}}.
\newblock \emph{\bibinfo{journal}{\mnras}} \textbf{\bibinfo{volume}{388}},
  \bibinfo{pages}{1729--1742} (\bibinfo{year}{2008}).

\bibitem{Nagakura+11}
\bibinfo{author}{{Nagakura}, H.}, \bibinfo{author}{{Ito}, H.},
  \bibinfo{author}{{Kiuchi}, K.} \& \bibinfo{author}{{Yamada}, S.}
\newblock \bibinfo{title}{{Jet Propagations, Breakouts, and Photospheric
  Emissions in Collapsing Massive Progenitors of Long-duration Gamma-ray
  Bursts}}.
\newblock \emph{\bibinfo{journal}{\apj}} \textbf{\bibinfo{volume}{731}},
  \bibinfo{pages}{80} (\bibinfo{year}{2011}).

\bibitem{Lindner+12}
\bibinfo{author}{{Lindner}, C.~C.}, \bibinfo{author}{{Milosavljevi{\'c}}, M.},
  \bibinfo{author}{{Shen}, R.} \& \bibinfo{author}{{Kumar}, P.}
\newblock \bibinfo{title}{{Simulations of Accretion Powered Supernovae in the
  Progenitors of Gamma-Ray Bursts}}.
\newblock \emph{\bibinfo{journal}{\apj}} \textbf{\bibinfo{volume}{750}},
  \bibinfo{pages}{163} (\bibinfo{year}{2012}).

\bibitem{Lopez-Camara+13}
\bibinfo{author}{{L{\'o}pez-C{\'a}mara}, D.}, \bibinfo{author}{{Morsony},
  B.~J.}, \bibinfo{author}{{Begelman}, M.~C.} \& \bibinfo{author}{{Lazzati},
  D.}
\newblock \bibinfo{title}{{Three-dimensional Adaptive Mesh Refinement
  Simulations of Long-duration Gamma-Ray Burst Jets inside Massive Progenitor
  Stars}}.
\newblock \emph{\bibinfo{journal}{\apj}} \textbf{\bibinfo{volume}{767}},
  \bibinfo{pages}{19} (\bibinfo{year}{2013}).

\bibitem{Batta&Lee14}
\bibinfo{author}{{Batta}, A.} \& \bibinfo{author}{{Lee}, W.~H.}
\newblock \bibinfo{title}{{Cooling-induced structure formation and evolution in
  collapsars}}.
\newblock \emph{\bibinfo{journal}{\mnras}} \textbf{\bibinfo{volume}{437}},
  \bibinfo{pages}{2412--2429} (\bibinfo{year}{2014}).

\bibitem{Mazzali+14}
\bibinfo{author}{{Mazzali}, P.~A.}, \bibinfo{author}{{McFadyen}, A.~I.},
  \bibinfo{author}{{Woosley}, S.~E.}, \bibinfo{author}{{Pian}, E.} \&
  \bibinfo{author}{{Tanaka}, M.}
\newblock \bibinfo{title}{{An upper limit to the energy of gamma-ray bursts
  indicates that GRBs/SNe are powered by magnetars}}.
\newblock \emph{\bibinfo{journal}{\mnras}} \textbf{\bibinfo{volume}{443}},
  \bibinfo{pages}{67--71} (\bibinfo{year}{2014}).

\bibitem{Maeda&Nomoto03}
\bibinfo{author}{{Maeda}, K.} \& \bibinfo{author}{{Nomoto}, K.}
\newblock \bibinfo{title}{{Bipolar Supernova Explosions: Nucleosynthesis and
  Implications for Abundances in Extremely Metal-Poor Stars}}.
\newblock \emph{\bibinfo{journal}{\apj}} \textbf{\bibinfo{volume}{598}},
  \bibinfo{pages}{1163--1200} (\bibinfo{year}{2003}).

\bibitem{Fryer+06}
\bibinfo{author}{{Fryer}, C.~L.}, \bibinfo{author}{{Young}, P.~A.} \&
  \bibinfo{author}{{Hungerford}, A.~L.}
\newblock \bibinfo{title}{{Explosive Nucleosynthesis from Gamma-Ray Burst and
  Hypernova Progenitors: Direct Collapse versus Fallback}}.
\newblock \emph{\bibinfo{journal}{\apj}} \textbf{\bibinfo{volume}{650}},
  \bibinfo{pages}{1028--1047} (\bibinfo{year}{2006}).

\bibitem{Maeda&Tominaga09}
\bibinfo{author}{{Maeda}, K.} \& \bibinfo{author}{{Tominaga}, N.}
\newblock \bibinfo{title}{{Nucleosynthesis of $^{56}$Ni in wind-driven
  supernova explosions and constraints on the central engine of gamma-ray
  bursts}}.
\newblock \emph{\bibinfo{journal}{\mnras}} \textbf{\bibinfo{volume}{394}},
  \bibinfo{pages}{1317--1324} (\bibinfo{year}{2009}).

\bibitem{Barnes+18}
\bibinfo{author}{{Barnes}, J.} \emph{et~al.}
\newblock \bibinfo{title}{{A GRB and Broad-lined Type Ic Supernova from a
  Single Central Engine}}.
\newblock \emph{\bibinfo{journal}{\apj}} \textbf{\bibinfo{volume}{860}},
  \bibinfo{pages}{38} (\bibinfo{year}{2018}).

\bibitem{Caballero2012}
\bibinfo{author}{{Caballero}, O.~L.}, \bibinfo{author}{{McLaughlin}, G.~C.} \&
  \bibinfo{author}{{Surman}, R.}
\newblock \bibinfo{title}{{Neutrino Spectra from Accretion Disks: Neutrino
  General Relativistic Effects and the Consequences for Nucleosynthesis}}.
\newblock \emph{\bibinfo{journal}{\apj}} \textbf{\bibinfo{volume}{745}},
  \bibinfo{pages}{170} (\bibinfo{year}{2012}).

\bibitem{Vlasov+17}
\bibinfo{author}{{Vlasov}, A.~D.}, \bibinfo{author}{{Metzger}, B.~D.},
  \bibinfo{author}{{Lippuner}, J.}, \bibinfo{author}{{Roberts}, L.~F.} \&
  \bibinfo{author}{{Thompson}, T.~A.}
\newblock \bibinfo{title}{{Neutrino-heated winds from millisecond
  protomagnetars as sources of the weak r-process}}.
\newblock \emph{\bibinfo{journal}{\mnras}} \textbf{\bibinfo{volume}{468}},
  \bibinfo{pages}{1522--1533} (\bibinfo{year}{2017}).

\bibitem{Heger2000a}
\bibinfo{author}{{Heger}, A.}, \bibinfo{author}{{Langer}, N.} \&
  \bibinfo{author}{{Woosley}, S.~E.}
\newblock \bibinfo{title}{{Presupernova Evolution of Rotating Massive Stars. I.
  Numerical Method and Evolution of the Internal Stellar Structure}}.
\newblock \emph{\bibinfo{journal}{\apj}} \textbf{\bibinfo{volume}{528}},
  \bibinfo{pages}{368--396} (\bibinfo{year}{2000}).

\bibitem{Bardeen1972}
\bibinfo{author}{{Bardeen}, J.~M.}, \bibinfo{author}{{Press}, W.~H.} \&
  \bibinfo{author}{{Teukolsky}, S.~A.}
\newblock \bibinfo{title}{{Rotating Black Holes: Locally Nonrotating Frames,
  Energy Extraction, and Scalar Synchrotron Radiation}}.
\newblock \emph{\bibinfo{journal}{\apj}} \textbf{\bibinfo{volume}{178}},
  \bibinfo{pages}{347--370} (\bibinfo{year}{1972}).

\bibitem{Bromberg2012a}
\bibinfo{author}{{Bromberg}, O.}, \bibinfo{author}{{Nakar}, E.},
  \bibinfo{author}{{Piran}, T.} \& \bibinfo{author}{{Sari}, R.}
\newblock \bibinfo{title}{{An Observational Imprint of the Collapsar Model of
  Long Gamma-Ray Bursts}}.
\newblock \emph{\bibinfo{journal}{\apj}} \textbf{\bibinfo{volume}{749}},
  \bibinfo{pages}{110} (\bibinfo{year}{2012}).

\bibitem{Sobacchi2017}
\bibinfo{author}{{Sobacchi}, E.}, \bibinfo{author}{{Granot}, J.},
  \bibinfo{author}{{Bromberg}, O.} \& \bibinfo{author}{{Sormani}, M.~C.}
\newblock \bibinfo{title}{{A common central engine for long gamma-ray bursts
  and Type Ib/c supernovae}}.
\newblock \emph{\bibinfo{journal}{\mnras}} \textbf{\bibinfo{volume}{472}},
  \bibinfo{pages}{616--627} (\bibinfo{year}{2017}).

\bibitem{Bhat2016}
\bibinfo{author}{{Narayana Bhat}, P.} \emph{et~al.}
\newblock \bibinfo{title}{{The Third Fermi GBM Gamma-Ray Burst Catalog: The
  First Six Years}}.
\newblock \emph{\bibinfo{journal}{\apjs}} \textbf{\bibinfo{volume}{223}},
  \bibinfo{pages}{28} (\bibinfo{year}{2016}).

\bibitem{Drout+17}
\bibinfo{author}{{Drout}, M.~R.} \emph{et~al.}
\newblock \bibinfo{title}{{Light curves of the neutron star merger
  GW170817/SSS17a: Implications for r-process nucleosynthesis}}.
\newblock \emph{\bibinfo{journal}{Science}} \textbf{\bibinfo{volume}{358}},
  \bibinfo{pages}{1570--1574} (\bibinfo{year}{2017}).

\bibitem{Tanvir+17}
\bibinfo{author}{{Tanvir}, N.~R.} \emph{et~al.}
\newblock \bibinfo{title}{{The Emergence of a Lanthanide-rich Kilonova
  Following the Merger of Two Neutron Stars}}.
\newblock \emph{\bibinfo{journal}{\apjl}} \textbf{\bibinfo{volume}{848}},
  \bibinfo{pages}{L27} (\bibinfo{year}{2017}).

\bibitem{Cote2018a}
\bibinfo{author}{{C{\^o}t{\'e}}, B.} \emph{et~al.}
\newblock \bibinfo{title}{{The Origin of r-process Elements in the Milky Way}}.
\newblock \emph{\bibinfo{journal}{\apj}} \textbf{\bibinfo{volume}{855}},
  \bibinfo{pages}{99} (\bibinfo{year}{2018}).

\bibitem{Barnes+16}
\bibinfo{author}{{Barnes}, J.}, \bibinfo{author}{{Kasen}, D.},
  \bibinfo{author}{{Wu}, M.-R.} \& \bibinfo{author}{{Mart{\'{\i}}nez-Pinedo},
  G.}
\newblock \bibinfo{title}{{Radioactivity and Thermalization in the Ejecta of
  Compact Object Mergers and Their Impact on Kilonova Light Curves}}.
\newblock \emph{\bibinfo{journal}{Astrophys. J.}}
  \textbf{\bibinfo{volume}{829}}, \bibinfo{pages}{110} (\bibinfo{year}{2016}).

\bibitem{Li+16}
\bibinfo{author}{{Li}, Y.}, \bibinfo{author}{{Zhang}, B.} \&
  \bibinfo{author}{{L{\"u}}, H.-J.}
\newblock \bibinfo{title}{{A Comparative Study of Long and Short GRBs. I.
  Overlapping Properties}}.
\newblock \emph{\bibinfo{journal}{\apjs}} \textbf{\bibinfo{volume}{227}},
  \bibinfo{pages}{7} (\bibinfo{year}{2016}).

\bibitem{Tchekhovskoy2015}
\bibinfo{author}{{Tchekhovskoy}, A.} \& \bibinfo{author}{{Giannios}, D.}
\newblock \bibinfo{title}{{Magnetic flux of progenitor stars sets gamma-ray
  burst luminosity and variability}}.
\newblock \emph{\bibinfo{journal}{\mnras}} \textbf{\bibinfo{volume}{447}},
  \bibinfo{pages}{327--344} (\bibinfo{year}{2015}).

\bibitem{Wanderman2015}
\bibinfo{author}{{Wanderman}, D.} \& \bibinfo{author}{{Piran}, T.}
\newblock \bibinfo{title}{{The rate, luminosity function and time delay of
  non-Collapsar short GRBs}}.
\newblock \emph{\bibinfo{journal}{\mnras}} \textbf{\bibinfo{volume}{448}},
  \bibinfo{pages}{3026--3037} (\bibinfo{year}{2015}).

\bibitem{Wanderman&Piran10}
\bibinfo{author}{{Wanderman}, D.} \& \bibinfo{author}{{Piran}, T.}
\newblock \bibinfo{title}{{The luminosity function and the rate of Swift's
  gamma-ray bursts}}.
\newblock \emph{\bibinfo{journal}{\mnras}} \textbf{\bibinfo{volume}{406}},
  \bibinfo{pages}{1944--1958} (\bibinfo{year}{2010}).

\bibitem{Berger14}
\bibinfo{author}{{Berger}, E.}
\newblock \bibinfo{title}{{Short-Duration Gamma-Ray Bursts}}.
\newblock \emph{\bibinfo{journal}{Annu. Rev. Astron. Astrophys.}}
  \textbf{\bibinfo{volume}{52}}, \bibinfo{pages}{43--105}
  (\bibinfo{year}{2014}).

\bibitem{Liang+07}
\bibinfo{author}{{Liang}, E.}, \bibinfo{author}{{Zhang}, B.},
  \bibinfo{author}{{Virgili}, F.} \& \bibinfo{author}{{Dai}, Z.~G.}
\newblock \bibinfo{title}{{Low-Luminosity Gamma-Ray Bursts as a Unique
  Population: Luminosity Function, Local Rate, and Beaming Factor}}.
\newblock \emph{\bibinfo{journal}{\apj}} \textbf{\bibinfo{volume}{662}},
  \bibinfo{pages}{1111--1118} (\bibinfo{year}{2007}).

\bibitem{Melandri2014}
\bibinfo{author}{{Melandri}, A.} \emph{et~al.}
\newblock \bibinfo{title}{{Diversity of gamma-ray burst energetics vs.
  supernova homogeneity: SN 2013cq associated with GRB 130427A}}.
\newblock \emph{\bibinfo{journal}{\aap}} \textbf{\bibinfo{volume}{567}},
  \bibinfo{pages}{A29} (\bibinfo{year}{2014}).

\bibitem{Arnould2007}
\bibinfo{author}{{Arnould}, M.}, \bibinfo{author}{{Goriely}, S.} \&
  \bibinfo{author}{{Takahashi}, K.}
\newblock \bibinfo{title}{{The r-process of stellar nucleosynthesis:
  Astrophysics and nuclear physics achievements and mysteries}}.
\newblock \emph{\bibinfo{journal}{\physrep}} \textbf{\bibinfo{volume}{450}},
  \bibinfo{pages}{97--213} (\bibinfo{year}{2007}).

\bibitem{Kistler+08}
\bibinfo{author}{{Kistler}, M.~D.}, \bibinfo{author}{{Y{\"u}ksel}, H.},
  \bibinfo{author}{{Beacom}, J.~F.} \& \bibinfo{author}{{Stanek}, K.~Z.}
\newblock \bibinfo{title}{{An Unexpectedly Swift Rise in the Gamma-Ray Burst
  Rate}}.
\newblock \emph{\bibinfo{journal}{\apjl}} \textbf{\bibinfo{volume}{673}},
  \bibinfo{pages}{L119--L122} (\bibinfo{year}{2008}).

\bibitem{Goldstein+16}
\bibinfo{author}{{Goldstein}, A.}, \bibinfo{author}{{Connaughton}, V.},
  \bibinfo{author}{{Briggs}, M.~S.} \& \bibinfo{author}{{Burns}, E.}
\newblock \bibinfo{title}{{Estimating Long GRB Jet Opening Angles and
  Rest-frame Energetics}}.
\newblock \emph{\bibinfo{journal}{\apj}} \textbf{\bibinfo{volume}{818}},
  \bibinfo{pages}{18} (\bibinfo{year}{2016}).

\bibitem{Perley+16}
\bibinfo{author}{{Perley}, D.~A.}, \bibinfo{author}{{Tanvir}, N.~R.},
  \bibinfo{author}{{Hjorth}, J.} \emph{et~al.}
\newblock \bibinfo{title}{{The Swift GRB Host Galaxy Legacy Survey. II.
  Rest-frame Near-IR Luminosity Distribution and Evidence for a Near-solar
  Metallicity Threshold}}.
\newblock \emph{\bibinfo{journal}{\apj}} \textbf{\bibinfo{volume}{817}},
  \bibinfo{pages}{8} (\bibinfo{year}{2016}).

\bibitem{Abolfathi+18}
\bibinfo{author}{{Abolfathi}, B.} \emph{et~al.}
\newblock \bibinfo{title}{{The Fourteenth Data Release of the Sloan Digital Sky
  Survey: First Spectroscopic Data from the Extended Baryon Oscillation
  Spectroscopic Survey and from the Second Phase of the Apache Point
  Observatory Galactic Evolution Experiment}}.
\newblock \emph{\bibinfo{journal}{\apjs}} \textbf{\bibinfo{volume}{235}},
  \bibinfo{pages}{42} (\bibinfo{year}{2018}).

\bibitem{Metzger+10}
\bibinfo{author}{{Metzger}, B.~D.} \emph{et~al.}
\newblock \bibinfo{title}{{Electromagnetic counterparts of compact object
  mergers powered by the radioactive decay of $r$-process nuclei}}.
\newblock \emph{\bibinfo{journal}{Mon. Not. R. Astron. Soc.}}
  \textbf{\bibinfo{volume}{406}}, \bibinfo{pages}{2650--2662}
  (\bibinfo{year}{2010}).

\bibitem{Tanaka+18}
\bibinfo{author}{{Tanaka}, M.} \emph{et~al.}
\newblock \bibinfo{title}{{Properties of Kilonovae from Dynamical and
  Post-merger Ejecta of Neutron Star Mergers}}.
\newblock \emph{\bibinfo{journal}{\apj}} \textbf{\bibinfo{volume}{852}},
  \bibinfo{pages}{109} (\bibinfo{year}{2018}).

\bibitem{Barnes&Kasen13}
\bibinfo{author}{{Barnes}, J.} \& \bibinfo{author}{{Kasen}, D.}
\newblock \bibinfo{title}{{Effect of a High Opacity on the Light Curves of
  Radioactively Powered Transients from Compact Object Mergers}}.
\newblock \emph{\bibinfo{journal}{Astrophys. J.}}
  \textbf{\bibinfo{volume}{775}}, \bibinfo{pages}{18} (\bibinfo{year}{2013}).

\bibitem{Tanaka&Hotokezaka13}
\bibinfo{author}{{Tanaka}, M.} \& \bibinfo{author}{{Hotokezaka}, K.}
\newblock \bibinfo{title}{{Radiative Transfer Simulations of Neutron Star
  Merger Ejecta}}.
\newblock \emph{\bibinfo{journal}{Astrophys. J.}}
  \textbf{\bibinfo{volume}{775}}, \bibinfo{pages}{113} (\bibinfo{year}{2013}).

\bibitem{Wollaeger+18}
\bibinfo{author}{{Wollaeger}, R.~T.} \emph{et~al.}
\newblock \bibinfo{title}{{Impact of ejecta morphology and composition on the
  electromagnetic signatures of neutron star mergers}}.
\newblock \emph{\bibinfo{journal}{\mnras}} \textbf{\bibinfo{volume}{478}},
  \bibinfo{pages}{3298--3334} (\bibinfo{year}{2018}).

\bibitem{Kasen+06}
\bibinfo{author}{{Kasen}, D.}, \bibinfo{author}{{Thomas}, R.~C.} \&
  \bibinfo{author}{{Nugent}, P.}
\newblock \bibinfo{title}{{Time-dependent Monte Carlo Radiative Transfer
  Calculations for Three-dimensional Supernova Spectra, Light Curves, and
  Polarization}}.
\newblock \emph{\bibinfo{journal}{\apj}} \textbf{\bibinfo{volume}{651}},
  \bibinfo{pages}{366--380} (\bibinfo{year}{2006}).

\bibitem{Kurucz&Bell95}
\bibinfo{author}{{Kurucz}, R.~L.} \& \bibinfo{author}{{Bell}, B.}
\newblock \emph{\bibinfo{title}{{Atomic line list}}} (\bibinfo{year}{1995}).

\bibitem{Bufano+12}
\bibinfo{author}{{Bufano}, F.} \emph{et~al.}
\newblock \bibinfo{title}{{The Highly Energetic Expansion of SN 2010bh
  Associated with GRB 100316D}}.
\newblock \emph{\bibinfo{journal}{\apj}} \textbf{\bibinfo{volume}{753}},
  \bibinfo{pages}{67} (\bibinfo{year}{2012}).

\bibitem{Villar+18}
\bibinfo{author}{{Villar}, V.~A.} \emph{et~al.}
\newblock \bibinfo{title}{{Spitzer Space Telescope Infrared Observations of the
  Binary Neutron Star Merger GW170817}}.
\newblock \emph{\bibinfo{journal}{\apjl}} \textbf{\bibinfo{volume}{862}},
  \bibinfo{pages}{L11} (\bibinfo{year}{2018}).

\bibitem{Wu+18}
\bibinfo{author}{{Wu}, M.-R.}, \bibinfo{author}{{Barnes}, J.},
  \bibinfo{author}{{Martinez-Pinedo}, G.} \& \bibinfo{author}{{Metzger}, B.~D.}
\newblock \bibinfo{title}{{Fingerprints of heavy element nucleosynthesis in the
  late-time lightcurves of kilonovae}}.
\newblock \emph{\bibinfo{journal}{ArXiv e-prints}}  (\bibinfo{year}{2018}).

\bibitem{Cote2017a}
\bibinfo{author}{{C{\^o}t{\'e}}, B.}, \bibinfo{author}{{O'Shea}, B.~W.},
  \bibinfo{author}{{Ritter}, C.}, \bibinfo{author}{{Herwig}, F.} \&
  \bibinfo{author}{{Venn}, K.~A.}
\newblock \bibinfo{title}{{The Impact of Modeling Assumptions in Galactic
  Chemical Evolution Models}}.
\newblock \emph{\bibinfo{journal}{\apj}} \textbf{\bibinfo{volume}{835}},
  \bibinfo{pages}{128} (\bibinfo{year}{2017}).

\bibitem{Komiya2016}
\bibinfo{author}{{Komiya}, Y.} \& \bibinfo{author}{{Shigeyama}, T.}
\newblock \bibinfo{title}{{Contribution of Neutron Star Mergers to the
  r-Process Chemical Evolution in the Hierarchical Galaxy Formation}}.
\newblock \emph{\bibinfo{journal}{\apj}} \textbf{\bibinfo{volume}{830}},
  \bibinfo{pages}{76} (\bibinfo{year}{2016}).

\bibitem{Cescutti2015}
\bibinfo{author}{{Cescutti}, G.}, \bibinfo{author}{{Romano}, D.},
  \bibinfo{author}{{Matteucci}, F.}, \bibinfo{author}{{Chiappini}, C.} \&
  \bibinfo{author}{{Hirschi}, R.}
\newblock \bibinfo{title}{{The role of neutron star mergers in the chemical
  evolution of the Galactic halo}}.
\newblock \emph{\bibinfo{journal}{\aap}} \textbf{\bibinfo{volume}{577}},
  \bibinfo{pages}{A139} (\bibinfo{year}{2015}).

\bibitem{Wehmeyer2015}
\bibinfo{author}{{Wehmeyer}, B.}, \bibinfo{author}{{Pignatari}, M.} \&
  \bibinfo{author}{{Thielemann}, F.-K.}
\newblock \bibinfo{title}{{Galactic evolution of rapid neutron capture process
  abundances: the inhomogeneous approach}}.
\newblock \emph{\bibinfo{journal}{\mnras}} \textbf{\bibinfo{volume}{452}},
  \bibinfo{pages}{1970--1981} (\bibinfo{year}{2015}).

\bibitem{Hirai2015}
\bibinfo{author}{{Hirai}, Y.} \emph{et~al.}
\newblock \bibinfo{title}{{Enrichment of r-process Elements in Dwarf Spheroidal
  Galaxies in Chemo-dynamical Evolution Model}}.
\newblock \emph{\bibinfo{journal}{\apj}} \textbf{\bibinfo{volume}{814}},
  \bibinfo{pages}{41} (\bibinfo{year}{2015}).

\bibitem{Ishimaru2015}
\bibinfo{author}{{Ishimaru}, Y.}, \bibinfo{author}{{Wanajo}, S.} \&
  \bibinfo{author}{{Prantzos}, N.}
\newblock \bibinfo{title}{{Neutron Star Mergers as the Origin of r-process
  Elements in the Galactic Halo Based on the Sub-halo Clustering Scenario}}.
\newblock \emph{\bibinfo{journal}{\apjl}} \textbf{\bibinfo{volume}{804}},
  \bibinfo{pages}{L35} (\bibinfo{year}{2015}).

\bibitem{Burris2000}
\bibinfo{author}{{Burris}, D.~L.} \emph{et~al.}
\newblock \bibinfo{title}{{Neutron-Capture Elements in the Early Galaxy:
  Insights from a Large Sample of Metal-poor Giants}}.
\newblock \emph{\bibinfo{journal}{\apj}} \textbf{\bibinfo{volume}{544}},
  \bibinfo{pages}{302--319} (\bibinfo{year}{2000}).

\bibitem{Battistini2016}
\bibinfo{author}{{Battistini}, C.} \& \bibinfo{author}{{Bensby}, T.}
\newblock \bibinfo{title}{{The origin and evolution of r- and s-process
  elements in the Milky Way stellar disk}}.
\newblock \emph{\bibinfo{journal}{\aap}} \textbf{\bibinfo{volume}{586}},
  \bibinfo{pages}{A49} (\bibinfo{year}{2016}).

\bibitem{Madau2017}
\bibinfo{author}{{Madau}, P.} \& \bibinfo{author}{{Fragos}, T.}
\newblock \bibinfo{title}{{Radiation Backgrounds at Cosmic Dawn: X-Rays from
  Compact Binaries}}.
\newblock \emph{\bibinfo{journal}{\apj}} \textbf{\bibinfo{volume}{840}},
  \bibinfo{pages}{39} (\bibinfo{year}{2017}).

\bibitem{Kopparapu2008}
\bibinfo{author}{{Kopparapu}, R.~K.} \emph{et~al.}
\newblock \bibinfo{title}{{Host Galaxies Catalog Used in LIGO Searches for
  Compact Binary Coalescence Events}}.
\newblock \emph{\bibinfo{journal}{\apj}} \textbf{\bibinfo{volume}{675}},
  \bibinfo{pages}{1459--1467} (\bibinfo{year}{2008}).

\bibitem{Li2011}
\bibinfo{author}{{Li}, W.} \emph{et~al.}
\newblock \bibinfo{title}{{Nearby supernova rates from the Lick Observatory
  Supernova Search - III. The rate-size relation, and the rates as a function
  of galaxy Hubble type and colour}}.
\newblock \emph{\bibinfo{journal}{\mnras}} \textbf{\bibinfo{volume}{412}},
  \bibinfo{pages}{1473--1507} (\bibinfo{year}{2011}).

\bibitem{Maoz2017}
\bibinfo{author}{{Maoz}, D.} \& \bibinfo{author}{{Graur}, O.}
\newblock \bibinfo{title}{{Star Formation, Supernovae, Iron, and
  {\ensuremath{\alpha}}: Consistent Cosmic and Galactic Histories}}.
\newblock \emph{\bibinfo{journal}{\apj}} \textbf{\bibinfo{volume}{848}},
  \bibinfo{pages}{25} (\bibinfo{year}{2017}).

\bibitem{Dominik2012}
\bibinfo{author}{{Dominik}, M.} \emph{et~al.}
\newblock \bibinfo{title}{{Double Compact Objects. I. The Significance of the
  Common Envelope on Merger Rates}}.
\newblock \emph{\bibinfo{journal}{\apj}} \textbf{\bibinfo{volume}{759}},
  \bibinfo{pages}{52} (\bibinfo{year}{2012}).

\bibitem{Chruslinska2018}
\bibinfo{author}{{Chruslinska}, M.}, \bibinfo{author}{{Belczynski}, K.},
  \bibinfo{author}{{Klencki}, J.} \& \bibinfo{author}{{Benacquista}, M.}
\newblock \bibinfo{title}{{Double neutron stars: merger rates revisited}}.
\newblock \emph{\bibinfo{journal}{\mnras}} \textbf{\bibinfo{volume}{474}},
  \bibinfo{pages}{2937--2958} (\bibinfo{year}{2018}).

\bibitem{Villar2017}
\bibinfo{author}{{Villar}, V.~A.} \emph{et~al.}
\newblock \bibinfo{title}{{The Combined Ultraviolet, Optical, and Near-infrared
  Light Curves of the Kilonova Associated with the Binary Neutron Star Merger
  GW170817: Unified Data Set, Analytic Models, and Physical Implications}}.
\newblock \emph{\bibinfo{journal}{\apjl}} \textbf{\bibinfo{volume}{851}},
  \bibinfo{pages}{L21} (\bibinfo{year}{2017}).

\bibitem{Suda2008}
\bibinfo{author}{{Suda}, T.} \emph{et~al.}
\newblock \bibinfo{title}{{Stellar Abundances for the Galactic Archeology
  (SAGA) Database --- Compilation of the Characteristics of Known Extremely
  Metal-Poor Stars}}.
\newblock \emph{\bibinfo{journal}{Pub. Astron. Soc. Japan}}
  \textbf{\bibinfo{volume}{60}}, \bibinfo{pages}{1159} (\bibinfo{year}{2008}).

\bibitem{Cote2018b}
\bibinfo{author}{{C{\^o}t{\'e}}, B.} \emph{et~al.}
\newblock \bibinfo{title}{{Neutron Star Mergers Might not be the Only Source of
  r-Process Elements in the Milky Way}}.
\newblock \emph{\bibinfo{journal}{ArXiv e-prints}}  (\bibinfo{year}{2018}).

\bibitem{Fong+13}
\bibinfo{author}{{Fong}, W.} \emph{et~al.}
\newblock \bibinfo{title}{{Short {GRB130603B}: {D}iscovery of a jet break in
  the optical and radio afterglows, and a mysterious late-time X-ray excess}}.
\newblock \emph{\bibinfo{journal}{Astrophys. J.}}
  \textbf{\bibinfo{volume}{780}}, \bibinfo{pages}{118} (\bibinfo{year}{2014}).

\bibitem{McMillan2011}
\bibinfo{author}{{McMillan}, P.~J.}
\newblock \bibinfo{title}{{Mass models of the Milky Way}}.
\newblock \emph{\bibinfo{journal}{\mnras}} \textbf{\bibinfo{volume}{414}},
  \bibinfo{pages}{2446--2457} (\bibinfo{year}{2011}).

\bibitem{Colgate.ea.1980_SN.lum.g_ray.dep}
\bibinfo{author}{{Colgate}, S.~A.}, \bibinfo{author}{{Petschek}, A.~G.} \&
  \bibinfo{author}{{Kriese}, J.~T.}
\newblock \bibinfo{title}{{The luminosity of type I supernovae}}.
\newblock \emph{\bibinfo{journal}{\apjl}} \textbf{\bibinfo{volume}{237}},
  \bibinfo{pages}{L81--L85} (\bibinfo{year}{1980}).

\end{thebibliography}
\end{document}